\documentclass[a4paper,12pt,dvipsnames]{report}

\usepackage[a4paper]{geometry}
\usepackage[stretch=10]{microtype} 
\usepackage{multicol} 
\usepackage{afterpage} 
\usepackage{pdfpages} 
\usepackage{xcolor} 
\usepackage{setspace} 

\usepackage{tocloft} 

\usepackage[utf8]{inputenc}
\usepackage[T1]{fontenc}
\usepackage[english]{babel}
\usepackage{indentfirst} 
\usepackage{lmodern} 

\usepackage{booktabs} 
\usepackage{array} 
\usepackage{multirow} 
\usepackage{tabularx} 
\usepackage{longtable} 

\usepackage{graphicx} 
\graphicspath{ {figures/} } 
\usepackage[font={small}]{caption} 
\usepackage{float} 
\usepackage{subcaption} 

\usepackage{amsmath}
\usepackage{amssymb}
\usepackage{mathtools}
\usepackage{bm} 
\usepackage{nicefrac} 
\usepackage{siunitx} 

\usepackage{todonotes} 
\usepackage{physics} 
\usepackage{url} 
\usepackage[makeroom]{cancel} 
\usepackage{chemformula} 
\allowdisplaybreaks[3] 
\usepackage{enumitem} 
\usepackage{lipsum} 

\usepackage[
	citestyle=numeric-comp, 
]{biblatex}

\newcommand \blankpage{
	\null
	\thispagestyle{empty}
	\newpage
} 

\def ≃{\approx} 
\def \ie{\textit{i.e.}} 
\def \eg{\textit{e.g.}} 

\begin{document}

\pagenumbering{gobble}

\begin{titlepage}
	\begin{center}
		{\setstretch{1.}
			\vspace*{1cm}
			
			\Huge
			\textbf{Padé Approximants and the analytic structure of the gluon and ghost propagators}
			
			\vspace{0.5cm}
			\LARGE
			
			\vspace{1.5cm}
			
			\textbf{Alexandre Fonseca Falcão}
			
			\vfill
			
			A thesis submitted for the degree of\\
			Master in Physics
			
			\vspace{0.8cm}
			
			\includegraphics[width=100pt]{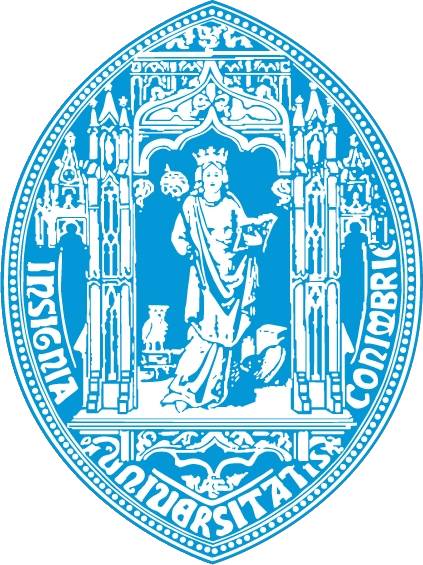}
			
			\vspace{0.8cm}
			\Large
			Department of Physics\\
			University of Coimbra\\
			Portugal\\
			\vspace{1em}
			July 2020
		}
	\end{center}
\end{titlepage}

\newpage
\blankpage{}

\newpage
\chapter*{\centering Abstract}
{\setstretch{1.1}
	In a Quantum Field Theory, the analytic structure of the 2-points correlation functions, \ie\ the propagators, encloses information about the properties of the corresponding quanta, particularly if they are or not confined. However, in Quantum Chromodynamics (QCD), we can only have an analytic solution in a perturbative picture of the theory. For the non-perturbative propagators, one resorts on numerical solutions of QCD that accesses specific regions of the Euclidean momentum space, as, for example, those computed via Monte Carlo simulations on the lattice. In the present work, we rely on Padé Approximants (PA) to approximate the numerical data for the gluon and ghost propagators, and investigate their analytic structures.
	
	In a first stage, the advantages of using PAs are explored when reproducing the properties of a function, focusing on its analytic structure. The use of PA sequences is tested for the perturbative solutions of the propagators, and a residue analysis is performed to help in the identification of the analytic structure. A technique used to approximate a PA to a discrete set of points is proposed and tested for some test data sets. Finally, the methodology is applied to the Landau gauge gluon and ghost propagators, obtained via lattice simulations. 
	
	The results identify a conjugate pair of complex poles for the gluon propagator, that is associated with the infrared structure of the theory. This is in line with the presence of singularities for complex momenta in theories where confinement is observed. Regarding the ghost propagator, a pole at $p^2=0$ is identified. For both propagators, a branch cut is found on the real negative $p^2$-axis, which recovers the perturbative analysis at high momenta.
	
	\vspace{2em}
	
	\noindent \textbf{Keywords:} Analytic structure, Padé Approximant, Gluon propagator, Ghost propagator, Lattice QCD.
}

\newpage
\blankpage{}

\newpage
\chapter*{\centering Resumo}
{\setstretch{1.1}
	Numa Teoria Quântica de Campos, a estrutura analítica das funções de correlação de 2 pontos, \ie, os propagadores, contêm diversas informações acerca das propriedades dos quanta da teoria, em particular se estes estão, ou não, confinados. No entanto, em Cromodinâmica Quântica (QCD), uma solução analítica é apenas possível num quadro perturbativo da teoria. A obtenção dos propagadores de uma forma não perturbativa pode ser feita com recurso a soluções numéricas da QCD para momentos definidos no espaço Euclidiano. Estas soluções podem ser conseguidas com base, por exemplo, em simulações de Monte Carlo na rede. Neste trabalho baseamo-nos em Aproximantes de Padé (PA) para analisar os propagadores do gluão e do campo fantasma, dessa forma obtidos na gauge de Landau, e investigamos a sua estrutura analítica.
	
	Numa primeira fase, são exploradas as vantagens do uso de PAs para reproduzir as propriedades de uma função, em especial a sua estrutura analítica. É testada a utilização de sequências de PAs nas soluções não perturbativas dos propagadores, sendo feita uma análise de resíduos como auxílio à identificação da estrutura analítica. É, também, proposta e testada uma nova técnica para aproximar um conjunto discreto de pontos a um PA, que é, por último, aplicada aos propagadores do gluão e do campo fantasma provindos de simulações na rede.
	
	Um par conjugado de polos complexos, associado à estrutura de infravermelho da teoria, é identificado no propagador do gluão, estanto de acordo com a presença de singularidades em momentos complexos em teorias nas quais se observa confinamento. Quanto ao propagador do campo fantasma, é identificado um polo em $p^2=0$. Em ambos os propagadores é identificada uma descontinuidade no eixo-$p^2$ real negativo, sendo, desta forma, recuperada a análise perturbativa a altos momentos.
	
	\vspace{1em}
	
	\noindent \textbf{Palavras-chave:} Estrutura analítica; Aproximante de Padé; Propagador do gluão, Propagador do campo fantasma, QCD na rede.
}

\newpage
\blankpage{}

\newpage
\chapter*{\centering Acknowledgements}
First and foremost, I wish to show my deepest gratitude to my supervisor, Professor Orlando Oliveira, for all the time devoted, the endless willingness, and the wise advice. I would also like to thank his research group, for having received and accompanied me over the last year.

A whole-heartedly thanks to my fellow colleagues, specially Guilherme and Maria, whose company is so spacial.

Finally, to Rúben, who read my thesis more times than I did myself, for being messy, but kind, and for being with me most of the time; to my family, who will always be there to catch me if I should fall; to my friends, with whom I share the maxima and minima of my life; and to Dirac and Duna, for having guided me through every single day of the past treacherous year, my biggest thank you.

\vspace{1em}

This work was supported with funds from the Portuguese National Budget through Fundação para a Ciência e Tecnologia under the project UIDB/04564/2020. The use of Lindgren has been provided under DECI-9 project COIMBRALATT. The use of Sisu has been provided under DECI-12 project COIMBRALATT2. I also acknowledge the Laboratory for Advanced Computing at the University of Coimbra (http://www.uc.pt/lca) for providing access to the HPC resource Navigator.

\newpage
\blankpage{}

\newpage
\pagenumbering{roman}
\setcounter{page}{9}
\tableofcontents{}

\newpage
\addcontentsline{toc}{chapter}{List of Figures}
\listoffigures

\newpage
\addcontentsline{toc}{chapter}{List of Tables}
\listoftables

\newpage
\addcontentsline{toc}{chapter}{List of Abbreviations}
\chapter*{List of Abbreviations}
\noindent \textbf{DE} Differential Evolution\par\vspace{1em}
\noindent \textbf{DSE} Dyson-Schwinger Equations\par\vspace{1em}
\noindent \textbf{PA} Padé Approximant\par\vspace{1em}
\noindent \textbf{QCD} Quantum Chromodynamics\par\vspace{1em}
\noindent \textbf{QED} Quantum Electrodynamics\par\vspace{1em}
\noindent \textbf{QFT} Quantum Field Theory\par\vspace{1em}
\noindent \textbf{RGZ} Refined Gribov-Zwanziger\par\vspace{1em}
\noindent \textbf{SA} Simulated Annealing\par\vspace{1em}
\noindent \textbf{SPM} Schlessinger Point Method\par\vspace{1em}

\newpage
\pagenumbering{arabic}

\chapter{Introduction}
The current theoretical picture of the electromagnetic interaction, a component of the electroweak part of the standard model, and the strong interaction, between quarks and gluons, boils down to Quantum Electrodynamics (QED) and Quantum Chromodynamics (QCD), respectively. Both are gauge theories associated with different gauge groups. QED is an abelian gauge theory associated with the symmetry group U(1), whilst QCD is a non-abelian gauge theory with the symmetry group SU(3). The fundamental quanta of QED, \eg\ the electron and the photon, are experimentally observed particles, whereas the quanta of QCD are not. Indeed, single particle states associated with quarks and gluons were never observed experimentally. It is believed that quarks and gluon states do not belong to the Hilbert space of the physical states. Therefore, quarks and gluons can only be present in Nature as components of other particles, \ie, they are confined particles.

In Quantum Field Theories (QFT), the 2-point correlation functions, \ie, the propagators, summarise the dynamical information of the theory. In the QED, that can be solved via perturbation theory, these propagators are well known, see, \eg, \cite{Ryder,Peskin}. Unfortunately, the same approach cannot be followed in QCD, where perturbative techniques can only be applied to the ultraviolet (UV) momentum region. Additionally, since these quanta cannot be experimentally observed alone, their behaviour and properties cannot be directly measured. Hence, the full knowledge of the gluon, quark and the unphysical ghost dynamics has to be acquired by means of theoretical \textit{ab initio} non-perturbative methods.

There are mainly two such methods that are commonly applied to investigate the non-perturbative regime of QCD: the Dyson-Schwinger Equations (DSE), and lattice regularised Monte Carlo simulations (lattice QCD). Both offer non-perturbative solutions for QCD in the whole range of momentum, but both have limitations. Although the DSE promise us an exact solution for the theory, an infinite system of coupled integral equations has to be solved, and a self-consistent truncation scheme needs to be applied in such a way that the important properties and quantities are not compromised. Regarding the lattice calculations, they are limited by the finite volume of the lattice. Notwithstanding, these non-perturbative methods offer valuable information in the infrared (IR) momentum region, a region of momenta that is not accessed with perturbation theory.

Most non-perturbative methods, including the ones above, are formulated in the Euclidean space. However, the real theory lives in the Minkowski space, where the observables are to be computed. Hence, the Wightman functions (Minkowski space correlation functions) must be obtained from the Schwinger functions (Euclidean space correlation functions) via a Wick rotation. This can only be done if the analytic structures of these functions are known.

\vspace{1em}

\begin{figure}[tp]
	\centering
	\includegraphics[width=.8\textwidth]{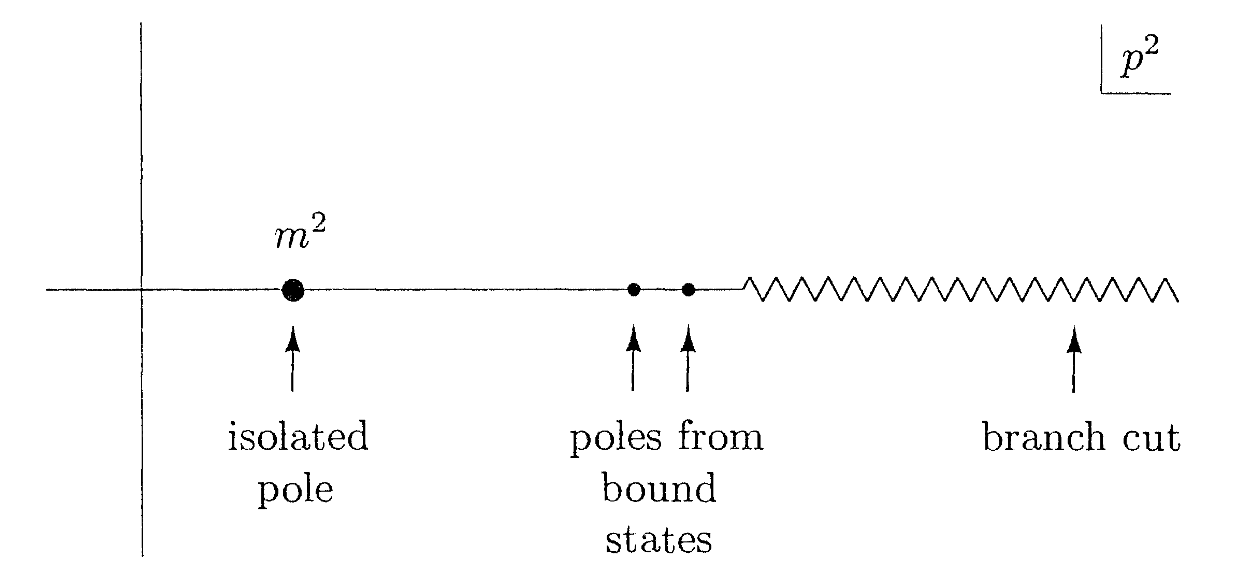}
	\caption[Analytic structure of a propagator, for a typical theory, obtained in the Minkowski space.]{Analytic structure of a propagator, for a typical theory, obtained in the Minkowski space. Image from \cite{Peskin}.} 
	\label{fig:TypAS}
\end{figure}

In general, the analytic structure, \ie, the set of zeros, singularities and branch cuts, of a propagator have a well defined physical interpretation. For example, in QED, the electron propagator has a singularity at the physical mass of this particle. For a typical theory, the analytic structure of a propagator is shown in Figure \ref{fig:TypAS}. In the complex $p^2$-plane, a pole that corresponds to the one particle state should appear, as well as a branch cut associated with two or more free particles; poles related to bound states also appear in the analytic structure \cite{Peskin}. All of these structures occur in the real $p^2$-axis. When a calculation is made, one can choose the integration path to go around these singularities. This also allows to perform a Wick rotation when going from the Euclidean space to the Minkowski space of momenta. However, this rotation is impracticable if complex singularities are present. While for four-dimensional QED, we do not find such singularities\footnote{To the author's best knowledge, complex singularities were found in QED only when formulated in lower dimensions, where it shows confinement, see, \eg, \cite{Maris1995,avkl2000}.}, in non-perturbative QCD it is a different story, since the propagators acquire a different analytic structure.

In fact, in a theory that displays confinement, which is believed to be the case of QCD, the analytic structure of the propagators of confined particles may have singularities that are not associated with physical states \cite{Maris1994}. This has to do with a violation of the local axioms of QFT by theories that exhibit confinement \cite{Binosi2020,Hayashi2020,Alkofer2001}.

Thereby, knowing the analytic structure of a propagator reveals to be crucial, not only because it gives information about the physical particle states of the theory, but also because it may bring new insight into the confinement mechanism. It is also indispensable to know the analytic structure when one attempts to go from the Euclidean space to the Minkowski space, whenever non-perturbative calculations are made. In this sense, many studies have been done with the main objective of finding the analytic structure of the full propagators for the QCD quanta, \ie, the gluon, the quark, and the unphysical ghost. Some predictions and studies around the existence of complex poles in the gluon propagator were made in, \eg, \cite{Sorella2011,Siringo2016,Stingl1986,Zwanziger1989,Alkofer2004,Hayashi2020}. Notably, the tree level solution for the propagator in the Refined Gribov-Zwanziger (RGZ) framework, that describes the lattice data extremely well, predicts the existence of a conjugate pair of complex poles at Euclidean momenta \cite{Dudal2010,Cucchieri2012}. These two complex poles were found using a global fit to lattice data, in \cite{Dudal2018}. Poles at similar positions in the complex $p^2$-plane were also found in \cite{Binosi2020}, using Padé Approximants (PA) to reconstruct the gluon and ghost propagators obtained via DSE and lattice simulations. Regarding the ghost propagator, its analytic structure seems to be similar to the one obtained perturbatively, \ie, with no complex poles, see, \eg, \cite{Binosi2020,Hayashi2020}.

On the other hand, additional studies, \eg\ \cite{Strauss2012}, found no evidence of such complex singularities. Further studies were undertaken in other types of theories to investigate the consequences of confinement in the analytic structure of the respective propagators, see \eg \cite{Maris1995,Roberts1992,avkl2000}.

In order to complement the already known results, this work focuses on the investigation of the analytic structure of the fundamental propagators in pure SU(3) Yang-Mills theory, obtained non-perturbatively via lattice calculations in the Landau gauge. No particular theoretical or empirical models to describe the lattice data will be considered. Instead, we rely on PAs to investigate the analytic structure of the QCD propagators, since it provides a general approach to study functions with singularities across the complex plane \cite{Hillion1977,Billi1994,Yamada2014,Boito2018}.

\vspace{1em}

This work is organised as follows: in the next chapter, we will start by introducing and defining Padé approximants in the context of the analytic continuation problem. The use of Padé approximants will, then, be tested for the perturbative propagators in the third chapter, in order to know how their analytic structures are reproduced, and how faithful this reproduction is. Considering that the full propagators come in the form of discrete sets of data points, a method of reconstructing the former with Padé approximants will be introduced and tested in the fourth chapter. In the fifth chapter, the obtained analytic structures for the gluon and ghost propagators will be investigated and discussed, followed by the final conclusions and ideas for possible future works.

\chapter{Elements of Padé Approximants} \label{Chap:PA}
The identification of the analytic structure of the gluon and ghost propagators requires the knowledge of the latter in the whole complex plane. However, the lattice simulations only provide the propagators in the real positive range of the Euclidean momenta.

In this chapter, we will look in detail at rational functions, particularly at the PA, and explore their use to identify the singularities and branch cuts for arbitrary complex momenta. A series of tests will be made, to examine the reliability of PAs in the reproduction of analytic structures.

\section{The numerical analytic continuation problem}
\begin{figure}[h]
	\centering
	\includegraphics[width=.6\textwidth]{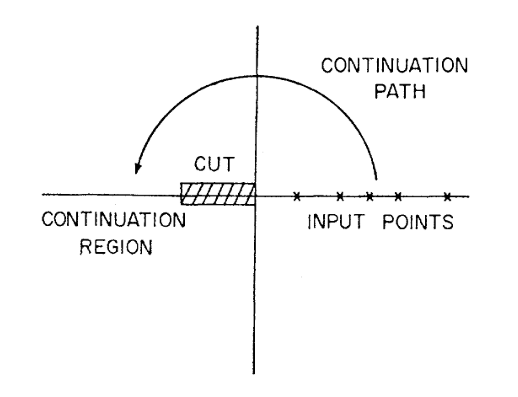}
	\caption[Graphical representation of an example of analytic continuation.]{Graphical representation of an example of analytic continuation, from \cite{Schlessinger1968}.}
	\label{fig:AnalyticContinuation}
\end{figure}

The numerical analytic continuation, \ie, the task of extending the domain of a function beyond the regime where the information is available, for example from a finite set of data points, is a known problem in Physics. The reconstruction of real-time correlations of spectral functions \cite{Tripolt2019}, the calculation of scattering amplitudes \cite{Schlessinger1968}, and situations, \eg \cite{Vidberg1977,Cillero2010,Binosi2020,Masjuan2010,Tripolt2017}, where the analytic continuation allows us to access information in different regions of momenta, while our knowledge is restricted to the physical one, are good examples of this problem. A graphical representation of the analytic continuation is shown in Figure \ref{fig:AnalyticContinuation}.

By adjusting a function to a set of data points, we can use the former to calculate its values in the whole domain of the function, and not only where the data is available. Yet, if we do not know the form of the function represented by the finite set of data, which one do we choose? Power series may seem a good solution. However, we will see that rational functions, particularly PAs, offer a more general and faithful approximation, making it more useful to the numerical continuation of the data to the complex plane.

\section{Why rational functions?}
In order to convince ourselves that rational functions are indeed richer structures by capturing the analytic properties of the approximated functions, let us consider the following example, taken from \cite{Yamada2014,Masjuan2010}.

\begin{figure}[tp]
	\centering
	\includegraphics[width=\textwidth]{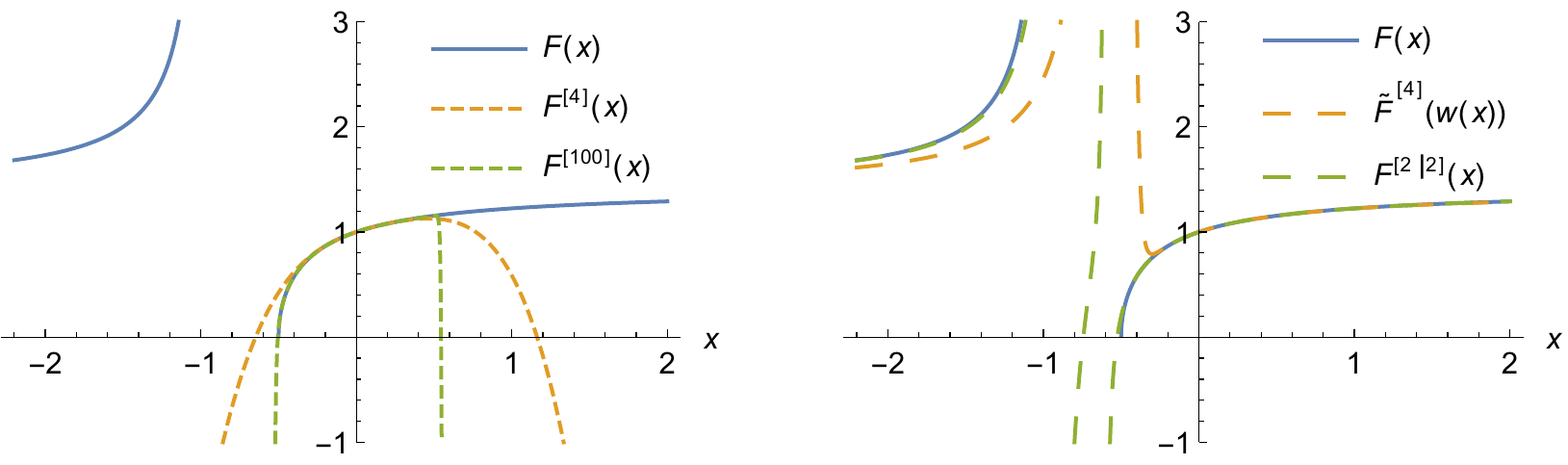}
	\caption[Comparison between the original function $F(x)$ and the truncated Taylor expansions $F^{[4]}(x)$ and $F^{[100]}(x)$; and between the original function $F(x)$, the rational function $\widetilde{F}^{[4]}(w(x))$, and the PA $F^{[2|2]}(x)$.]{\textbf{Left: }Comparison between the original function $F(x)$ and the truncated Taylor expansions $F^{[4]}(x)$ and $F^{[100]}(x)$. \textbf{Right: }Comparison between the original function $F(x)$, the rational function $\widetilde{F}^{[4]}(w(x))$, and the PA $F^{[2|2]}(x)$. For $x>0$, the three curves overlap.}
	\label{fig:TaylorVsRational}
\end{figure}

Consider the function $F(x)$, given by
\begin{equation}
F(x)=\sqrt{\frac{1+2x}{1+x}},
\label{eq:Fx}
\end{equation}
represented graphically in Figure \ref{fig:TaylorVsRational}, that has the following expansion in power series, around $x=0$ \footnote{Without loss of generality, the expansion is made around the origin. The problem ahead is independent of the expansion point.}
\begin{equation}
F(x)=\sum_{n=0}^\infty a_n x^n =1+\frac{x}{2}-\frac{5x^2}{8}+\frac{13x^3}{16}-\frac{141x^4}{128}+\mathcal{O}(x^5),
\label{eq:TaylorFx}
\end{equation}
where $a_n$ are the Taylor coefficients of $F(x)$.

By denoting the Taylor expansion truncated at order $N$ as the partial sum
\begin{equation}
F^{[N]}(x) \equiv \sum_{n=1}^N a_n x^n,
\end{equation}
we have, for $N=4$,
\begin{equation}
F^{[4]}(x)=1+\frac{x}{2}-\frac{5x^2}{8}+\frac{13x^3}{16}-\frac{141x^4}{128}.
\end{equation}

Let us suppose that we only have access to the partial sum $F^{[4]}$ and that we have no idea of the exact function that originated it. If we want to compute the value of $F(x)$ at the origin, we have $F^{[4]}(0)=1$, which concurs perfectly with the exact result of $F(0)$. On the other hand, if we try to calculate the value of $F(x)$ at $x=0.5$ via $F^{[4]}(x)$, the approximation is less accurate. In fact, $F^{[4]}(0.5)≃1.1265$, while $F(0.5)=2\sqrt{3}/3≃1.1547$. Naturally, if we stray away from the expansion point and beyond the radius of convergence, in this case $r=1/2$, the approximation fails drastically, even using expansions truncated at higher orders, as it can be seen in Figure \ref{fig:TaylorVsRational}.

In fact, computing $F(\infty)=\lim_{x\to\infty}F(x)$, with only the Taylor expansion, is a hopeless task. In this limit, $F^{[N]}(x)$ diverges for any $N\in\mathbb{N}$. Nonetheless, the original function $F(x)$ does not diverge when $x\to\infty$, $\lim_{x\to\infty}F(x)=\sqrt{2}$. How could we achieve this value with only $F^{[4]}(x)$? A cunning trick to transform the expansion in one which will let us estimate the value of $F(\infty)$ is: to perform the change of variables $x\equiv w/(1-2w)$; to define
\begin{equation}
\widetilde{F}(w) \equiv F(x(w))=(1-w)^{-\frac{1}{2}};
\end{equation}
to re-expand it in $w$,
\begin{equation}
\widetilde{F}(w)=\sum_{n=0}^\infty b_n w^n =1+\frac{w}{2}+\frac{3w^2}{8}+\frac{5w^3}{16}+\frac{35w^4}{128}+\mathcal{O}(w^5);
\label{eq:TaylorFw}
\end{equation}
and, in a similar way to $F(x)$, to define the truncated Taylor expansion of $\widetilde{F}(w)$ as the partial sum
\begin{equation}
\widetilde{F}^{[N]}(w)\equiv\sum_{n=0}^N b_n w^n.
\end{equation}

By doing this, the limit $x\to\infty$ is translated into $w\to1/2$. For this value of $w$, the Taylor expansion of (\ref{eq:TaylorFw}) converges, and we are able to approximate the value of $\widetilde{F}(1/2)$ and, therefore, of $F(\infty)$. Let us, then, do so for the first lowest $N$ of $\widetilde{F}^{[N]}(1/2)$. The resulting partial sums are shown in Table \ref{tab:Fkoo}. The sequence of values shown in Table \ref{tab:Fkoo} converges to $\sqrt{2}≃1.4241$.

\begin{table}[t]
	\centering
	\begin{tabular}{c|cccccc}
		\toprule
		$N$ & 0 & 1 & 2 & 3 & 4 & $\cdot\cdot\cdot$ \\
		\midrule
		$F^{[N]}(1/2)$ & 1 & 1.25 & 1.34375 & 1.38281 & 1.39990 & $\cdot\cdot\cdot$ \\
		\bottomrule
	\end{tabular}
	\caption{Results for the lowest partial sums of Eq. (\ref{eq:TaylorFw}) with $w\to 1/2$.}
	\label{tab:Fkoo}
\end{table}

Now, let us go back and rewrite $\widetilde{F}^{[4]}(w)$ in terms of $x$ \footnote{Note that $\widetilde{F}^{[4]}(w(x))$ and $F^{[4]}(x)$ are not the same. The first comes from the expansion of $\widetilde{F}(w)$ in $w$, while the latter comes from the expansion of the original function before the change of variables.},
\begin{equation}
\widetilde{F}^{[4]}(w(x))= \frac{1+(17/2)x+(219/8)x^2+(637/16)x^3+(2867/128)x^4}{(1+2x)^4}.
\label{eq:Ft4x}
\end{equation}
Clearly, this is not a power expansion, but a rational function. We already saw that an  approximation like (\ref{eq:Ft4x}) allows us to estimate not only the value of $F(x)$ near the origin, but also in the limit $x\to\infty$. By graphically comparing $\widetilde{F}^{[4]}(w(x))$ with the original function (Figure \ref{fig:TaylorVsRational}) an improvement can be seen, which was brought by the use of a rational function. Despite the fact that $\widetilde{F}^{[4]}(w(x))$ is defined in $x\in[-1,-1/2]$, where $F(x)$ is undefined (visible in Figure \ref{fig:TaylorVsRational}), the overall behaviour of the original function can be reproduced.

If the use of rational functions can considerably improve the approximate description of a function, how do we build one? Surely, it is not of our interest to find the right change of variables for every function we come across. A particular type of rational functions is the PA. The idea of PAs is to use the first Taylor coefficients of a given function to build a ratio of polynomials, \ie, a rational function. A simple PA is
\begin{equation}
P(x)=\frac{a_0+a_1x}{1+b_1x}.
\end{equation}
In this case, the goal is to fix the unknowns $a_0$, $a_1$ and $b_1$ in such a way that the first three coefficients of the Taylor expansion of $P(x)$ match the first three Taylor coefficients of the function to be approximated. For the function $F(x)$, defined in (\ref{eq:Fx}), we find
\begin{equation}
P(x)=\frac{1+(7/4)x}{1+(5/4)x}= 1+\frac{x}{2}-\frac{5x^2}{8}+\frac{25x^3}{32}-\frac{125x^4}{128}+\mathcal{O}(x^5).
\end{equation}
When comparing it with (\ref{eq:TaylorFx}), it can be seen that the first three coefficients are exactly the same (but not the remaining ones). If we now use the limit of $P(x)$ to estimate the value of $F(\infty)$, we get $\lim_{x\to\infty}P(x)=1.4$, which is a better determination than any in Table \ref{tab:Fkoo}. By requiring the matching of the first five Taylor coefficients, we obtain the PA
\begin{equation}
P(x)=\frac{1+(13/4)x+(41/16)x^2}{1+(11/4)x+(29/16)x^2},
\end{equation}
and $\lim_{x\to\infty}P(x)=41/29≃1.4137$. If we continue to higher numbers of matching Taylor coefficients, the precision in the estimation of $F(\infty)$ is increased. Indeed, for eleven matching coefficients we reach a precision of $\sim\num{e-8}$.

Graphically (Figure \ref{fig:TaylorVsRational}), the precision of the approximation via PA is evident. The reproduction of the divergence at $x=-1$ is worth mentioning. This capability of reproducing divergences without damaging the overall behaviour of a function makes the PA a valuable tool.

\section{The Padé Approximant}
In order to use the PA as a tool, a rigorous definition must be made, as in \cite{Yamada2014,Masjuan2010,Boito2018}. For a more formal and complete definition see, \eg, \cite{George1970}. Thus, let us consider a function $f(z)$ that has a series expansion\footnote{Without loss of generality, an expansion around the origin is considered.} in the complex plane
\begin{equation}
f(z)=\sum_{n=0}^\infty c_n z^n,
\end{equation}
where $c_n$ are its Taylor coefficients. Let us also denote $f^{[N]}(z)$ as the respective truncated Taylor expansion of order $N$,
\begin{equation}
f^{[N]}(z)\equiv\sum_{n=0}^N c_n z^n.
\end{equation}

A Padé Approximant of order $[L|M]$ is defined as the ratio of two polynomials $Q_L(z)$ and $R_M(z)$, of orders $L$ and $M$ respectively,
\begin{equation}
P^{[L|M]}(z)\equiv\frac{Q_L(z)}{R_M(z)}=\frac{q_0+q_1z+q_2z^2+...+q_Lz^L}{1+r_1z+r_2z^2+...+r_Mz^M}.
\label{eq:PA}
\end{equation}
As it is usually done, the normalisation $r_0=1$ is considered. The coefficients $q_0,...,q_L$ and $r_1,...,r_M$ will be called \textit{Padé coefficients}.

The PA of the function $f(z)$ is denoted by $f^{[L|M]}(z)$, and is built such that the Taylor expansion of $f^{[L|M]}(z)$ reproduces exactly the first $L+M+1$ Taylor coefficients of $f(z)$. In this sense, we say that the PA has a \textit{contact} of order $L+M$ with the expansion of $f(z)$, and the difference between the PA and the original function satisfies
\begin{equation}
f(z)-f^{[L|M]}(z)=\mathcal{O}(z^{L+M+1}).
\end{equation}
When it exists, the PA is unique for any $L$ and $M$.

As we will see later, sequences of PAs are extremely important (and fundamental in the scope of the present work), for it is their stability that gives us the confidence on the outcome of the approximations. Sequences with $L=M+J$ are called \textit{near-diagonal} when $J\neq0$, and \textit{diagonal} when $J=0$.

Despite the already seen advantages of using PAs, there is a downside: unlike the Taylor series, there exists no general convergence theories for PAs. Nevertheless, there are some complete convergence theorems for some particular cases. A good resume of the existing convergence theorems and their details can be found, for example, in \cite{Masjuan2010}.

\section{Analytic structure of a PA} \label{Sec:ASofPA}
Let us now focus on the analytic structure of functions and on how well the PA can reproduce it. In \cite{Yamada2014}, a series of general examples with test functions are carried out with this objective. Here, we will aim our attention to some of them before we move further into more specific tests.

We begin to consider the following complex test functions:
\begin{align}
\label{eq:f1}
f_1(z) &= e^{-z}, \\
\label{eq:f2}
f_2(z) &= \left(\frac{z-2}{z+2}\right)e^{-z}, \\
\label{eq:f3}
f_3(z) &= e^{-z/(1+z)}, \\
\label{eq:f4}
f_4(z) &= \sqrt{\frac{1+2z}{1+z}}.
\end{align}

The analytic structures of each of the functions above are represented, in the complex plane, in Figures \ref{fig:OriginalVsPade1} to \ref{fig:OriginalVsPade4}. These representations, made with the use of the software \textit{Mathematica} \cite{Mathematica}, are built in such a way that the poles, zeros, and branch cuts are enhanced. To do so, the argument of $f_i(z)$ is represented, instead of its value. This allows us to use the following key to read figures made in this way.

\begin{figure}[H]
	\centering
	\includegraphics[width=.8\textwidth]{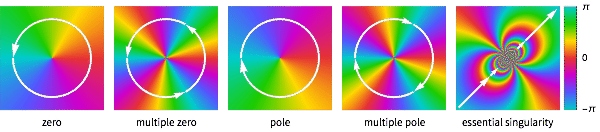}
	\caption[Key for the identification of poles, zeros and essential singularities in representations of complex functions.]{Key for the identification of poles, zeros and essential singularities in representations of complex functions, where it is used a cyclic colour function over the argument of the represented function. Image from \cite{ComplexPlot}.}
	\label{fig:Key}
\end{figure}

\noindent The branch cut, it is identified by a black dashed line.

\begin{figure}[tp]
	\centering
	\includegraphics[width=\textwidth]{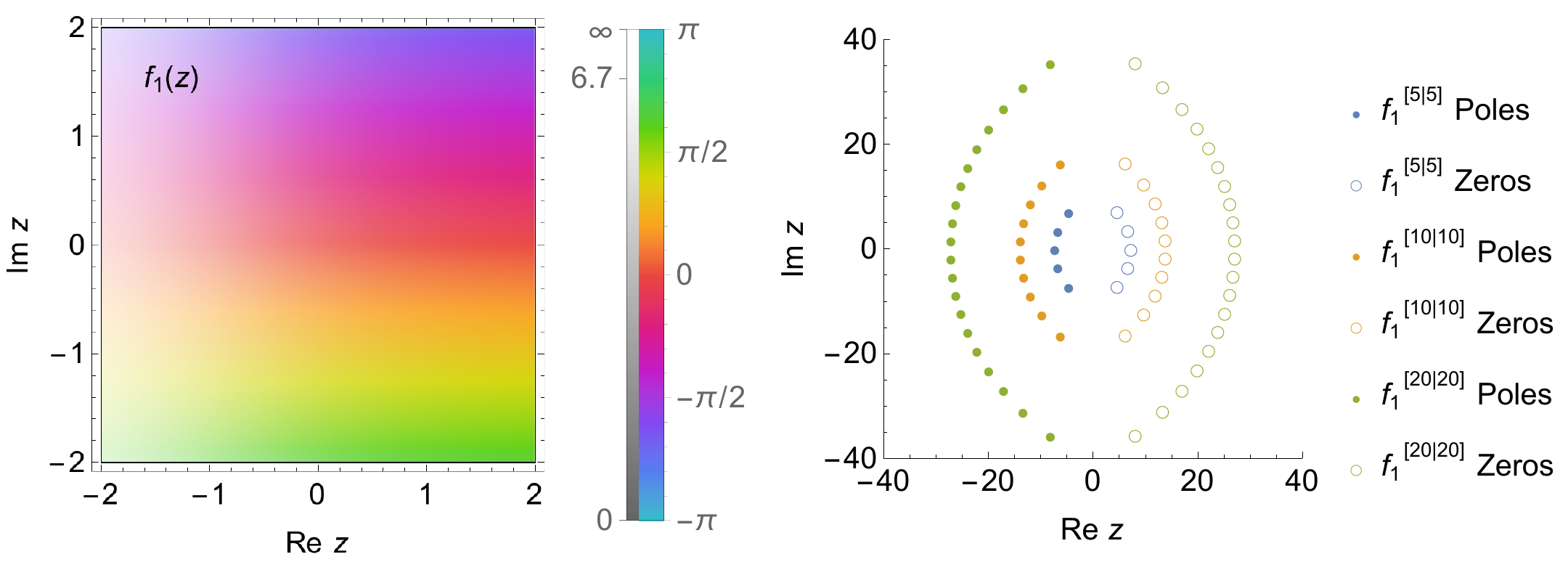}
	\caption[{Representation of the test functions' analytic structure $f_1(z)$, and distribution, in the complex plane, of poles and zeros for the sequence of diagonal PAs of orders $[5|5]$, $[10|10]$ and $[20|20]$.}]{\textbf{Left: }Representation of the test functions' analytic structure $f_1(z)$. The key for the structure identification is in Figure \ref{fig:Key}. \textbf{Right: }Distribution, in the complex plane, of poles and zeros for the sequence of diagonal PAs of orders $[5|5]$, $[10|10]$ and $[20|20]$, for the test function $f_1(z)$.}
	\label{fig:OriginalVsPade1}
\end{figure}

\begin{figure}[tp]
	\centering
	\includegraphics[width=\textwidth]{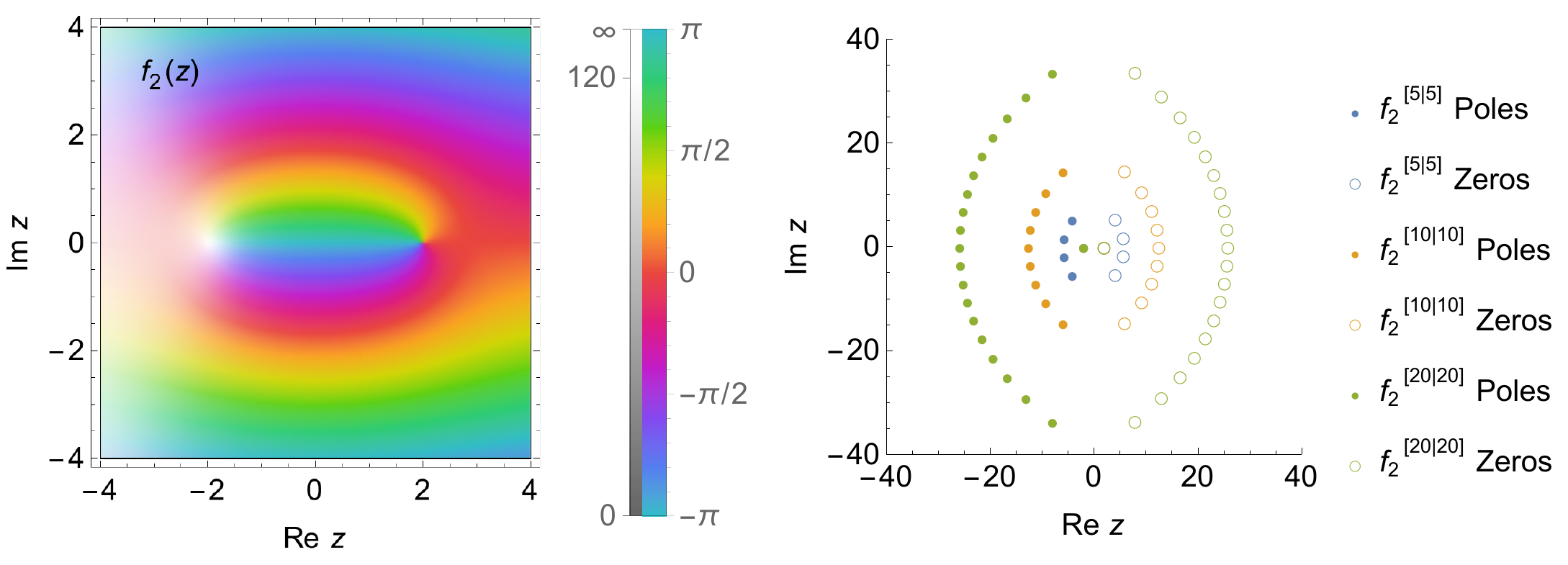}
	\caption[{Representation of the test functions' analytic structure $f_2(z)$, and distribution, in the complex plane, of poles and zeros for the sequence of diagonal PAs of orders $[5|5]$, $[10|10]$ and $[20|20]$.}]{\textbf{Left: }Representation of the test functions' analytic structure $f_2(z)$. The key for the structure identification is in Figure \ref{fig:Key}. \textbf{Right: }Distribution, in the complex plane, of poles and zeros for the sequence of diagonal PAs of orders $[5|5]$, $[10|10]$ and $[20|20]$, for the test function $f_2(z)$. The pole at $z=-2$ and the zero at $z=2$  appear at the same position for the three represented orders, and are, therefore, overlapped.}
	\label{fig:OriginalVsPade2}
\end{figure}

\begin{figure}[tp]
	\centering
	\includegraphics[width=\textwidth]{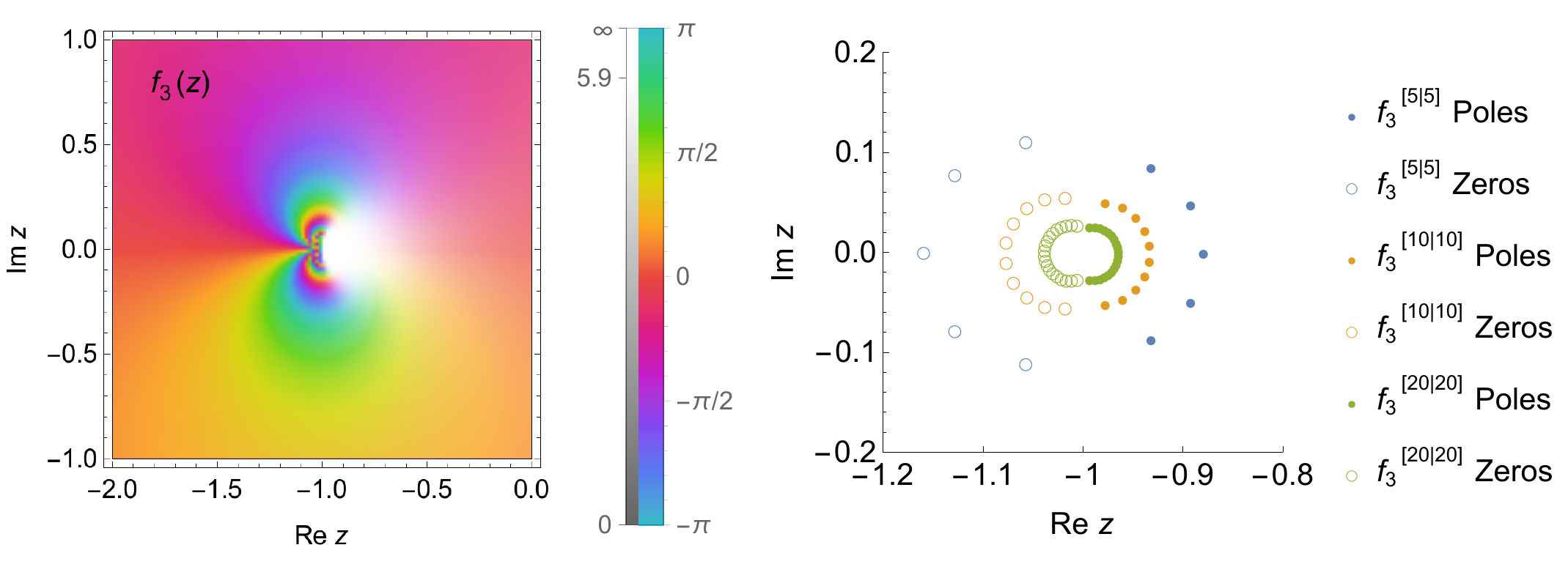}
	\caption[{Representation of the test functions' analytic structure $f_3(z)$, and distribution, in the complex plane, of poles and zeros for the sequence of diagonal PAs of orders $[5|5]$, $[10|10]$ and $[20|20]$.}]{\textbf{Left: }Representation of the test functions' analytic structure $f_3(z)$. The key for the structure identification is in Figure \ref{fig:Key}. \textbf{Right: }Distribution, in the complex plane, of poles and zeros for the sequence of diagonal PAs of orders $[5|5]$, $[10|10]$ and $[20|20]$, for the test function $f_3(z)$.}
	\label{fig:OriginalVsPade3}
\end{figure}

\begin{figure}[tp]
	\centering
	\includegraphics[width=\textwidth]{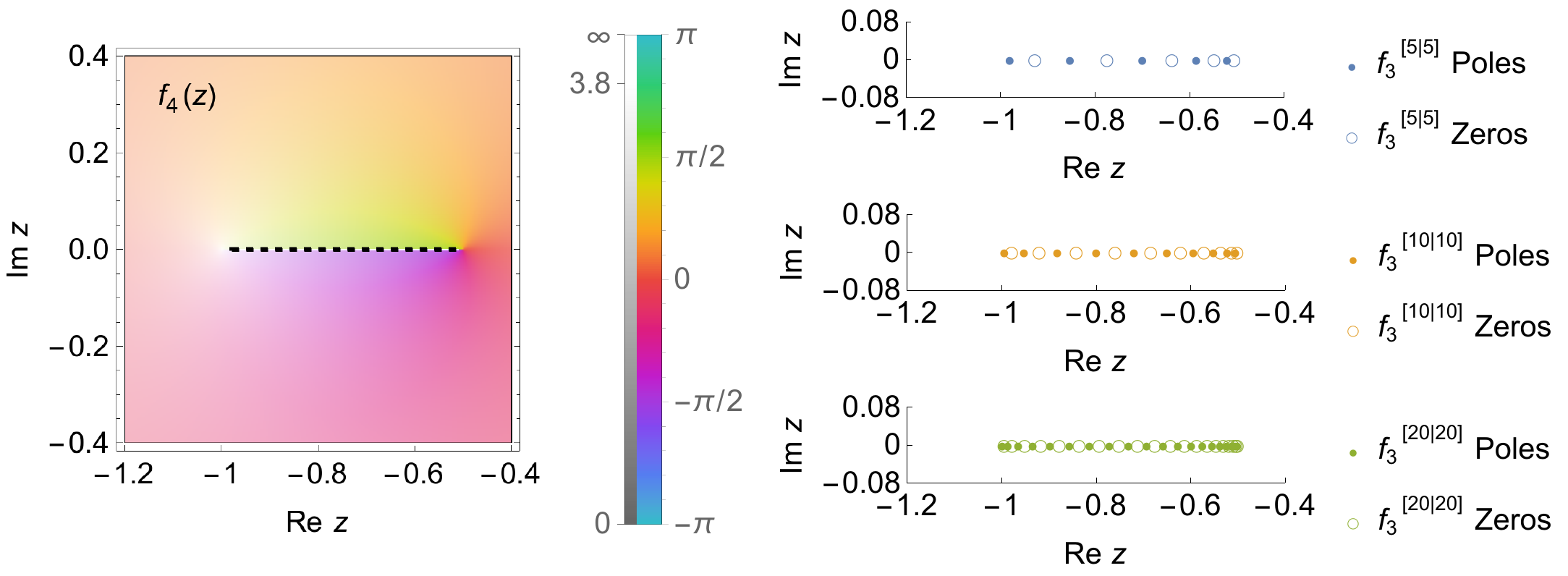}
	\caption[{Representation of the test functions' analytic structure $f_4(z)$, and distribution, in the complex plane, of poles and zeros for the sequence of diagonal PAs of orders $[5|5]$, $[10|10]$ and $[20|20]$.}]{\textbf{Left: }Representation of the test functions' analytic structure $f_4(z)$. The key for the structure identification is in Figure \ref{fig:Key}. \textbf{Right: }Distribution, in the complex plane, of poles and zeros for the sequence of diagonal PAs of orders $[5|5]$, $[10|10]$ and $[20|20]$, for the test function $f_4(z)$.}
	\label{fig:OriginalVsPade4}
\end{figure}

By looking at expressions (\ref{eq:f1}) to (\ref{eq:f4}) and Figures \ref{fig:OriginalVsPade1} to \ref{fig:OriginalVsPade4}, we can conclude that:
\begin{itemize}
	\item $f_1(z)$ has no singularities for $|z|<\infty$;
	\item $f_2(z)$ has a zero at $z=2$ and a simple pole at $z=-2$;
	\item $f_3(z)$ has an essential singularity at $z=-1$;
	\item $f_4(z)$ has a branch cut along a line on [-1,-1/2].
\end{itemize}

After computing the PA for a given function, its analytic structure can be extrapolated from the PA's own analytic structure. As a fraction between polynomials, the zeros of a PA correspond to the roots of its numerator - $Q_L(z)$ in (\ref{eq:PA}) -, and its poles correspond to the roots of its denominator - $R_M(z)$ in (\ref{eq:PA}). Note that these are the only structures present in the analytic structure of a PA. As a consequence, the reconstruction of the original function's analytic structure only relies on the distribution of poles and zeros of the associated PA.

For some functions, an analytic expression of the respective PA can be found. For example, for $f_1(z)$ we have
\begin{equation}
Q_L(z)=\sum_{k=0}^L\frac{(2L-k)!L!}{(2L)!k!(L-k)!}(-z)^k,
\end{equation}
\begin{equation}
R_M(z)=\sum_{k=0}^M\frac{(2M-k)!M!}{(2M)!k!(M-k)!}z^k.
\end{equation}
However, in general, a PA can be found numerically\footnote{For all the numerical calculations in the present work, the software \textit{Mathematica} \cite{Mathematica} was used.} for a given order $[L|M]$. Once it is done, the poles and zeros can be obtained and represented graphically in the complex plane, and the analytic structures can be compared. For our test functions, some orders of diagonal PAs were calculated. The corresponding distributions of poles and zeros are presented in Figures \ref{fig:OriginalVsPade1} to \ref{fig:OriginalVsPade4}.

For $f_1(z)$, the distribution of poles and zeros of the obtained PAs is symmetric around the origin, as seen in Figure \ref{fig:OriginalVsPade1}. However, their position strongly depends on the order of the PA used. Indeed, by increasing the PA's order, its distribution seems to spread and move towards infinity, leaving no structure behind.

The same happens with $f_2(z)$ (Figure \ref{fig:OriginalVsPade2}), except that, in this case, a pole and a zero appear in the expected positions, $z=-2$ and $z=2$ respectively. These pole and zero are stable, and their positions, in the complex plane, seem to be independent of the order of approximation. This behaviour indicates that the original poles and zeros are identified by stable ones, in the complex plane, in a PA sequence. On the other hand, unstable poles and zeros do not correspond to any characteristic of the analytic structure of $f_2(z)$.

A similar, but opposite, behaviour to the one seen for $f_1(z)$ and $f_2(z)$ can be observed in the distribution of poles and zeros for $f_3(z)$. However, instead of spreading, the poles and zeros gather around the position where the essential singularity should be (see Figure \ref{fig:OriginalVsPade3}).

Lastly, the branch cut of $f_4(z)$ is reproduced by the PA in the form of an alternating sequence of poles and zeros, between $z=-1$ and $z=-1/2$. In this case, the distance between nearby zeros and poles decreases when the order of the PA is increased.

Additional numerical tests were performed in \cite{Yamada2014}. It was found that ``spurious poles'' may appear with the increase of the order of the approximation, due to insufficient numerical accuracy. Some of the poles have an associated zero that cancels its contribution to the analytic structure. These pole-zero pairings, often called \textit{Froissart doublets}, are artefacts of the approximation, and accumulate in structures around the origin. These have to be identified as such, in order to properly identify the correct analytic structure. Later on, we will introduce a way to remove these unwanted poles by performing a residue analysis (Section \ref{Sec:ResidueAnalysis}).

We can, now, draw some conclusions regarding the reproduction of the analytic structure of a function:
\begin{itemize}
	\item For a function with or without singularities, distributions of poles and zeros may appear. As the PA's order is increased, if they spread to infinity or are overall unstable, they are not associated with the analytic structure of the original function, but are artefacts of the method;
	\item For a function with an essential singularity, the structures of poles and zeros tighten around the position of the singularity for increasing orders of approximation;
	\item Poles and zeros that are stable throughout the PA sequence may be correctly identified as being part of the original function's analytic structure;
	\item A branch cut can be identified by a PA as a sequence of alternating poles and zeros, for which the distance between nearby poles and zeros decreases when the order of the PA is increased;
	\item The increase of the order of approximation may cause the emergence of Froissart doublets (pole-zero pairings), which do not contribute to the analytic structure.
\end{itemize}

\chapter{Preliminary tests} \label{Chap:AnalyticTests}
In the previous chapter, we explored the use of PAs to identify the analytic structure of a function based on the distribution of poles and zeros. Notwithstanding, the functions that we are dealing with in our study are not so simple as the ones covered there, and more specific tests have to be performed. These will allow to identify a set of tools for a proper identification of poles and branch cuts.

\section{The perturbative result for the gluon propagator}
The results from the renormalisation group improved perturbation theory for the gluon propagator and its dressing function are given, respectively, by
\begin{equation}
D_{gl}(p^2)=\frac{1}{p^2}\left[\frac{11N_f\alpha_s}{12\pi}\ln\left(\frac{p^2}{\Lambda^2}\right)+1\right]^{-\gamma},
\label{Eq:Prop}
\end{equation}
\begin{equation}
d_{gl}(p^2)\equiv p^2 D_{gl}(p^2) =\left[\frac{11N_f\alpha_s}{12\pi}\ln\left(\frac{p^2}{\Lambda^2}\right)+1\right]^{-\gamma}.
\label{Eq:Dress}
\end{equation}
Following \cite{Dudal2018}, in the numerical tests we will use $\alpha_s=0.3837$, $\Lambda=0.425~\si{GeV}$ and $\gamma=13/22$. These results offer us a valuable opportunity to study the reliability of using PAs to study the QCD propagators, since we expect an equivalent behaviour in the UV limit. Throughout this chapter we look at the Equations (\ref{Eq:Prop}) and (\ref{Eq:Dress}) and study them as test functions to understand the behaviour and validity of the PA approach.

\begin{figure}[tp]
	\centering
	\includegraphics[width=\textwidth]{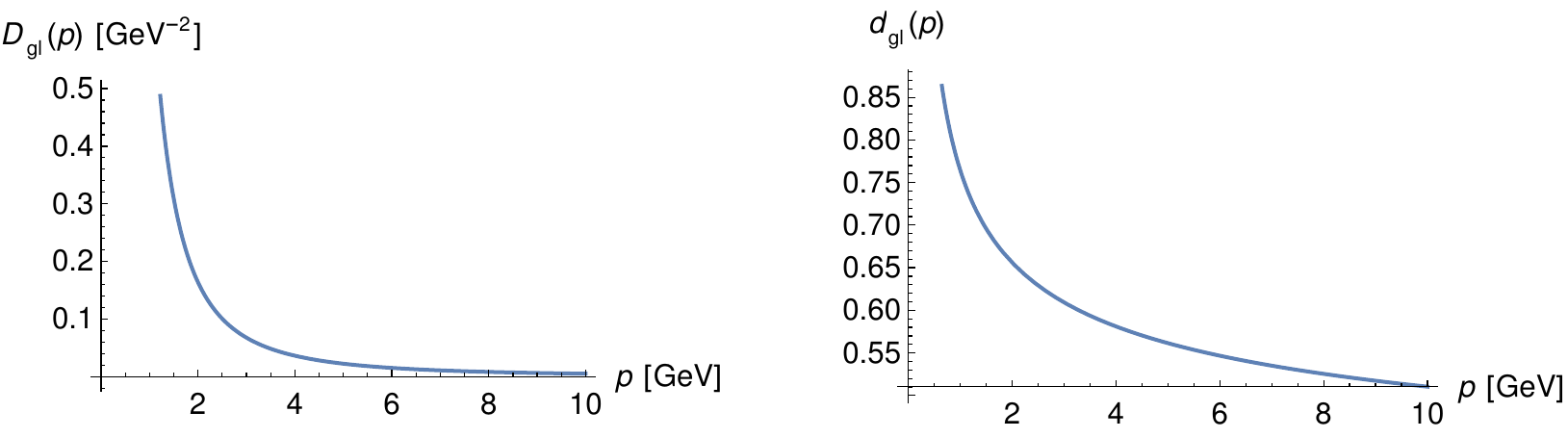}
	\caption[Graphical representation of the gluon propagator $D_{gl}(p^2)$, and of its dressing function $d_{gl}(p^2)$.]{Graphical representation of the gluon propagator $D_{gl}(p^2)$ (left), and of its dressing function $d_{gl}(p^2)$ (right).}
	\label{fig:PropAndDress}
\end{figure}

\begin{figure}[tp]
	\centering
	\includegraphics[width=\textwidth]{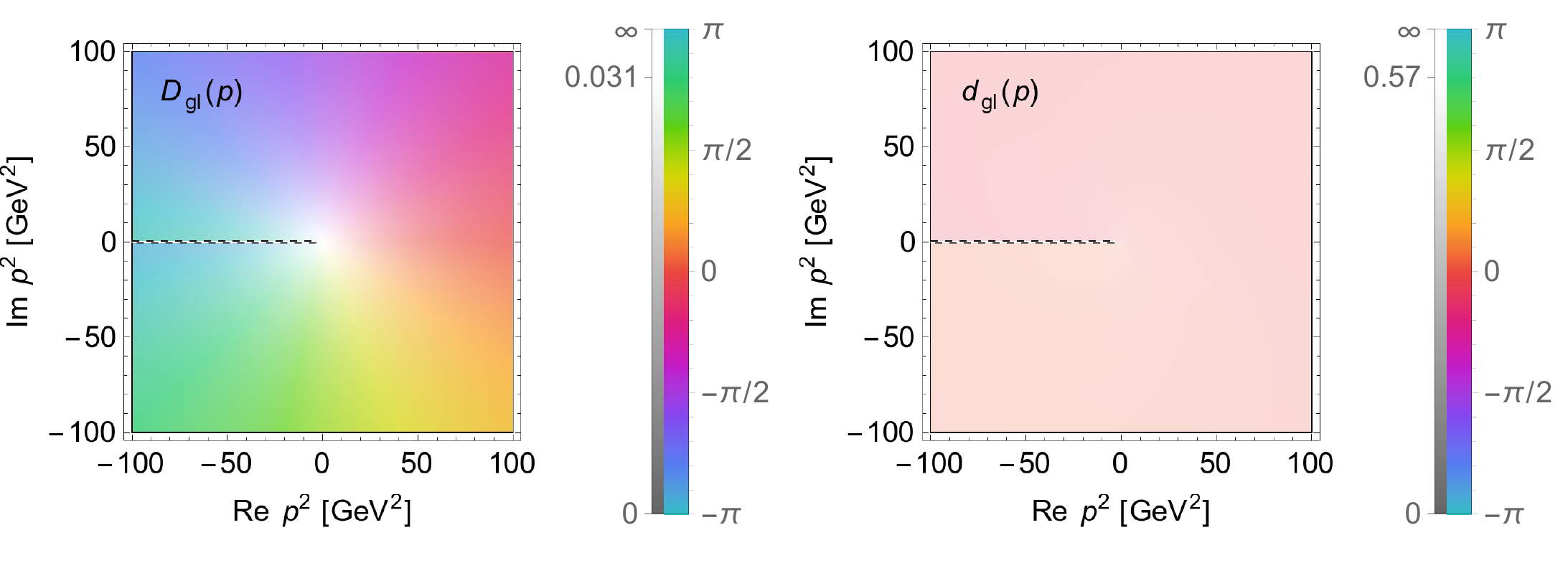}
	\caption[Analytic structures of the gluon propagator $D_{gl}(p^2)$, and of its dressing function $d_{gl}(p^2)$.]{Analytic structures of the gluon propagator $D_{gl}(p^2)$ (left), and of its dressing function $d_{gl}(p^2)$ (right). The key for the structure identification is in Figure \ref{fig:Key}.}
	\label{fig:PropAndDressAS}
\end{figure}

To give us a visual idea of the behaviour of these test functions, a graphical representation of (\ref{Eq:Prop}) and (\ref{Eq:Dress}) is shown in Figure \ref{fig:PropAndDress}. As for the respective analytic structures, they are represented in the complex $p^2$-plane in Figure \ref{fig:PropAndDressAS}. For the propagator, we see a simple pole at the origin, created by the factor $1/p^2$, as well as a branch cut on the whole real negative $p^2$-axis, from the logarithm. On the other hand, only the branch cut in the real negative $p^2$-axis appears in the analytic structure of the dressing function. These are the structures we want to reproduce using PA sequences.

\section{Relation between $L$ and $M$} \label{Sec:LM}
The first step in the construction of a PA sequence is to establish the best relation between the orders $L$ and $M$ of the polynomials, in (\ref{eq:PA}), since this relation is what dictates the limit behaviour of a PA.

We know that the propagator, as well as the dressing function, have a dependence only on $p^2$. For this reason, we may impose that only the coefficients associated with even powers of momentum have nonzero values. Hence, for simplicity, we will build our PAs in order of $p^2$ and not $p$, \ie,
\begin{align}
d_{gl}(p^2)\to d_{gl}^{[L|M]}(p^2) = \frac{Q_L(p^2)}{R_M(p^2)}&= \frac{q_0+q_1p^2+q_2(p^2)^2+...+q_L(p^2)^L}{1+r_1p^2+r_2(p^2)^2+...+r_M(p^2)^M} \nonumber \\
&= \frac{q_0+q_1p^2+q_2p^4+...+q_Lp^{2L}}{1+r_1p^2+r_2p^4+...+r_Mp^{2M}}.
\end{align}
The same happens to the propagator,
\begin{equation}
D_{gl}(p^2)\to D_{gl}^{[L|M]}(p^2).
\end{equation}

By looking at the representation of the dressing function (Figure \ref{fig:PropAndDress}), we see that it slowly goes to zero for high values of $p$. It does so as $[\ln p^2]^{-13/22}$, and, thus, the right choice seems to be a relation that reproduces a similar behaviour at large momenta. Unfortunately, a ratio of polynomials cannot describe exactly a logarithmic function over a wide range of its arguments and, so, we have to look for the best approach.

\begin{figure}[tp]
	\centering
	\includegraphics[width=\textwidth]{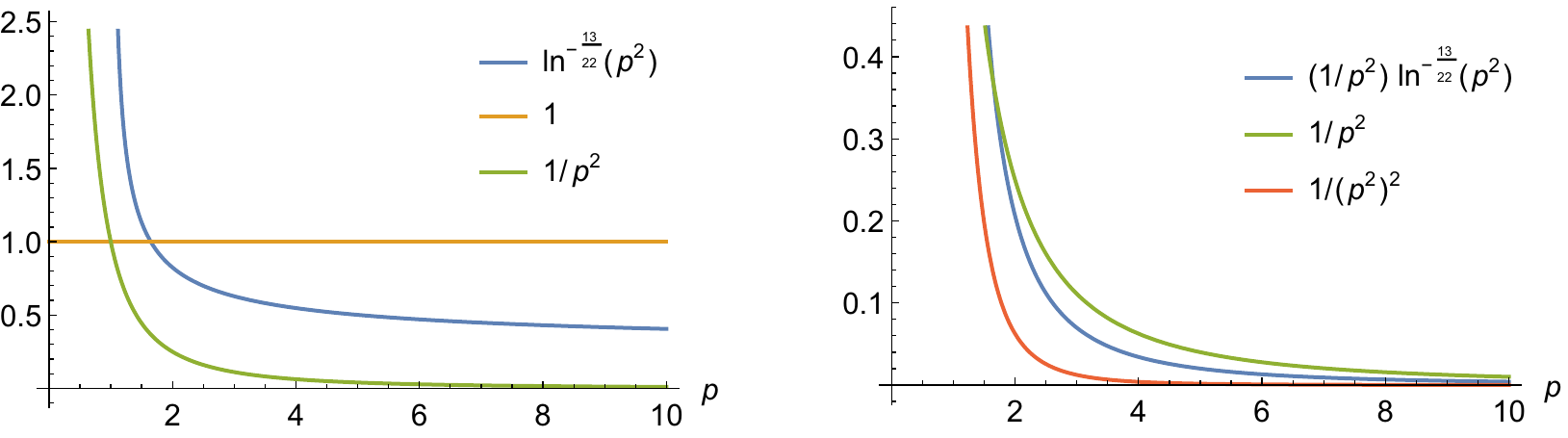}
	\caption[{Graphical representation of the asymptotic behaviour of $D_{gl}(p^2)$ and $d_{gl}(p^2)$, together with PAs of orders $[N|N]$, $[N-1|N]$, and $[N-2|N]$.}]{Graphical representation of the asymptotic behaviour of $D_{gl}(p^2)$ (left), and $d_{gl}(p^2)$ (right), together with PAs of orders $[N|N]$ (constant function $1$), $[N-1|N]$ (function $1/p^2$), and $[N-2|N]$ (function $1/(p^2)^2$).}
	\label{fig:RelationLM}
\end{figure}

In Figure \ref{fig:RelationLM}, the functions $[\ln p^2]^{-13/22}$ and $(1/p^2)[\ln p^2]^{-13/22}$ are shown together with some simple PAs. For relatively high values of momentum ($p\sim 10~\si{GeV}$), the dressing function seems to tend to $0$ between 1 and $1/p^2$ \footnote{Here, the value 1 is an example of a constant value. In fact, for any constant $c>0$, there is a value $p_\text{min}$ such that $[\ln p^2]^{-13/22}<c,~\forall_{p>p_\text{min}}$. Furthermore, it is verified that\[\exists_{p_\text{min}>0}:~c>[\ln p^2]^{-13/22}>\frac{1}{p^2},~\forall_{p>p_\text{min}}.\]}. A similar analysis for the propagator can be made. This time, for high values of momentum, the propagator goes to zero with $(1/p^2)[\ln p^2]^{-13/22}$, between $1/p^2$ and $1/(p^2)^2$, as seen in Figure \ref{fig:RelationLM} \footnote{Formally, it is verified that\[\exists_{p_\text{min}>0}:~\frac{1}{p^2}>\frac{1}{p^2}[\ln p^2]^{-13/22}>\frac{1}{(p^2)^2},~\forall_{p>p_\text{min}}.\]}.

A criterion to determine the best sequence for both cases, by choosing the suitable relation between $L$ and $M$ to use, will be discussed in Section \ref{Sec:Hint}.

\section{The expansion point}
By definition, the PA of a function is built using its Taylor expansion, and so, it depends on the point around which it is made. Throughout the examples in Chapter \ref{Chap:PA}, we showed little concern for this matter, making all the expansions around the origin. However, we are now confronted with functions that are not defined at the origin, \eg, the logarithm. For this reason, we have to find the expansion point that enables us to make the best approximation.

\begin{figure}[tp]
	\centering
	\includegraphics[width=\textwidth]{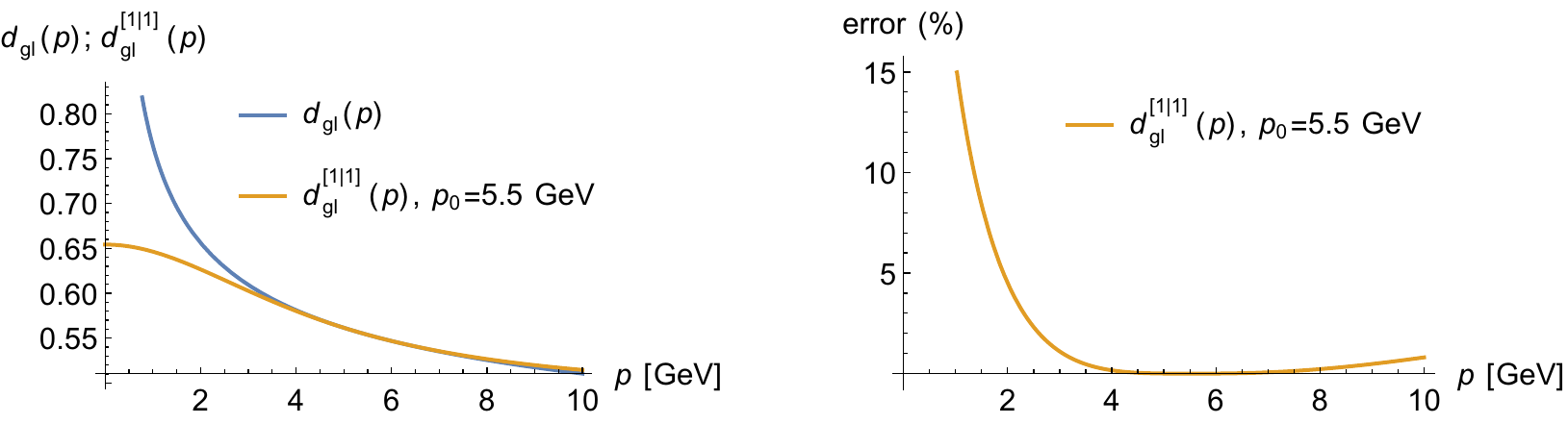}
	\caption[{Representation of $d_{gl}(p^2)$, together with the respective PA of order $[1|1]$ using, as expansion point, $p_0=5.5~\si{GeV}$, and the approximation error.}]{Representation of $d_{gl}(p^2)$, together with the respective PA of order $[1|1]$ using, as expansion point, $p_0=5.5~\si{GeV}$ (left), and the approximation error (right).}
	\label{fig:p05.5}
\end{figure}

\begin{figure}[tp]
	\centering
	\includegraphics[width=\textwidth]{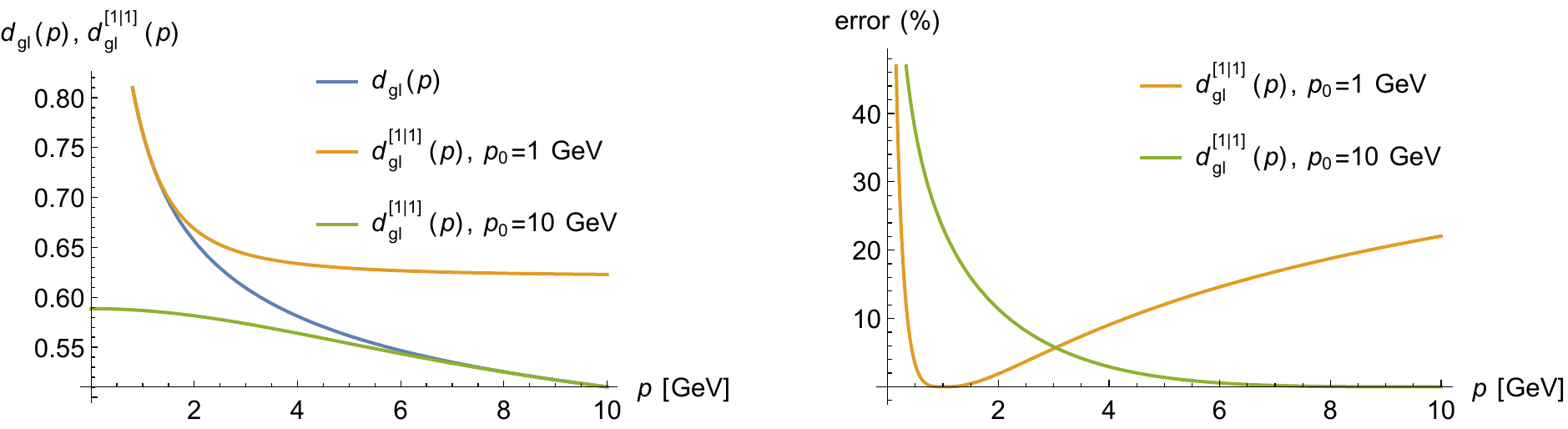}
	\caption[{Representation of $d_{gl}(p^2)$, together with the respective PAs of order $[1|1]$ using, as expansion points, $p_0=1~\si{GeV}$ and $p_0=10~\si{GeV}$, and the approximation error.}]{Representation of $d_{gl}(p^2)$, together with the respective PAs of order $[1|1]$ using, as expansion points, $p_0=1~\si{GeV}$ and $p_0=10~\si{GeV}$ (left), and the approximation error (right).}
	\label{fig:p01and10}
\end{figure}

Since we are interested in values of $p$ between $\sim 1~\si{GeV}$ and $\sim 10~\si{GeV}$, we require a good precision in the reproduction of the original function in this range of momentum. In this sense, we choose the central point $p_0=5.5~\si{GeV}$, and examine the precision of the obtained PA along $p$. Figure \ref{fig:p05.5} shows $d_{gl}(p^2)$, together with the respective PA of order $[1|1]$, and the percent error of approximation. Other interesting points to expand around are the endpoints of the considered interval, \ie, $p_0=1~\si{GeV}$ and $p_0=10~\si{GeV}$. The PA of order $[1|1]$, and the percent error for these expansion points are shown in Figure \ref{fig:p01and10}.

By comparing the obtained errors, we see that if we choose the expansion point to be near an endpoint we gain precision around it, but loose it at the opposite side. Furthermore, by analysing the error curves, we conclude that the error is at its lowest at $p=p_0$, reaching its maximum values at the endpoints. Thus, a good expansion point should be one that lowers the error on both endpoints. For this reason, we just need to analyse the error at the endpoints, since we know that there will be no higher values in between.

\begin{figure}[tp]
	\centering
	\includegraphics[width=.8\textwidth]{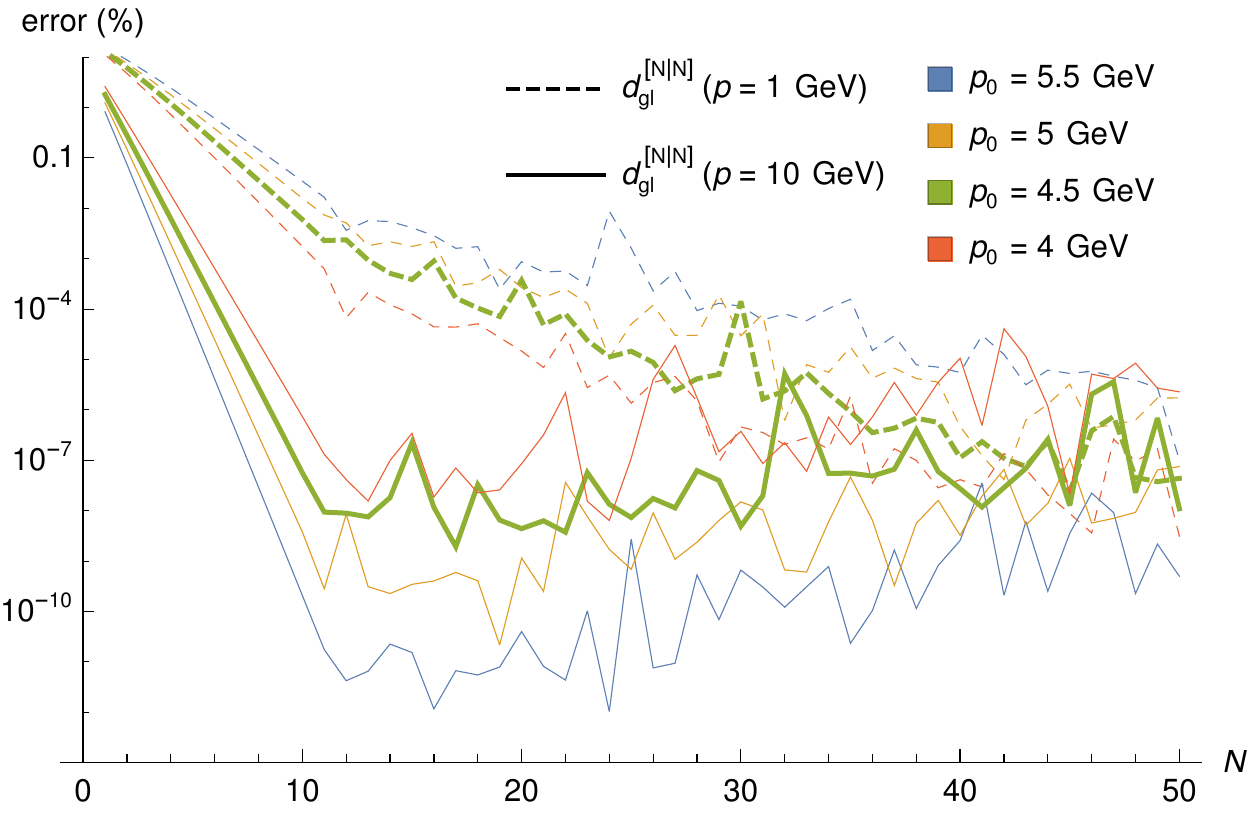}
	\caption{Evolution of the approximation error at the endpoints $p=1~\si{GeV}$ and $p=10~\si{GeV}$ with $N$, using $p_0=5.5,~5,~4.5~\text{and}~4~\si{GeV}$ as the expansion point, for the dressing function.}
	\label{fig:p0choice}
\end{figure}

Let us, now, examine how the errors at $p=1~\si{GeV}$ and $p=10~\si{GeV}$ evolve when we increase the order of approximation $[N|N]$ in a diagonal sequence with $p_0=5.5~\si{GeV}$, represented by blue lines in Figure \ref{fig:p0choice}. We see that, for any value of $N$, the error at $p=1~\si{GeV}$ is the highest of the two, making it the maximum error reached in the interval $[1,10]~\si{GeV}$. The decrease of the error for both values of momentum is more pronounced for lower orders of approximation, up to $N\sim11$. From then on, the decrease in the maximum error is slower. Nonetheless, it is assured that the approximation should not present errors higher than $\sim0.02\%$ for orders of approximation with $N$ greater than 11.

Nonetheless, we notice a discrepancy between the errors at $p=1~\si{GeV}$ and $p=10~\si{GeV}$. We can lower the error at $p=1~\si{GeV}$, without letting the error on the opposite side grow too much, by changing the expansion point. The evolution of the errors at $p=1~\si{GeV}$ and $p=10~\si{GeV}$, is represented in  Figure \ref{fig:p0choice}, for four different values of the expansion point, $p_0=5.5,~5,~4.5~\text{and}~4~\si{GeV}$. This representation shows that the best compromise is obtained for $p_0=4.5~\si{GeV}$, allowing us to have the lowest maximum error for increasing values of $N$. Thus, during the next tests, PAs made around $p_0=4.5~\si{GeV}$ will be considered for both the dressing function and the propagator, since a very similar result can be obtained for the latter\footnote{Although this value may not be globally the best one - a deeper analysis could be made -, it does not have to be very precise. Variations up to $0.5~\si{GeV}$ in $p_0$ translate on minimal magnitude variations of the errors, as Figure \ref{fig:p0choice} suggests.}.

\section{``Padé's hint''} \label{Sec:Hint}
With the expansion point already chosen, we are finally able to calculate PAs. Let us go back and continue the discussion of Section \ref{Sec:LM}, on the relations between $L$ and $M$, and calculate, \eg, the PAs of  orders $[3|3]$ (with $J=0$) and $[3|4]$ (with $J=-1$) for the dressing function. The Padé coefficients are displayed, respectively, in Tables \ref{tab:dress3.3} and \ref{tab:dress3.4}.

\begin{table}[t]
	\begin{subtable}{.5\textwidth}
		\centering
		\begin{tabular}{ccc}
			\toprule
			$i$ & $q_i$ & $r_i$ \\
			\midrule
			0 & \num{5.7e-1} & $-$ \\
			1 & \num{4.3e-2} & \num{7.9e-2} \\
			2 & \num{8.7e-4} & \num{1.7e-3} \\
			3 & {\color{BrickRed}\num{3.7e-6}} & {\color{BrickRed}\num{8.3e-6}} \\
			\bottomrule
		\end{tabular}
		\caption{}
		\label{tab:dress3.3}
	\end{subtable}
	\begin{subtable}{.5\textwidth}
		\centering
		\begin{tabular}{ccc}
			\toprule
			$i$ & $q_i$ & $r_i$ \\
			\midrule
			0 & \num{5.7e-1} & $-$ \\
			1 & \num{4.9e-2} & \num{9.0e-2} \\
			2 & \num{1.2e-3} & \num{2.4e-3} \\
			3 & {\color{BrickRed}\num{8.3e-6}} & {\color{BrickRed}\num{1.8e-5}} \\
			4 & $-$ & \num{3.0e-9} \\
			\bottomrule
		\end{tabular}
		\caption{}
		\label{tab:dress3.4}
	\end{subtable}
	\caption[{Padé coefficients obtained for the orders of approximation $[3|3]$ and $[3|4]$ of the dressing function $d_{gl}(p^2)$.}]{Padé coefficients obtained for the orders of approximation $[3|3]$ (a) and $[3|4]$ (b) of the dressing function $d_{gl}(p^2)$.}
\end{table}

Notice the last coefficients\footnote{Here, the last coefficients are understood as the coefficients of the terms of highest orders.} of each polynomial in the tables mentioned above, from which a careful analysis grants us an important result. In Table  \ref{tab:dress3.3}, the last coefficients, $q_3$ and $r_3$, are both of the same order of magnitude. On the other hand, if we compare the last coefficients in Table \ref{tab:dress3.4}, we see that $r_4$ is three orders of magnitude smaller than $q_3$. However, still in Table \ref{tab:dress3.4}, $r_3$ is just one order of magnitude higher than $q_3$. In general, for any PA of order $[L|M]$ of the dressing function, the following relations between the Padé coefficients can be verified\footnote{The first relation in (\ref{eq:dressCoefficients}), and later in (\ref{eq:propCoefficients}), are considered to be true for differences with a maximum of two orders of magnitude, \ie, $|\log(q_L/r_M)|\lesssim2$.}:
\begin{equation}
\begin{cases}q_L\sim r_M,\quad L=M\\q_L\ll r_M,\quad L>M\\q_L\gg r_M,\quad L<M\end{cases}.
\label{eq:dressCoefficients}
\end{equation}

\begin{table}[t]
	\begin{subtable}{.5\textwidth}
		\centering
		\begin{tabular}{ccc}
			\toprule
			$i$ & $q_i$ & $r_i$ \\
			\midrule
			0 & \num{2.8e-2} & $-$ \\
			1 & \num{1.8e-3} & \num{1.2e-1} \\
			2 & \num{3.1e-5} & \num{4.6e-3} \\
			3 & {\color{BrickRed}\num{1.0e-7}} & \num{6.6e-5} \\
			4 & $-$ & {\color{BrickRed}\num{2.3e-7}} \\
			\bottomrule
		\end{tabular}
		\caption{}
		\label{tab:prop3.4}
	\end{subtable}
	\begin{subtable}{.5\textwidth}
		\centering
		\begin{tabular}{ccc}
			\toprule
			$i$ & $q_i$ & $r_i$ \\
			\midrule
			0 & \num{2.8e-2} & $-$ \\
			1 & \num{2.1e-3} & \num{1.3e-1} \\
			2 & \num{4.5e-5} & \num{5.7e-3} \\
			3 & {\color{BrickRed}\num{2.6e-7}} & \num{1.0e-4} \\
			4 & $-$ & {\color{BrickRed}\num{5.5e-7}} \\
			5 & $-$ & \num{7.4e-11} \\
			\bottomrule
		\end{tabular}
		\caption{}
		\label{tab:prop3.5}
	\end{subtable}
	\caption[{Padé coefficients obtained for the orders of approximation $[3|4]$ and $[3|5]$ of the propagator $D_{gl}(p^2)$.}]{Padé coefficients obtained for the orders of approximation $[3|4]$ (a) and $[3|5]$ (b) of the propagator $D_{gl}(p^2)$.}
\end{table}

Regarding the coefficients for the propagator, we observe that, for the order $[3|4]$, in Table \ref{tab:prop3.4}, the last coefficients are of the same order of magnitude, whereas for the order $[3|5]$, in Table \ref{tab:prop3.5}, the coefficients that are of the same order of magnitude are $q_3$ and $r_4$. Even if we had not discussed earlier, in Section \ref{Sec:LM}, that the relation between $L$ and $M$ should be $M=L-1$ or $M=L-2$, we could arrive to the same conclusion with a diagonal PA. A quick calculation for a PA of order $[3|3]$ shows exactly this: $q_2=\num{1.4e-5}$, $q_3=\num{3.2e-9}$ and $r_3=\num{2.9e-5}$. In this sense, similar relations to (\ref{eq:dressCoefficients}) can be established for the PAs of the propagator,
\begin{equation}
\begin{cases}q_L\sim r_M,\quad L=M-1\\q_L\ll r_M,\quad L>M-1\\q_L\gg r_M,\quad L<M-1\end{cases}.
\label{eq:propCoefficients}
\end{equation}

The relations (\ref{eq:dressCoefficients}) and (\ref{eq:propCoefficients}) seem to indicate that, in a certain way, the PA is sensitive to the relation between $L$ and $M$ that better reproduces the original function. For the dressing function, the PA ``hints'' that the most faithful approximation is achieved for approximants where $M=L$, while for the propagator the PA ``advises'' us to use $L=M-1$. This ability of the PA to tell us the proper relation between $L$ and $M$, and to ``correct it'' if a different relation is considered, will be of great importance later on.

\section{Poles and zeros distribution}
We are now in a position to study the distributions of poles and zeros within PA sequences, and to compare them with the expected analytic structures (Figure \ref{fig:PropAndDressAS}). As it was already mentioned in Section \ref{Sec:ASofPA}, the zeros of a PA are given by the roots of its numerator and the poles by the roots of its denominator.

\begin{figure}[tp]
	\centering
	\includegraphics[width=\textwidth]{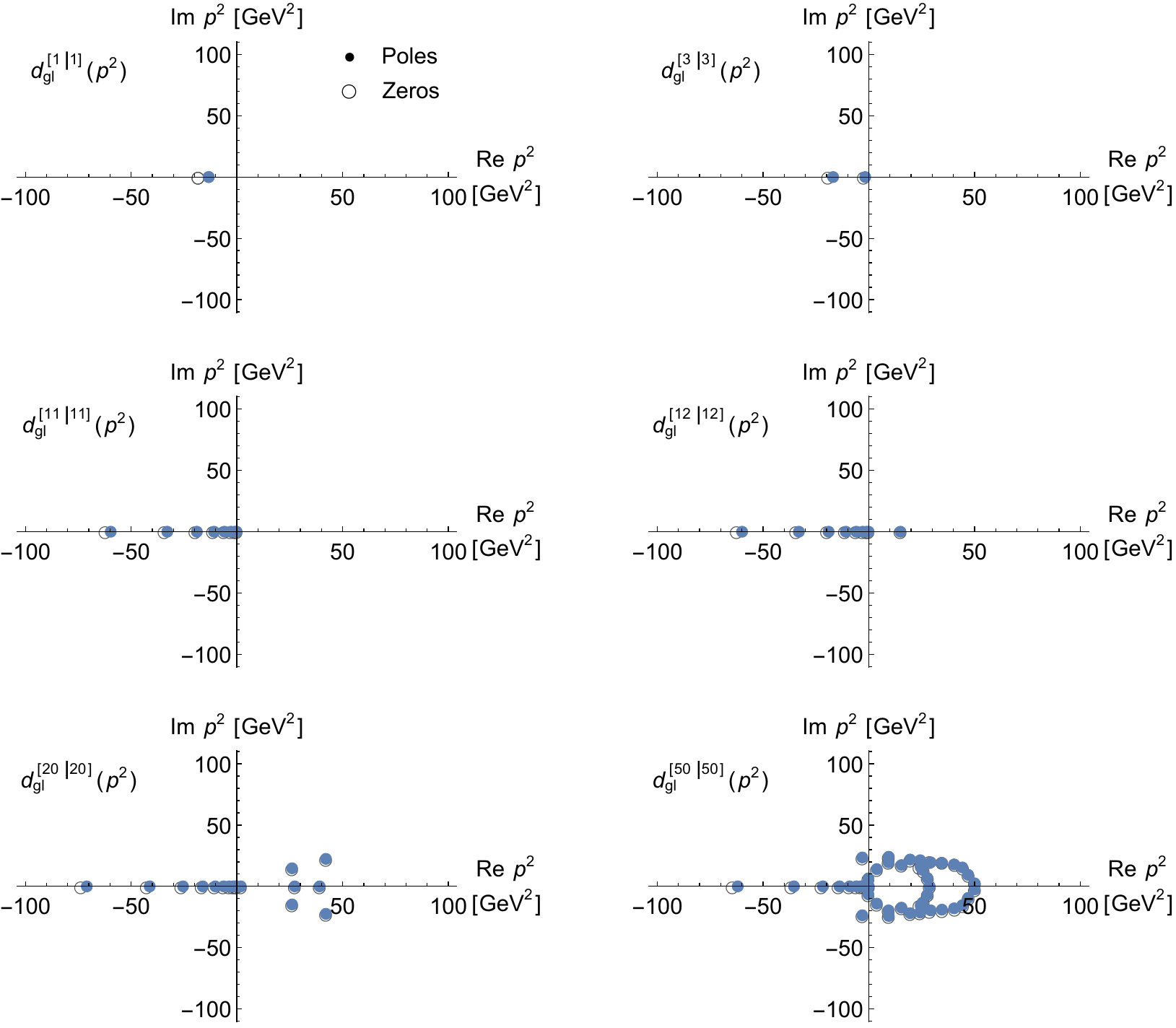}
	\caption{Distribution of poles and zeros for some values of $N$ within the diagonal PA sequence of order $[N|N]$, for the perturbative gluon dressing function.}
	\label{fig:dNN}
\end{figure}

In Figure \ref{fig:dNN}, the distribution of poles and zeros are represented, in the complex $p^2$-plane, for some values of $N$ within the diagonal PA sequence of order $[N|N]$, for the dressing function.

Form $N=1$ to $N=11$ we see an accumulation of alternating poles and zeros on the real negative $p^2$-axis. Following the conclusions of Section \ref{Sec:ASofPA}, this represents the original branch cut, associated with a branch point at $p^2=0$. Beginning at $N=12$, poles and zeros start to emerge in the rest of the complex plane, mostly on its right side. These new poles and zeros come in pairs, \ie, in the same position, and, therefore, the zeros cancel the pole's contributions to the total function.

\vspace{1em}

\begin{figure}[tp]
	\centering
	\includegraphics[width=\textwidth]{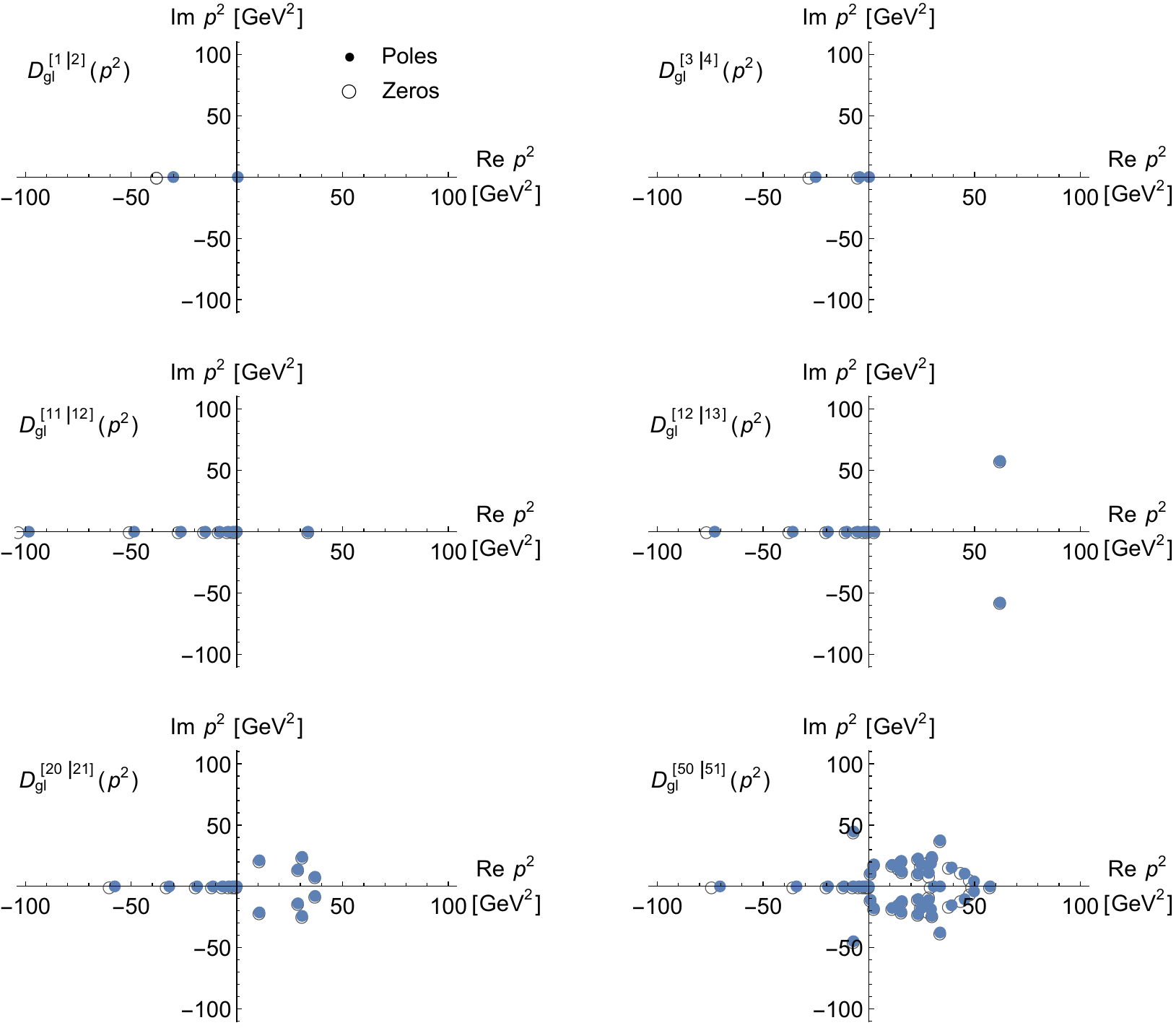}
	\caption{Distribution of poles and zeros for some values of $N$ within the near-diagonal PA sequence of order $[N-1|N]$, for the perturbative gluon propagator.}
	\label{fig:DN1N}
\end{figure}

An analogous behaviour is seen for the near-diagonal PA sequence of order $[N-1|N]$ for the propagator, in Figure \ref{fig:DN1N}. However, this time we face an additional problem: the pole from the factor $1/p^2$ is in the same position as the branch point, so we cannot distinguish them by simply representing the distribution of poles and zeros.

In the next section we will try to solve this problem, by carrying out a residue analysis.

\section{Residue analysis} \label{Sec:ResidueAnalysis}
Consider, for example, the distribution of poles and zeros of the PA of order $[50|50]$, for the dressing function (Figure \ref{fig:dNN}). Amidst so many poles and zeros, how can we separate the real ones from those that are artefacts of the method? We need a mathematical tool to decide if a pole is part of the analytic structure, or if it is just cancelled by a zero, forming a Froissart doublet. The use of the residues to sort out the undesired poles is suggested in \cite{Yamada2014}.

\begin{figure}[tp]
	\centering
	\includegraphics[width=\textwidth]{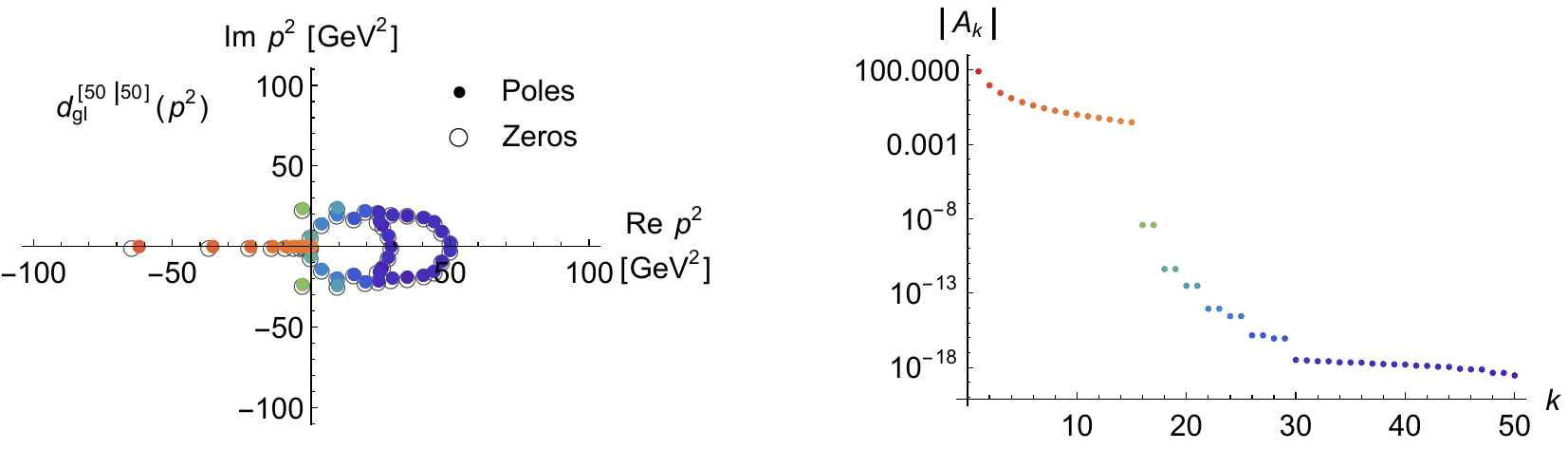}
	\caption[{Distribution of poles and zeros for the PA of order $[50|50]$ of $d_{gl}(p^2)$, and the absolute value of the residues $|A_k|$ for each pole $k$.}]{Distribution of poles and zeros for the PA of order $[50|50]$ of $d_{gl}(p^2)$ (left), and the absolute value of the residues $|A_k|$ for each pole $k$ (right). The values of $|A_k|$ are arranged in descending order. The colour code used in the left graphic corresponds to the one used in the right graphic, for the residues' absolute value.}
	\label{fig:dRes50}
\end{figure}

In Figure \ref{fig:dRes50}, the distribution of poles and zeros obtained from the order $[50|50]$ PA is represented along with the absolute values of the respective residues\footnote{Recalling the definition of residue of a complex function $f(z)$ at $z=z_0$, it is defined as the coefficient of $(z-z_0)^{-1}$ in the respective Laurent expansion \cite{Arfken}.} $|A_k|$ for each pole $k$ \footnote{Throughout this work, the residues were numerically computed using the software \textit{Mathematica} \cite{Mathematica}.}. The existence of two distinct levels is clear. The first one, in red tones, with residue values above $\num{e-2}$, corresponds to the poles lying on the real negative $p^2$-axis. The second one, in green/blue tones, corresponds to the poles belonging to the pole-zero pairings. We can clear these last poles out of the representation by performing a \textit{cut} in the residues at $|A_k|=\num{e-2}$, \ie, by only representing the poles $k$ for which $|A_k|\gtrsim\num{e-2}$.

\begin{figure}[tp]
	\centering
	\includegraphics[width=.8\textwidth]{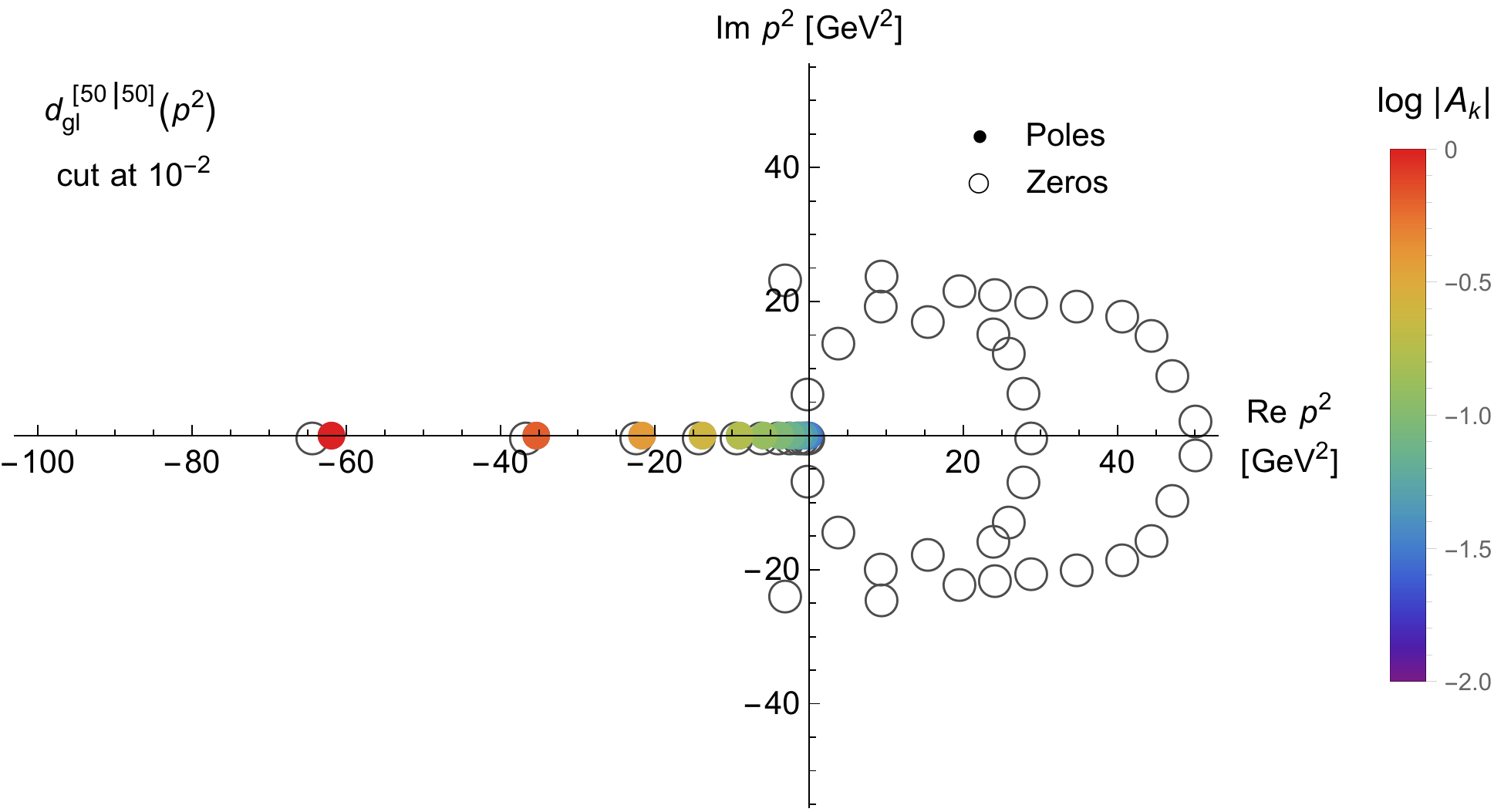}
	\caption[{Distribution of poles and zeros for the PA of order $[50|50]$ of $d_{gl}(p^2)$, with a cut in the residues at $|A_k|=\num{e-2}$.}]{Distribution of poles and zeros for the PA of order $[50|50]$ of $d_{gl}(p^2)$, with a cut in the residues at $|A_k|=\num{e-2}$. The colour scheme codes the residue's absolute value of each pole.}
	\label{fig:dRes50cut}
\end{figure}

For the PA of order $[50|50]$, the distribution of poles and zeros with a cut at $|A_k|=\num{e-2}$ is represented in Figure \ref{fig:dRes50cut}. We see that only the poles on the real negative $p^2$-axis remain after the cut. In this way, the  branch cut of the dressing function is faithfully reproduced by the poles that remain after performing the cuts in the residues.

\begin{figure}[tp]
	\centering
	\includegraphics[width=.8\textwidth]{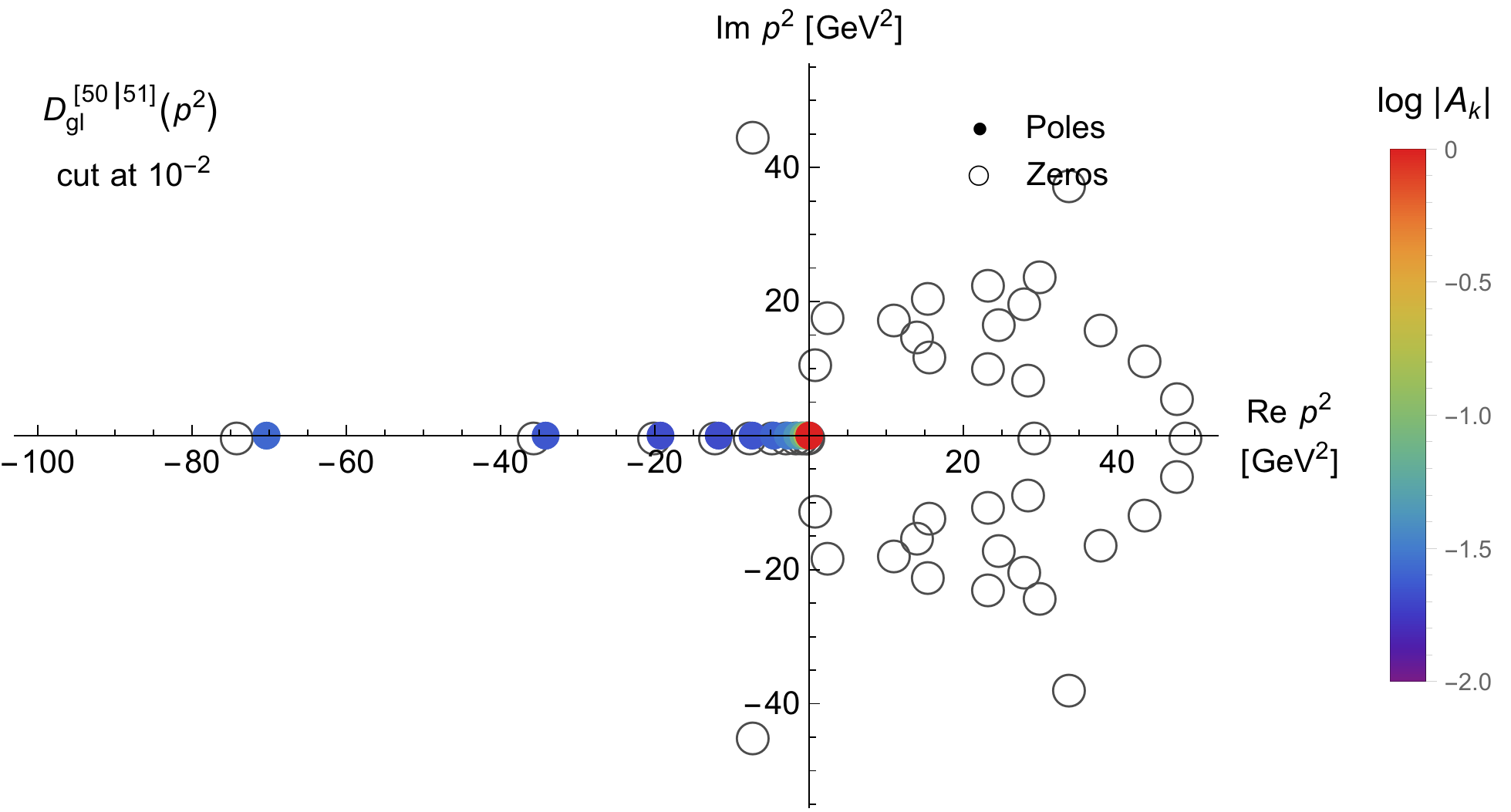}
	\caption[{Distribution of poles and zeros for the PA of order $[50|51]$ of $D_{gl}(p^2)$, with a cut in the residues at $|A_k|=\num{e-2}$.}]{Distribution of poles and zeros for the PA of order $[50|51]$ of $D_{gl}(p^2)$, with a cut in the residues at $|A_k|=\num{e-2}$. The colour scheme codes the residue's absolute value of each pole.}
	\label{fig:DpRes50cut}
\end{figure}

Let us, now, compare the distribution of poles and zeros for the PA of order $[50|50]$ for the dressing function with the distribution of poles and zeros for the PA of order $[50|51]$ of the propagator, represented in Figure \ref{fig:DpRes50cut}, already with the cut in the residues. With an attentive look, we observe that, while for the dressing function the absolute value of the residues decreases as the poles get nearer the branch point, the opposite happens to the propagator, thus suggesting the presence of a pole at the origin, as expected.

\section{Adding mass generation terms}
In order to investigate the reproduction of poles and branch cuts in other locations of the complex $p^2$-plane, we can change their positions by adding mass terms, $m_1^2$ and $m_2^2$, to $D_{gl}(p^2)$ and $d_{gl}(p^2)$,
\begin{equation}
D_{gl}(p^2)=\frac{1}{p^2+m_2^2}\left[\frac{11N_f\alpha_s}{12\pi}\ln\left(\frac{p^2+m_1^2}{\Lambda^2}\right)+1\right]^{-\gamma},
\label{Eq:PropM}
\end{equation}
\begin{equation}
d_{gl}(p^2)=\left[\frac{11N_f\alpha_s}{12\pi}\ln\left(\frac{p^2+m_1^2}{\Lambda^2}\right)+1\right]^{-\gamma}.
\label{Eq:DressM}
\end{equation}
Thereby, the branch points are moved from the origin to $p^2=-m_1^2$, both in $D_{gl}(p^2)$ and $d_{gl}(p^2)$, and the pole of $D_{gl}(p^2)$ will now appear at $p^2=-m_2^2$.

\begin{figure}[tp]
	\centering
	\includegraphics[width=\textwidth]{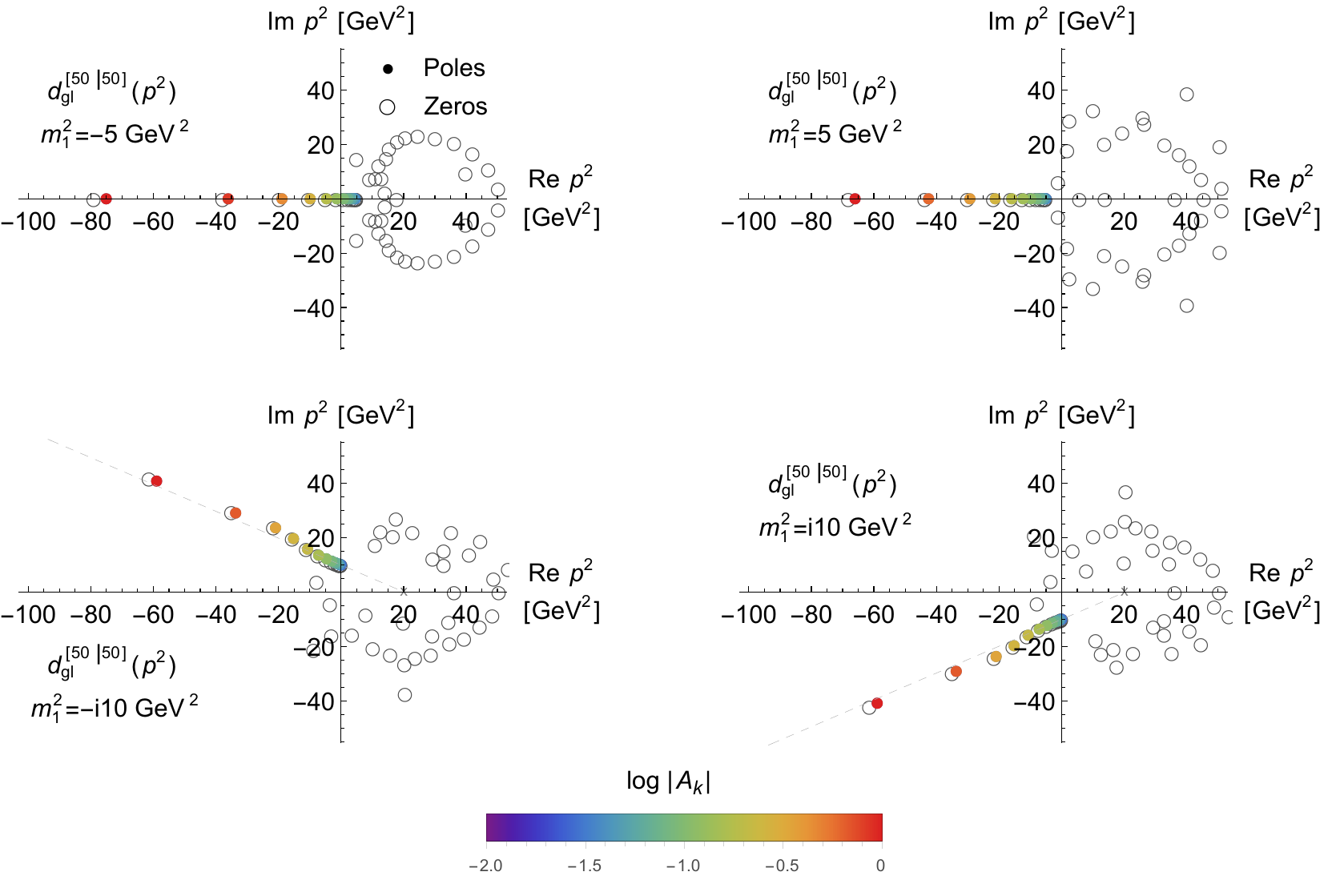}
	\caption[{Distribution of poles and zeros for the elements of the PA sequences of order $[N|N]$ with $N=50$ of $d_{gl}(p^2)$, obtained for $m_1^2=-5,~5,~-i10$ and $i10~\si{GeV^2}$.}]{Distribution of poles and zeros for the elements of the PA sequences of order $[N|N]$ with $N=50$ of $d_{gl}(p^2)$, obtained for $m_1^2=-5,~5,~-i10~\text{and}~i10~\si{GeV^2}$. The cut in the residues has been done at $|A_k|=\num{e-2}$, by following the residue analysis of Section \ref{Sec:ResidueAnalysis}. The colour scheme codes the residue's absolute value of each pole.}
	\label{fig:BCdance}
\end{figure}

As an example, the mass term $m_1^2$ in $d_{gl}(p^2)$ is set to four different values: $-5$, $5$, $-i10$ and $i10~\si{GeV^2}$. These should cause a translation of the branch point in the complex $p^2$-plane to $p^2=5,~-5,~i10~\text{and}~-i10~\si{GeV^2}$, respectively. However, these changes should not alter the direction of the branch cut, which is, according to Figure \ref{fig:PropAndDressAS}, parallel to the real $p^2$-axis and goes from the branch point to the left side of the plane. In Figure \ref{fig:BCdance}, the poles and zeros distributions obtained for the approximants in the PA sequences of order $[N|N]$ with $N=50$ \footnote{Despite the fact that only one element from each PA sequence is presented here, the positions of the important poles in the distribution of poles and zeros, according to the residue analysis, are very stable throughout the sequences, and, therefore, they are not shown.} are represented for the four values of $m_1^2$.

In Figure \ref{fig:BCdance}, we see that, for $m_1^2=-5~\text{and}~5~\si{GeV^2}$, the branch cut and the branch point are in the expected positions and, thus, correctly identified. Notwithstanding, for $m_1^2=-i10~\text{and}~i10~\si{GeV^2}$, the branch cut is not so well reproduced\footnote{This is made based on the convention that the branch cut from the logarithm makes an angle of $\pi$ with the real axis in the complex plane.}. Despite the correct identification of the branch point, the identified branch cut is not parallel to the real $p^2$-axis, as anticipated. In fact, the grey dashed lines (Figure \ref{fig:BCdance}) reveal that the branch cut has the direction that connects the branch point to the expansion point\footnote{The same result can be obtained by moving the expansion point through the complex $p^2$-plane.}.

\vspace{1em}

\begin{figure}[tp]
	\centering
	\includegraphics[width=\textwidth]{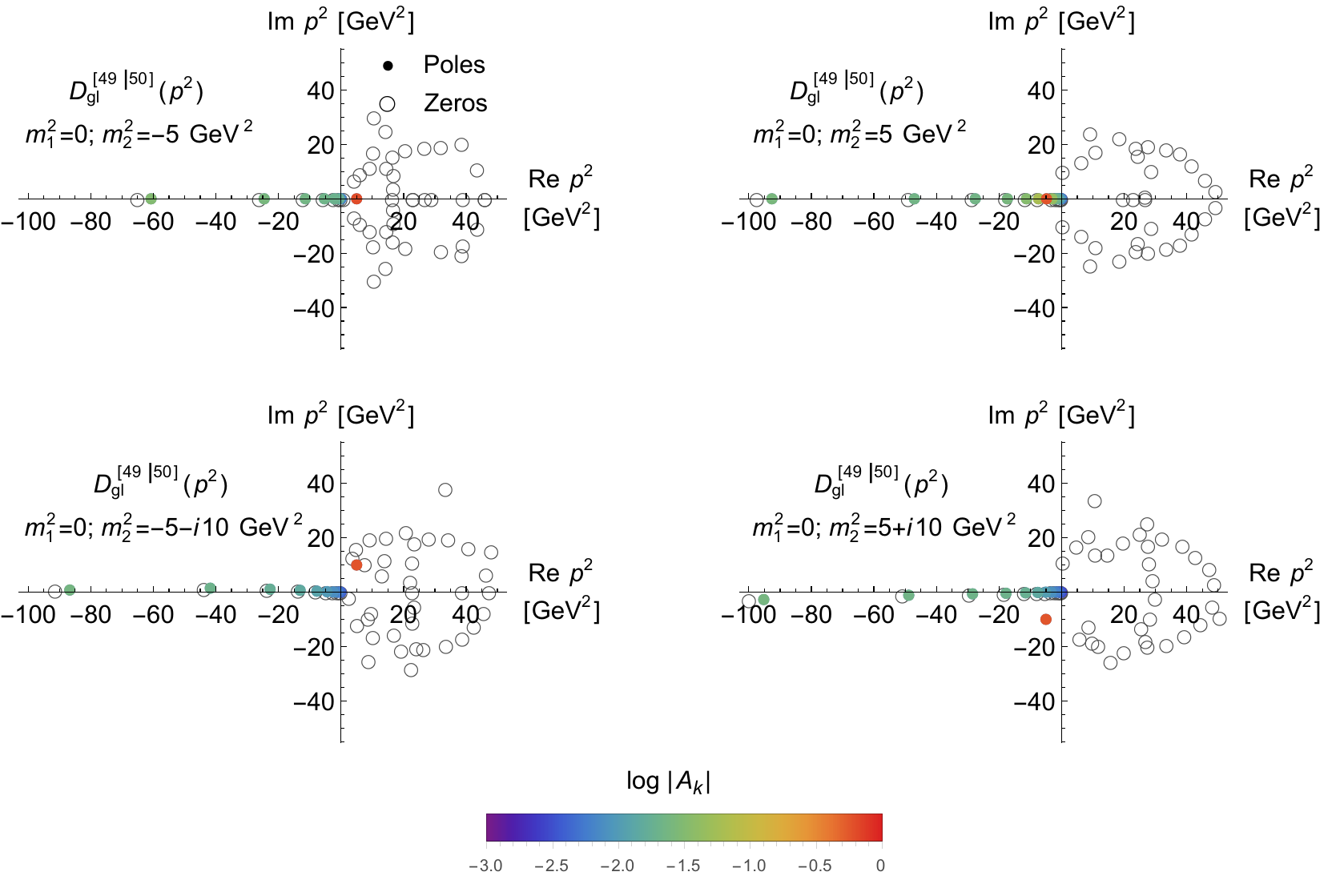}
	\caption[{Distribution of poles and zeros for the elements of the PA sequences of order $[N-1|N]$ of $D_{gl}(p^2)$ with $N=50$, obtained for $m_2^2=-5,~5,~-5-i10~\text{and}~5+10~\si{GeV^2}$ and $m_1^2=0$.}]{Distribution of poles and zeros for the elements of the PA sequences of order $[N-1|N]$ of $D_{gl}(p^2)$ with $N=50$, obtained for $m_2^2=-5,~5,~-5-i10~\text{and}~5+10~\si{GeV^2}$ and $m_1^2=0$. The cut in the residues has been done at $|A_k|=\num{e-2}$, by following the residues analysis of Section \ref{Sec:ResidueAnalysis}. The colour scheme codes the residue's absolute value of each pole.}
	\label{fig:Pdance}
\end{figure}

Lastly, we need to examine the pole's and branch cut's behaviour for the propagator when the mass term $m_2^2$ is introduced. In Figure \ref{fig:Pdance}, the distribution of poles and zeros obtained from the approximants within the PA sequences of order $[N-1|N]$ with $N=50$ \footnote{Again, only one element from each PA sequence is presented here, due to the stability of the important poles and structures that emerge in the poles and zeros distributions throughout the sequences.} are represented for four different values of $m_2^2$: $-5,~5,~-5-i10~\text{and}~5+10~\si{GeV^2}$.

A quick analysis of Figure \ref{fig:Pdance} shows that, for the four different values of $m_2^2$, the pole and the branch point are in the expected positions. Thus, the position of these singularities in the analytic structure of the propagator are correctly identified by the distribution of poles and zeros of the PA.

\section{The perturbative result for the ghost propagator}
Regarding the ghost, the expression for its propagator and for its dressing function, obtained perturbatively, are very similar to the ones of the gluon. Indeed, the only difference between them is the anomalous dimension value, which is $\delta=9/44$ for the ghost.

The tests made throughout the sections above were repeated for the ghost propagator and identical results were obtained. For this reason, the conclusions drawn in this chapter are valid for both particles.

\chapter{Approximating a discrete set of points} \label{Chap:Discrete}
The tests performed in the last chapter, move us a step closer in our journey to find the analytic structure of the gluon and ghost non-perturbative propagators. To do so, we have to approximate the lattice data to PAs. This is exactly when we reach a \textit{cul-de-sac}. Until now, the tests were all done with full knowledge of the original function's analytic expression. Now, we are confronted with a discrete set of data points, with associated statistical errors.

In this chapter, we introduce a methodology to approximate a discrete set of data points to a PA. The method will be tested against data sets generated from a collection of simple functions, before applying it to the lattice data for the gluon and ghost propagators.

\section{A ``simple fit''} \label{Sec:SimpleFit}
The problem of building a PA from a discrete set of points, and not based on the Taylor expansion of a function, is not new. In fact, several methods used to approximate a set of data points by a rational function were already explored and employed in Physics, in situations of numerical analytic continuation. The \textit{Norm Method} and the \textit{Moment Method} - both require solving a system of equations -, and the \textit{Point Method} (SPM) - in which the coefficients can be determined recursively -, all introduced by Schlessinger \cite{Schlessinger1968}, are examples of the usage of PAs, \eg, in \cite{Haymaker1970,Tripolt2019,Binosi2020}. Some of these methods were tested in the context of this work, but with very poor results concerning the quality of approximation. Moreover, none of these methods take into account the statistical errors of the data. For these reasons, a different approach must be considered.

Essentially, we want to reproduce a function $f(x)$ based on a number $K$ of data points $\{(x_1,y_1),...,(x_K,y_K)\}$, with a statistical error $\sigma_i$ associated with each value $y_i$. To do so, these data points will be approximated by a ratio of polynomials, \ie, a PA. The Padé coefficients can be computed by minimising an objective function that measures the deviation of the data points to the approximated function. Throughout the current work, we will use the chi squared $\chi^2$ as the objective function, which is defined as
\begin{equation}
\chi^2\equiv\sum_{i=1}^K\left(\frac{y_i-f(x_i)}{\sigma_i}\right)^2,
\end{equation}
where $\sigma_i$ is the statistical error associated with the value $y_i$. The quality of approximation can be measured from the reduced chi squared, given by
\begin{equation}
\widetilde{\chi}^2\equiv\frac{\chi^2}{\text{degrees of freedom}},
\end{equation}
where the number of degrees of freedom is given by the difference between the number $K$ of points to be approximated and the number of Padé coefficients to be calculated. A good approximation to the data translates into a reduced chi squared close to unit, \ie, $\widetilde{\chi}^2\sim1$.

From a mathematical point of view, we are dealing with a non-linear global optimisation problem of determining the absolute minimum of $\chi^2$. Global optimisation problems are non-trivial, and there is no available method that solves such class of problems for any function. Herein, the minimisation was done numerically using \textit{Mathematica} \cite{Mathematica}. The minimisation methods used in this work for this global optimisation problem were the \textit{Differential Evolution} (DE) method \cite{DifferentialEvolution}, and the \textit{Simulated Annealing} (SA) method \cite{SimulatedAnnealing}\footnote{The \textit{Levenberg Marquardt} method \cite{LevenbergMarquardt} was also used, but with very unstable results, compared with the other two methods.}.
Both are stochastic optimisation methods. The first one, the DE, relies on the maintenance of a population of points, from which a new one is generated based on random processes. This population evolves to explore the search space, in order to escape possible local minima. The convergence is achieved when the best points of two consecutive populations are inside the chosen tolerance. As for the SA method, it is inspired by the physical/metallurgic process of annealing. At each iteration, a new point in the research space is randomly generated close to the previous one, replacing it if a lower value is reached. However, the algorithm allows it to be exchanged for a point with a higher value, with a probability that follows a Boltzmann distribution, giving the possibility of escaping local minima. As the number of iterations increases, the probability of a replacement for a point with a higher value decreases, simulating a decrease in temperature. The process ends when the maximum number of iterations is reached, or the method converges to a point within the chosen tolerance.

In order to test the reliability of this methodology, we first apply it to test data sets, generated from given functions.

\section{Reproduction of the analytic structure}
The functions that will be used to generate the test data sets are:
\begin{align}
f_1(p^2) &= \ln(p^2+m^2), \\
f_2(p^2) &= \frac{1}{p^2+m^2}, \\
f_3(p^2) &= D_{gl}(p^2+m^2),
\end{align}
where $D_{gl}(p^2)$ is the perturbative gluon propagator given by (\ref{Eq:Prop}) and, in all cases, $p^2$ is dimensionless. For the mass terms, the cases with $m^2=0$ and $m^2=0.5$ will be explored. These functions give rise to the kind of singularities that are expected to be found in the analysis of the lattice data. Their analytic structures are:
\begin{itemize}
	\item a pole at $p^2=-m^2$, for $f_1(p^2)$;
	\item a branch cut parallel to the real axis with the branch point at $p^2=-m^2$, for $f_2(p^2)$;
	\item a pole at $p^2=-m^2$ and a branch cut parallel to the real axis with the branch point also at $p^2=-m^2$, for $f_3(p^2)$.
\end{itemize}

\vspace{1em}

For each function $f_i(p^2)$, a set of $K$ points $(p_j,f_i(p_j^2))$, with $j=1,...,K$, is generated by randomly choosing $K$ values of $p_j$ in the chosen interval and calculating $f_i(p_j^2)$ for each one. The statistical errors can be simulated by replacing the value of $f_i(p_j^2)$, at each $p_j$, by a random value near the original one, \ie, $f_i(p_j^2)\to y_j=f_i(p_j^2)(1+\varepsilon \rho)$, where $\rho$ is a random number that follows a Gaussian distribution centred in 0 and with a standard deviation of 1, and $\varepsilon$ is the desired percent error. Hence, the statistical error associated with $y_j$ is $\sigma_j=\varepsilon f_i(p_j^2)$. When it comes to the minimisation, each point $(p_j,y_j)$ will contribute more or less to it based on the associated weight, given by $1/\sigma_j^2$.

The lattice data for the gluon and the ghost propagators that will be used has, in all cases, more than a hundred data points in the range $p\in[0,8]~\si{GeV}$, with statistical errors between $\sim1\%$ and $\sim0.1\%$. In this sense, test data sets of 100 points distributed in the range $p\in[0,8]$, with $\varepsilon=1\%$, were generated.

Following the conclusions of Chapter \ref{Chap:AnalyticTests}, near-diagonal PA sequences of order $[N-1|N]$, with $N$ going from 1 to 20, will be calculated, and the PAs themselves will take be $p^2$ as the independent variable.

\subsection{Identifying a branch cut} 
\begin{figure}[tp]
	\centering
	\includegraphics[width=\textwidth]{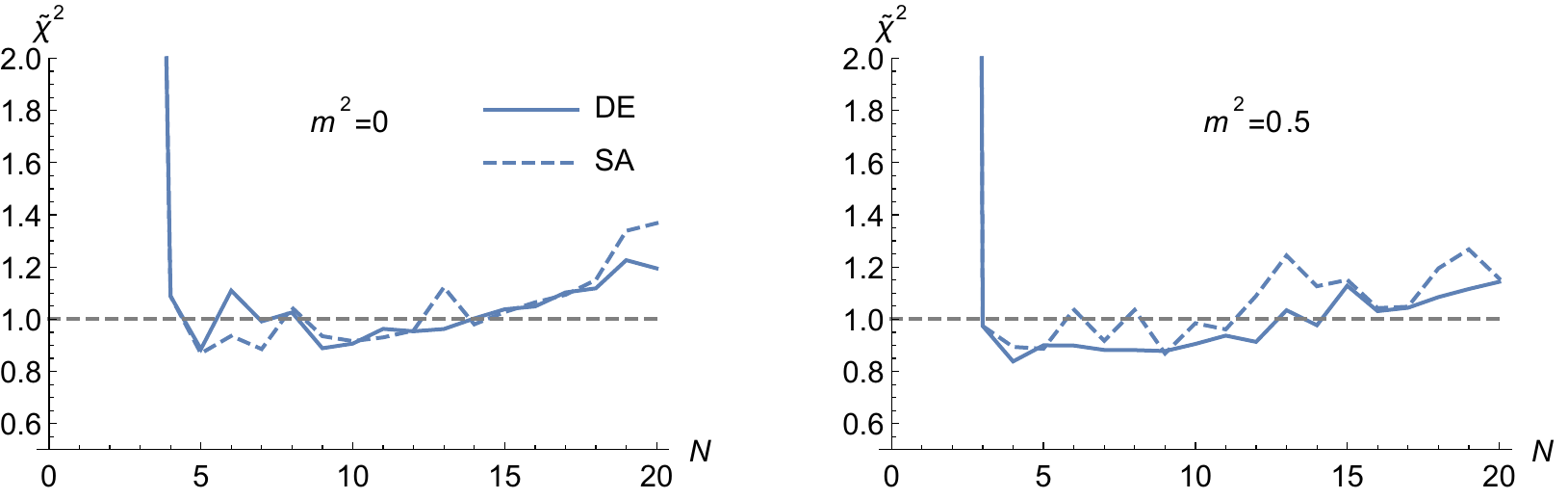}
	\caption{Achieved values of $\widetilde{\chi}^2$ for the test data generated from $f_1(p^2)$, using $m^2=0$ and $m^2=0.5$, and for both methods of minimisation, DE and SA.}
	\label{fig:Tt1chi2red}
\end{figure}

\begin{figure}[tp]
	\centering
	\includegraphics[width=.85\textwidth]{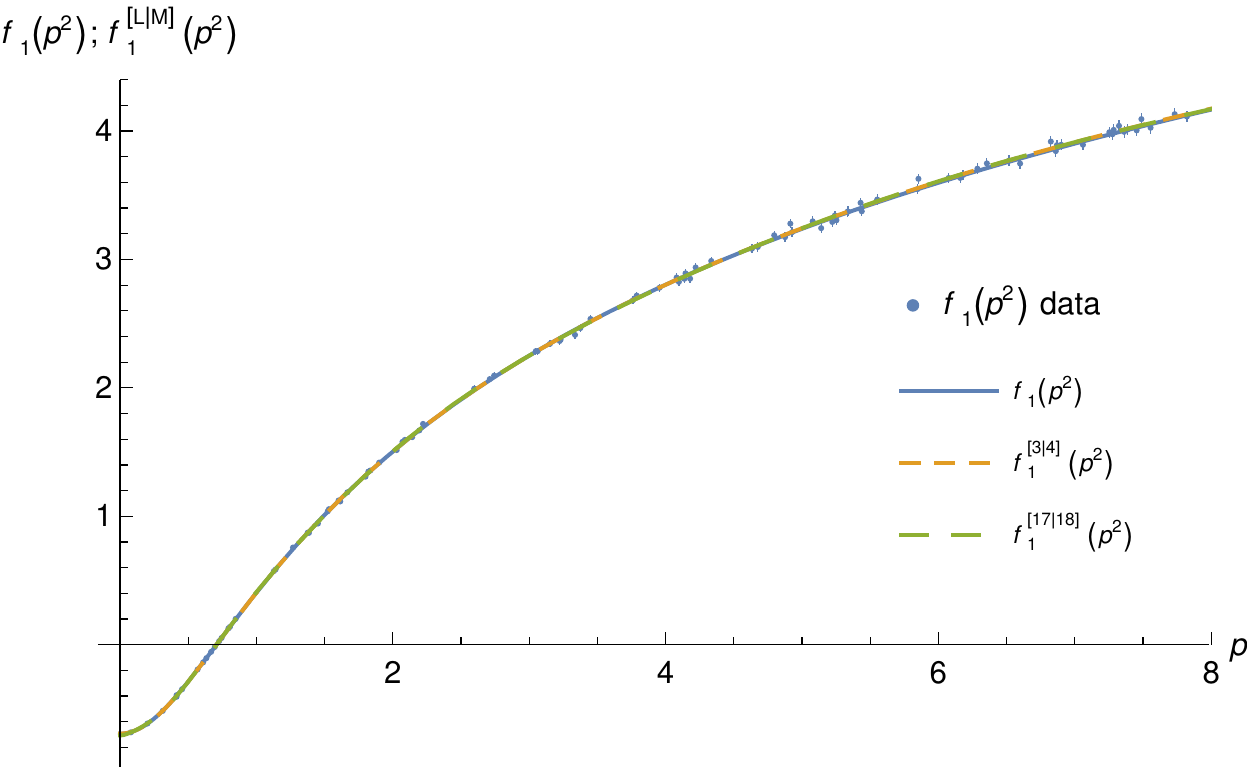}
	\caption{Representation of the test data points generated from the $f_1(p^2)$, which is also represented, using $m^2=0.5$ and DE, together with the obtained approximants of orders $[3|4]$ and $[17|18]$.}
	\label{fig:Tt1curve}
\end{figure}

In Figure \ref{fig:Tt1chi2red}, the achieved values of $\widetilde{\chi}^2$ for $f_1(p^2)$, using $m^2=0$ and $m^2=0.5$, are represented for the two minimisation methods, DE and SA. The obtained values, which are close to unit, show the quality of the minimisation, and also show that the data is well adjusted by PAs. In Figure \ref{fig:Tt1curve}, we can visualise an example that illustrates how well the generated data is adjusted, by representing simultaneously the function $f_1(p^2)$, the respective generated data, and some PAs obtained within the calculated sequence for, for example, $m^2=0.5$ and DE\footnote{In general, a very good adjustment can be observed in graphical representations beginning at low orders of approximation, usually $N\sim3$. This type of representation will not be presented for more cases, to avoid overloading this work with unnecessary figures. The approximations' quality will be evaluated only through the analysis of $\widetilde{\chi}^2$.}.

\begin{figure}[tp]
	\centering
	\includegraphics[width=\textwidth]{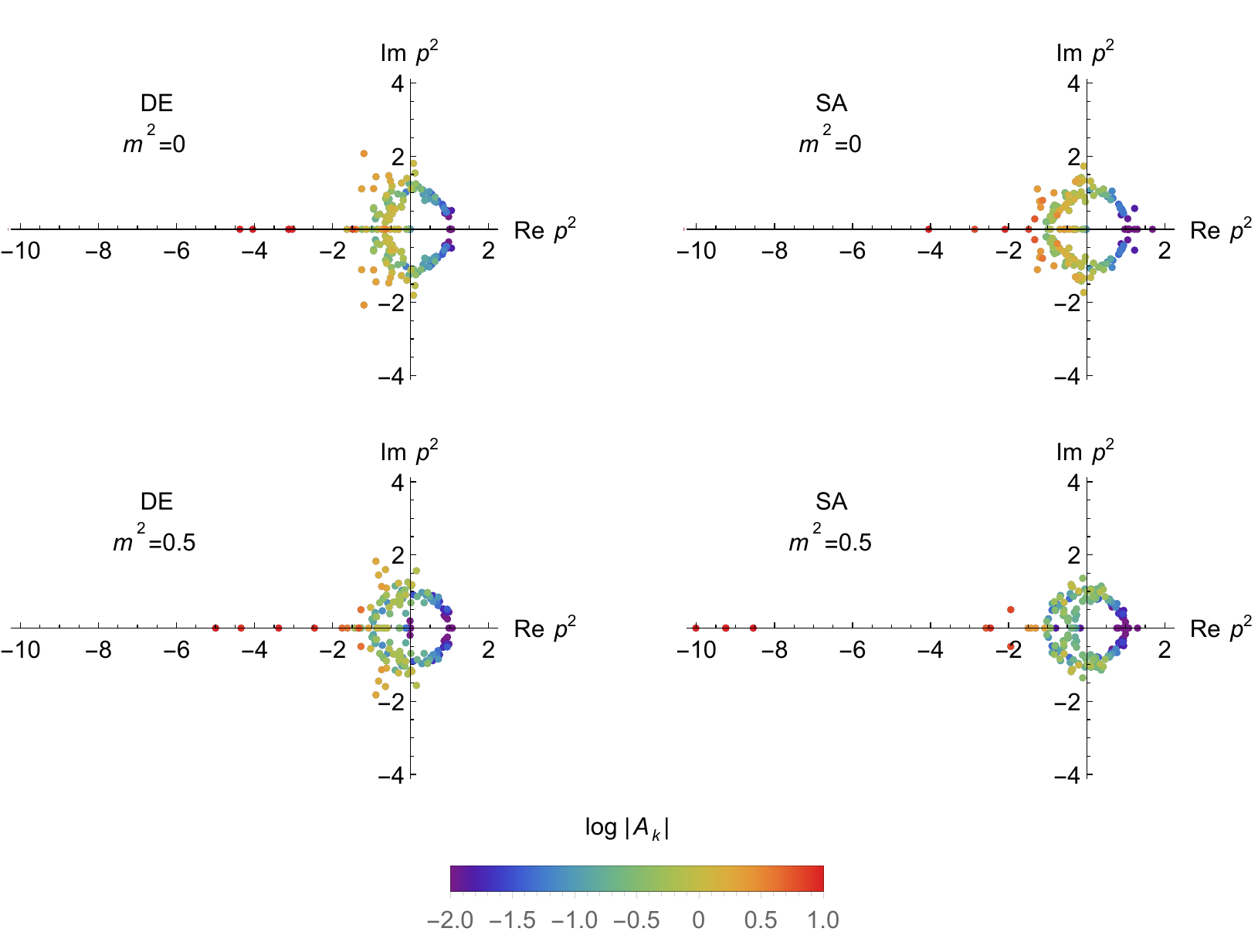}
	\caption[{All-poles representation obtained for the test data generated from $f_1(p^2)$, using $m^2=0$ and $m^2=0.5$, and for both methods of minimisation, DE and SA.}]{All-poles representation obtained for the test data generated from $f_1(p^2)$, using $m^2=0$ and $m^2=0.5$, and for both methods of minimisation, DE and SA. The colour scheme codes the residue's absolute value of each pole.}
	\label{fig:Tt1all}
\end{figure}

A summary of the results can be made by representing simultaneously, in the complex $p^2$-plane, the poles obtained for all values of $N$. In Figure \ref{fig:Tt1all}, this \textit{all-poles representation} is made for $f_1(p^2)$, using $m^2=0$ and $m^2=0.5$, and for both methods of minimisation, DE and SA. A circumference of unit radius, formed by poles with residues between $\sim\num{e-2}$, in purple, and $\sim\num{e0}$, in yellow/orange, is seen around the origin. In fact, this type of structures are an artefact of the approximation method, created by poles of the Froissart doublets, already studied in \cite{Yamada2014}. The fact that the poles on the right-hand side of the complex $p^2$-plane have, in general, smaller residues, and are much less scattered than the ones on the left-hand side, reflects the difficulty on identifying the analytic structure beyond the region where the data is defined. We also see an accumulation of poles with high residue on the real negative $p^2$-axis, which suggests the presence of a branch cut. However, we cannot be deceived by this all-poles representation, as it only serves as an overview of the structures that emerge in the PA sequence. An analysis of the evolution of the poles and zeros distributions with $N$ is required.

\begin{figure}[p]
	\centering
	\includegraphics[width=.9\textwidth]{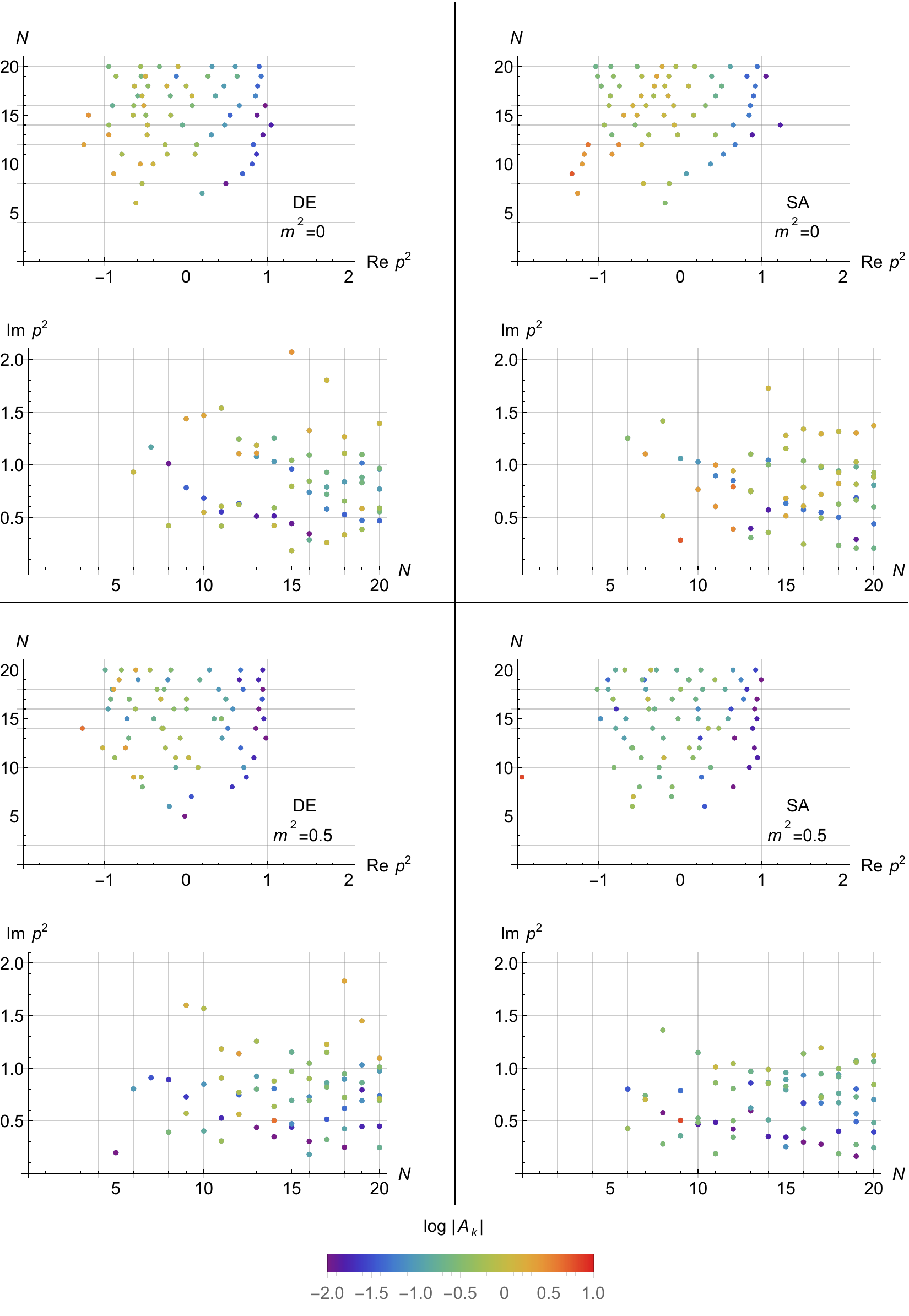}
	\caption[{Evolution of the off-axis poles and zeros with $N$ in the PA sequence, obtained from the test data generated from $f_1(p^2)$, using $m^2=0$ and $m^2=0.5$, and for both methods of minimisation, DE and SA.}]{Evolution of the off-axis poles and zeros with $N$ in the PA sequence, obtained from the test data generated from $f_1(p^2)$, using $m^2=0$ and $m^2=0.5$, and for both methods of minimisation, DE and SA. The colour scheme codes the residue's absolute value of each pole.}
	\label{fig:Tt1offaxis}
\end{figure}

The evolution, with $N$, of the off-axis poles, \ie, poles that appear at $\text{Re}(p^2)\neq0$, is represented in Figure \ref{fig:Tt1offaxis}, for $f_1(p^2)$ with $m^2=0$ and $m^2=0.5$ and for both DE and SA. The relevant poles\footnote{When PAs are used to approximate a discrete set of data points with associated statistical errors, a residue analysis is impracticable, as it was presented in Section \ref{Sec:ResidueAnalysis}, since the levels in the absolute values of the residues become indistinguishable. Thus, from now on, the residue analysis is done based on the relative value of the poles' residues, \ie, poles with higher residues are considered to be more relevant.}, in orange/red tones, are not stable throughout the PA sequence for any case, since they are mixed with the poles from the Froissart doublets in the circumference of unit radius. Following the conclusions of Chapter \ref{Chap:PA}, these can be considered spurious poles.

\begin{figure}[p]
	\centering
	\includegraphics[width=.9\textwidth]{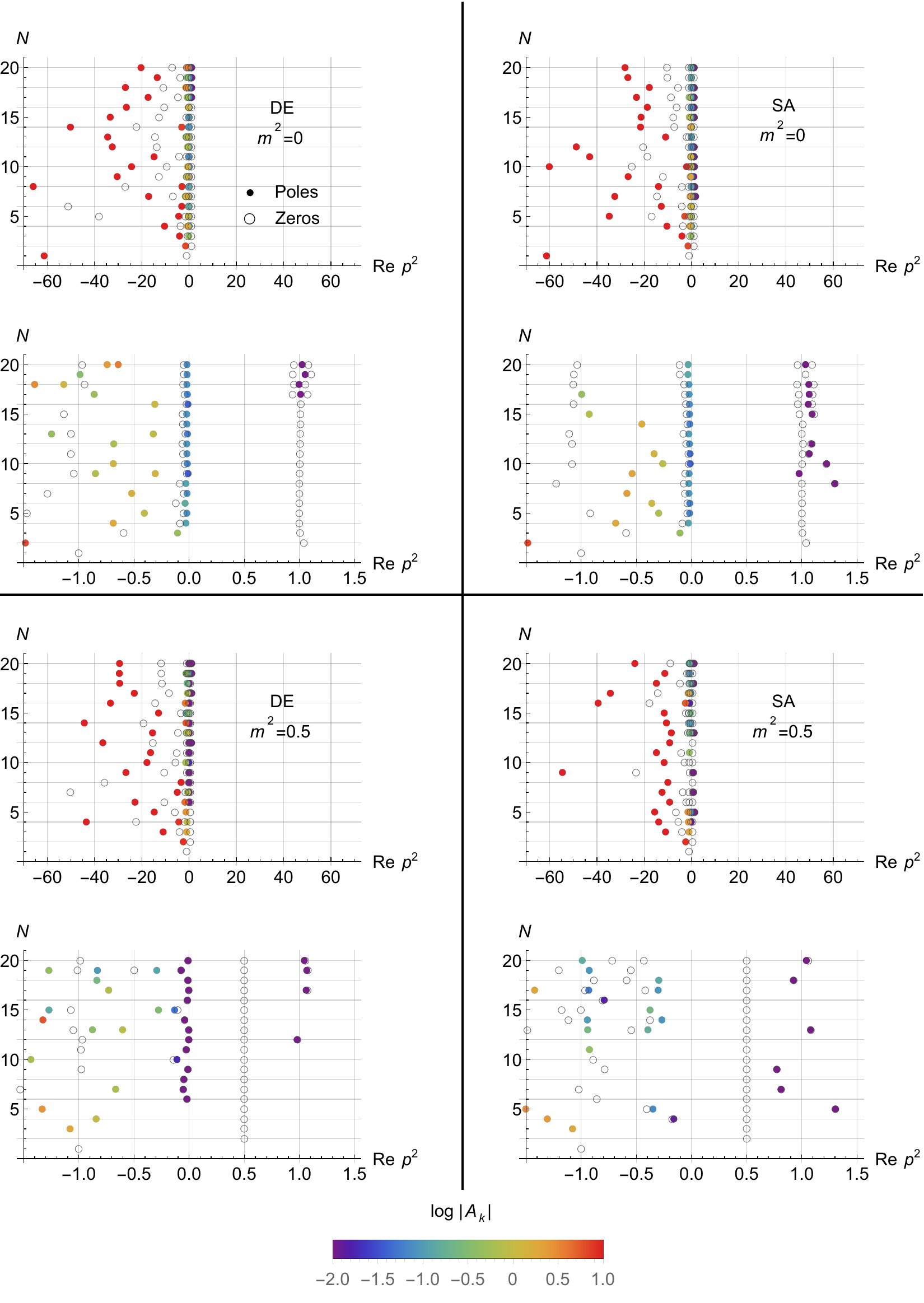}
	\caption[{Evolution of the on-axis poles and zeros with $N$ in the PA sequence, obtained from the test data generated from $f_1(p^2)$, using $m^2=0$ and $m^2=0.5$, and for both methods of minimisation, DE and SA.}]{Evolution of the on-axis poles and zeros with $N$ in the PA sequence, obtained from the test data generated from $f_1(p^2)$, using $m^2=0$ and $m^2=0.5$, and for both methods of minimisation, DE and SA. The colour scheme codes the residue's absolute value of each pole.}
	\label{fig:Tt1onaxis}
\end{figure}

The evolution of the on-axis poles and zeros, \ie, poles and zeros that appear on the real $p^2$-axis, is shown in Figure \ref{fig:Tt1onaxis}, for $f_1(p^2)$ with $m^2=0$ and $m^2=0.5$ and for both DE and SA. For all orders of approximation the branch cut is well reproduced by alternating poles and zeros on the real negative $p^2$-axis, in the same way as the branch cuts were reproduced in Chapters \ref{Chap:PA} and \ref{Chap:AnalyticTests}. However, whereas for the case with $m^2=0$ the branch point is well identified, the alternating poles and zeros begin at the origin, for $m^2=0.5$ the position of the branch point is not so clear. Nonetheless, there is a difference between the results for the two minimisation methods: for the DE method, unimportant poles, according to the respective residues, appear near the origin; while for the SA method this is not seen, for which the nearest poles, on the real negative semiaxis, are at $p^2\sim-0.3$.

\subsection{Looking at a single pole} 
\begin{figure}[tp]
	\centering
	\includegraphics[width=\textwidth]{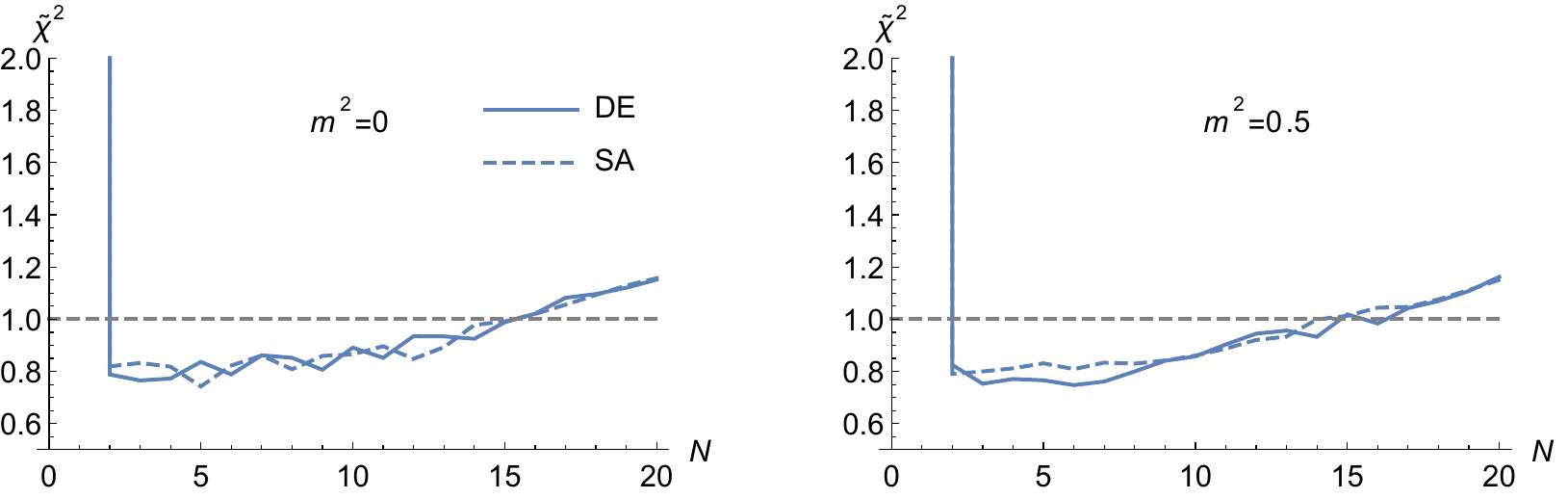}
	\caption{Achieved values of $\widetilde{\chi}^2$ for the test data generated from $f_2(p^2)$, using $m^2=0$ and $m^2=0.5$, and for both methods of minimisation, DE and SA.}
	\label{fig:Tt2chi2red}
\end{figure}

In Figure \ref{fig:Tt2chi2red}, the $\widetilde{\chi}^2$ achieved in the minimisation for $f_2(p^2)$, using $m^2=0$ and $m^2=0.5$, is represented for both minimisation methods. Again, the values of $\widetilde{\chi}^2$ show that the PAs provide a good approximation to the data and the function.

\begin{figure}[tp]
	\centering
	\includegraphics[width=\textwidth]{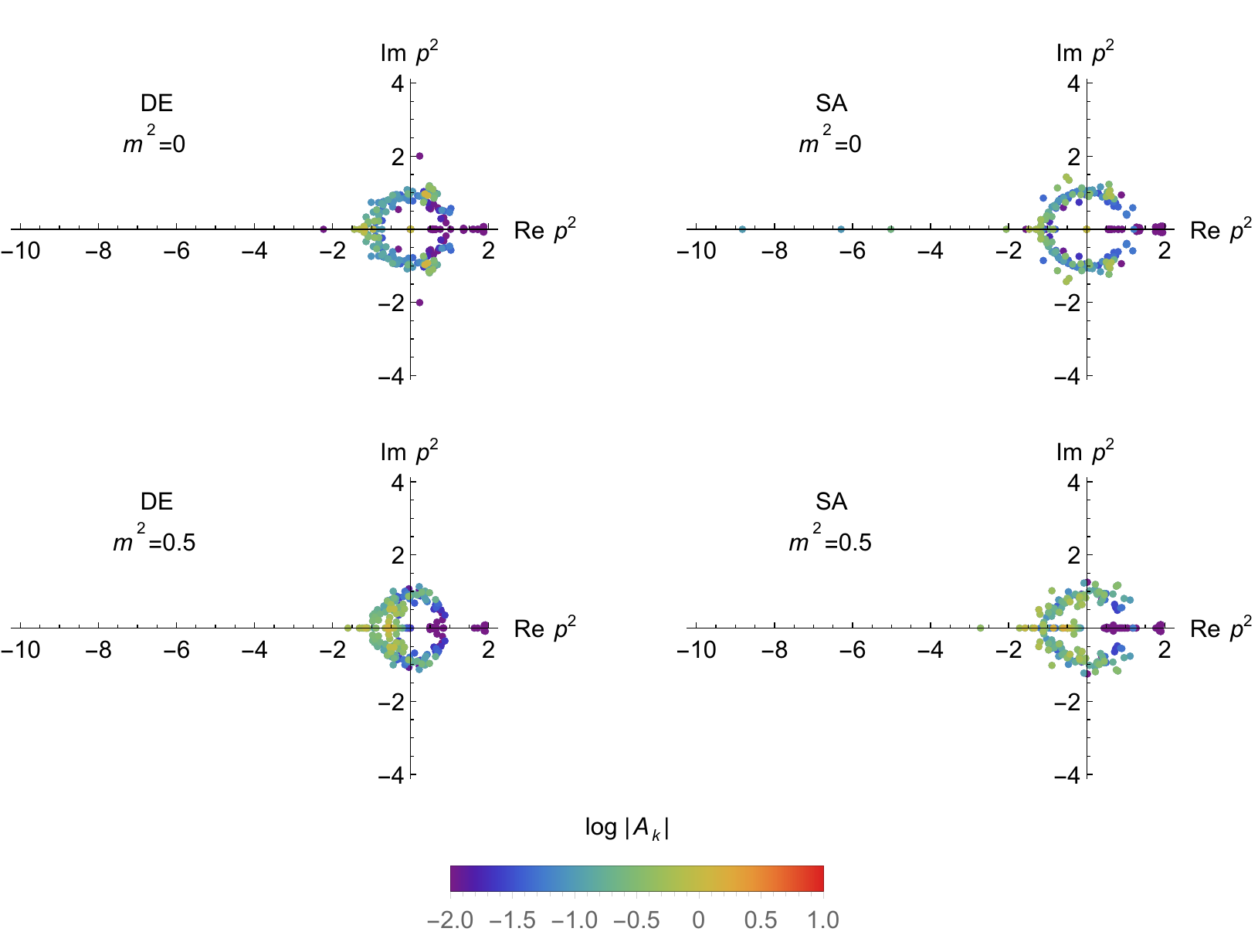}
	\caption[{All-poles representation obtained for the test data generated from $f_2(p^2)$, using $m^2=0$ and $m^2=0.5$, and for both methods of minimisation, DE and SA.}]{All-poles representation obtained for the test data generated from $f_2(p^2)$, using $m^2=0$ and $m^2=0.5$, and for both methods of minimisation, DE and SA. The colour scheme codes the residue's absolute value of each pole.}
	\label{fig:Tt2all}
\end{figure}

For $f_2(p^2)$, the same artefact that was present for $f_1(p^2)$ (the circumference of poles around the origin) also appears in the all-poles representation, in Figure \ref{fig:Tt2all}, for each value of $m^2$, and for both minimisation methods.

\begin{figure}[p]
	\centering
	\includegraphics[width=.9\textwidth]{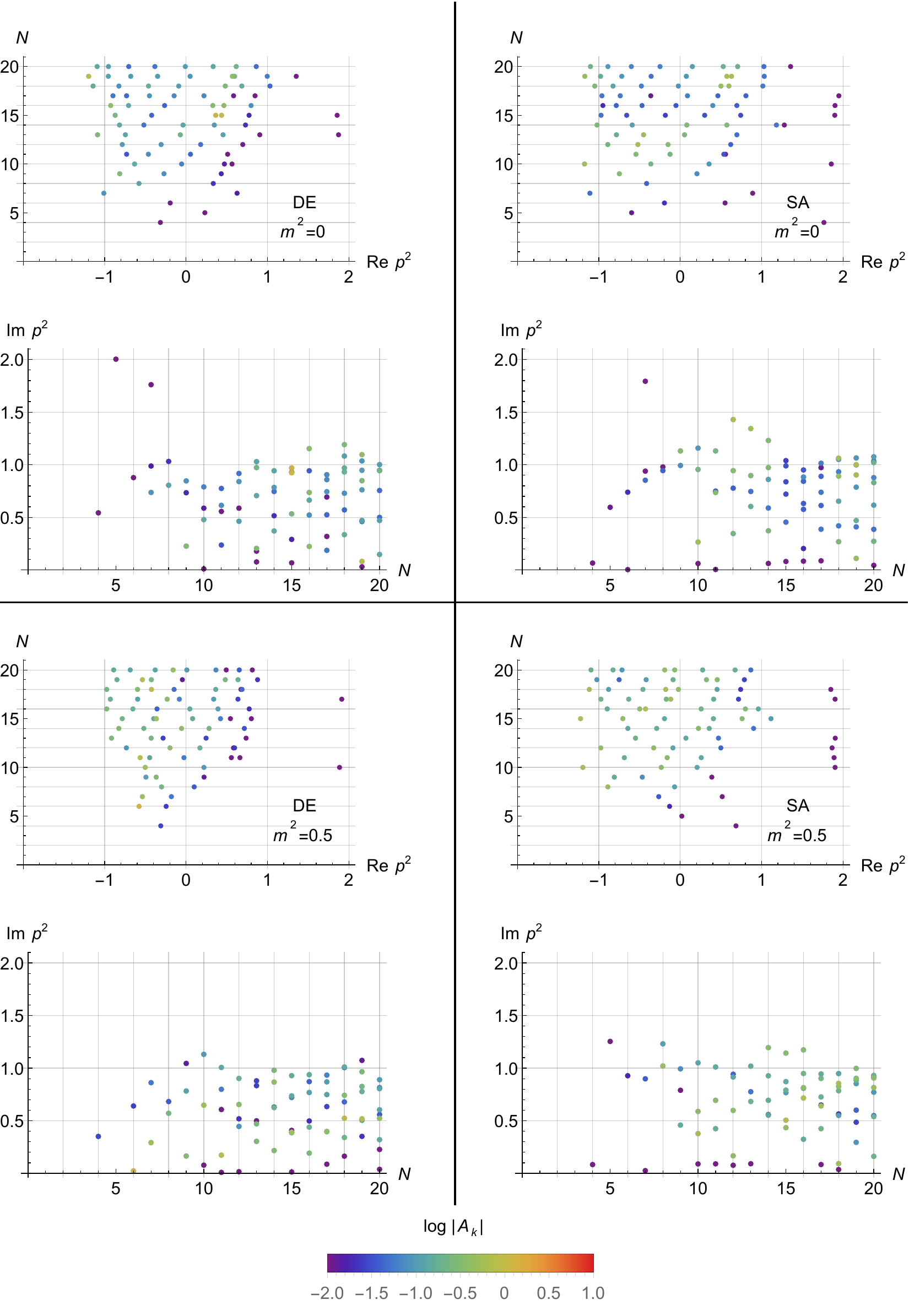}
	\caption[{Evolution of the off-axis poles and zeros with $N$ in the PA sequence, obtained from the test data generated from $f_2(p^2)$, using $m^2=0$ and $m^2=0.5$, and for both methods of minimisation, DE and SA.}]{Evolution of the off-axis poles and zeros with $N$ in the PA sequence, obtained from the test data generated from $f_2(p^2)$, using $m^2=0$ and $m^2=0.5$, and for both methods of minimisation, DE and SA. The colour scheme codes the residue's absolute value of each pole.}
	\label{fig:Tt2offaxis}
\end{figure}

The evolution of the off-axis poles, in Figure \ref{fig:Tt2offaxis}, shows that no important poles appear at complex $p^2$.

\begin{figure}[p]
	\centering
	\includegraphics[width=.9\textwidth]{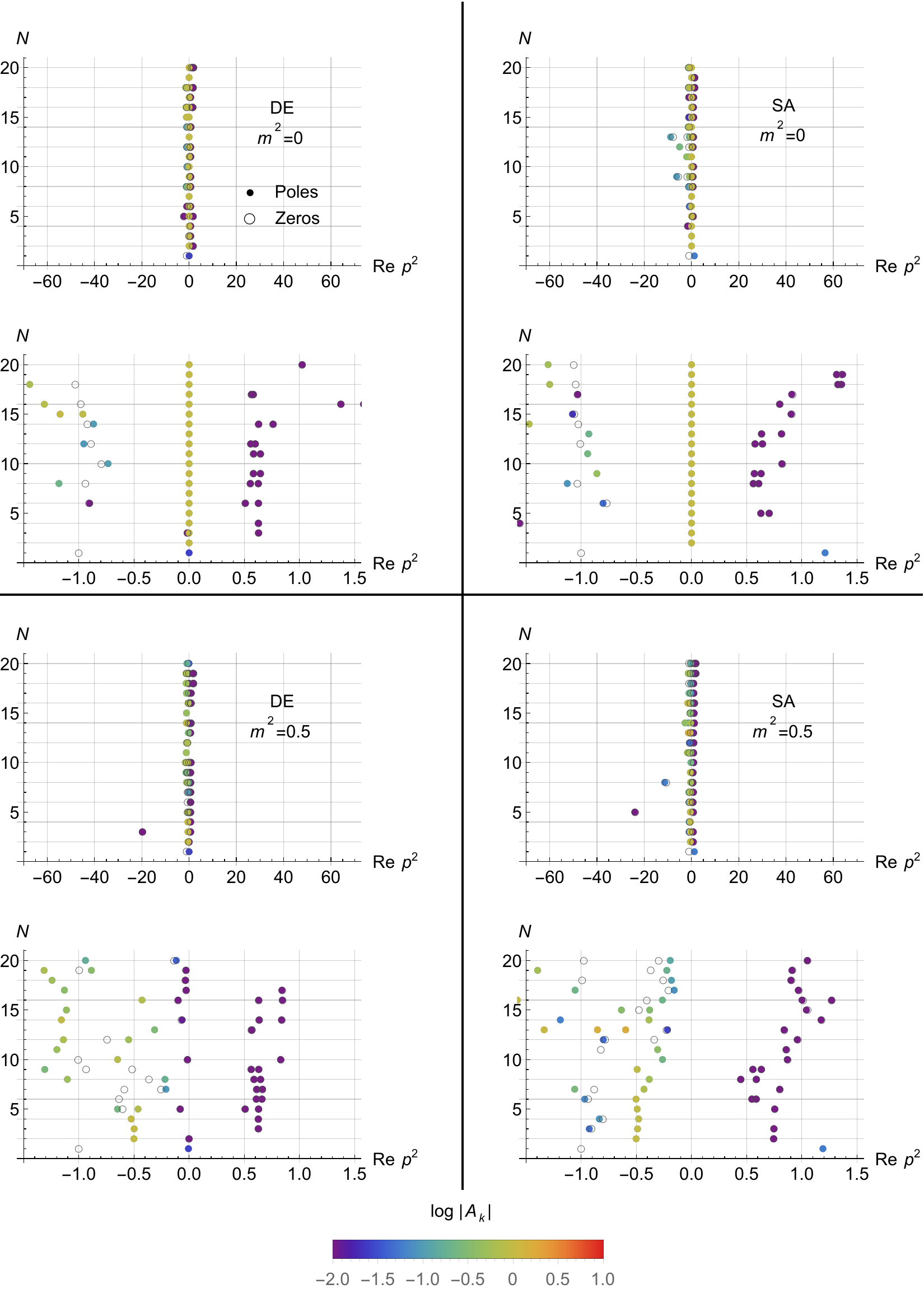}
	\caption[{Evolution of the on-axis poles and zeros with $N$ in the PA sequence, obtained from the test data generated from $f_2(p^2)$, using $m^2=0$ and $m^2=0.5$, and for both methods of minimisation, DE and SA.}]{Evolution of the on-axis poles and zeros with $N$ in the PA sequence, obtained from the test data generated from $f_2(p^2)$, using $m^2=0$ and $m^2=0.5$, and for both methods of minimisation, DE and SA. The colour scheme codes the residue's absolute value of each pole.}
	\label{fig:Tt2onaxis}
\end{figure}

The expected pole at $p^2=-m^2$ can be seen in the evolution of the on-axis poles and zeros, in Figure \ref{fig:Tt2onaxis}. For $m^2=0$, both methods successfully identify the pole at the origin. Indeed, a very stable pole is seen at $p^2=0$ throughout the whole PA sequence. On the contrary, for $m^2=0.5$, the pole at $p^2=-0.5$ is far more unstable. For the DE method, a good precision in the pole's position is only achieved for $N\in [3,5]$, while for the SA the pole's position can be considered stable up to $N=9$, with some minor deviations. For higher orders of approximation we observe that the poles obtained with the SA method deviate toward the origin, followed by a decrease in the respective residue.

\subsection{Identification of the analytic structure for the perturbative propagator} 
\begin{figure}[tp]
	\centering
	\includegraphics[width=\textwidth]{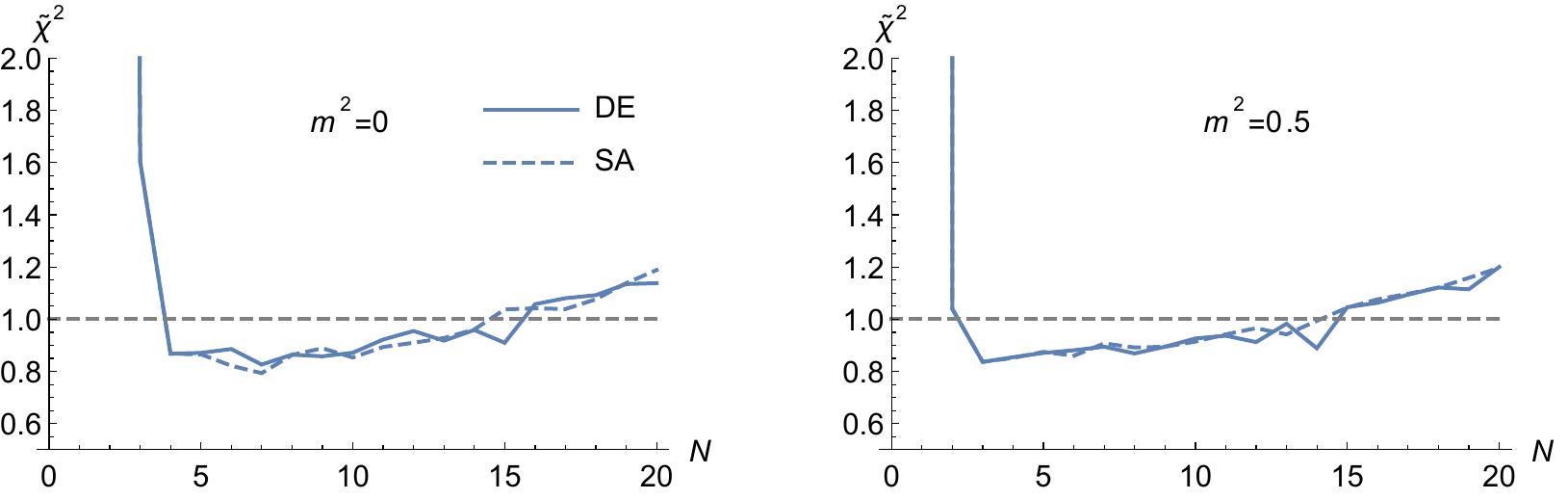}
	\caption{Achieved values of $\widetilde{\chi}^2$ for the test data generated from $f_3(p^2)$, using $m^2=0$ and $m^2=0.5$, and for both methods of minimisation, DE and SA.}
	\label{fig:Tt3chi2red}
\end{figure}

Finally, for the perturbative gluon propagator $f_3(p^2)$, the generated data was, once more, well adjusted by the PAs within the sequence, as can be seen by the values of $\widetilde{\chi}^2$ in Figure \ref{fig:Tt3chi2red}.

\begin{figure}[tp]
	\centering
	\includegraphics[width=\textwidth]{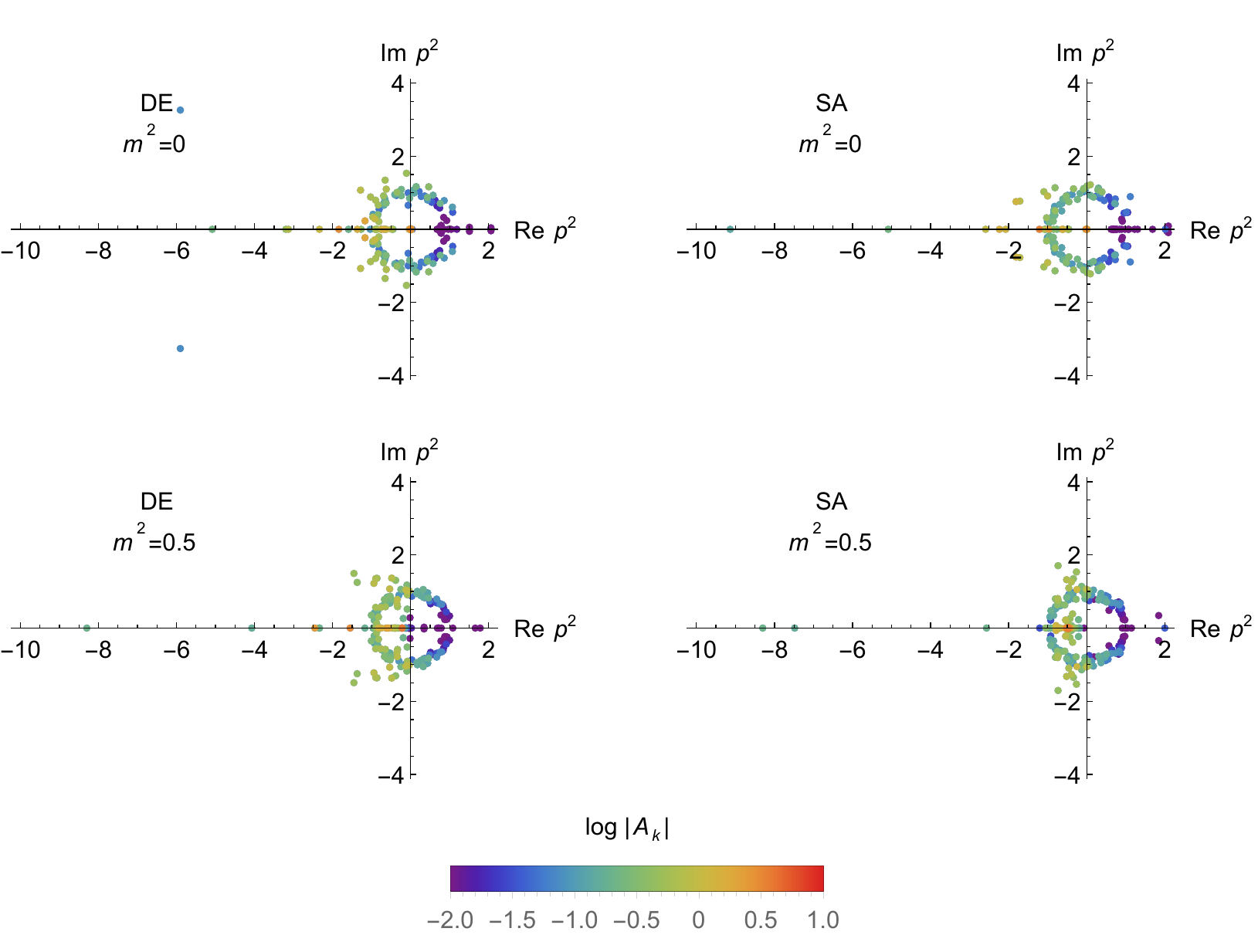}
	\caption[{All-poles representation obtained for the test data generated from $f_3(p^2)$, using $m^2=0$ and $m^2=0.5$, and for both methods of minimisation, DE and SA.}]{All-poles representation obtained for the test data generated from $f_3(p^2)$, using $m^2=0$ and $m^2=0.5$, and for both methods of minimisation, DE and SA. The colour scheme codes the residue's absolute value of each pole.}
	\label{fig:Tt3all}
\end{figure}

\begin{figure}[p]
	\centering
	\includegraphics[width=.9\textwidth]{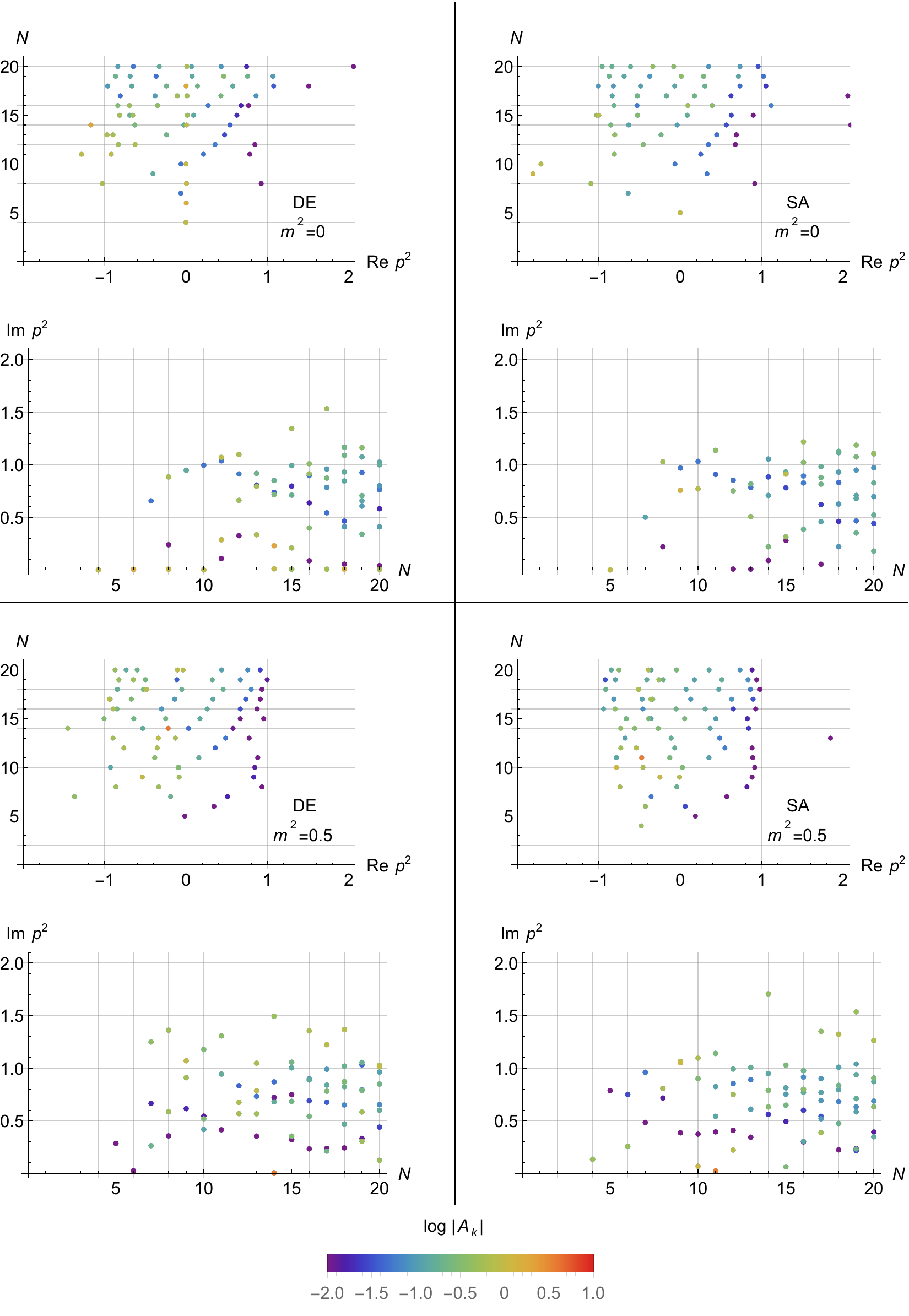}
	\caption[{Evolution of the off-axis poles and zeros with $N$ in the PA sequence, obtained from the test data generated from $f_3(p^2)$, using $m^2=0$ and $m^2=0.5$, and for both methods of minimisation, DE and SA.}]{Evolution of the off-axis poles and zeros with $N$ in the PA sequence, obtained from the test data generated from $f_3(p^2)$, using $m^2=0$ and $m^2=0.5$, and for both methods of minimisation, DE and SA. The colour scheme codes the residue's absolute value of each pole.}
	\label{fig:Tt3offaxis}
\end{figure}

A quick look to the all-pole representation for $f_3(p^2)$, in Figure \ref{fig:Tt3all}, shows that, for the third time, the circumference of poles around the origin is present for all cases, reinforcing the fact that it is an artefact of the approximation. As for $f_1(p^2)$ and $f_2(p^2)$, there are no relevant poles in the complex $p^2$-plane apart from the ones on the real negative $p^2$-axis, as can be observed in the representation of the off-axis poles, in Figure \ref{fig:Tt3offaxis}.

\begin{figure}[p]
	\centering
	\includegraphics[width=.9\textwidth]{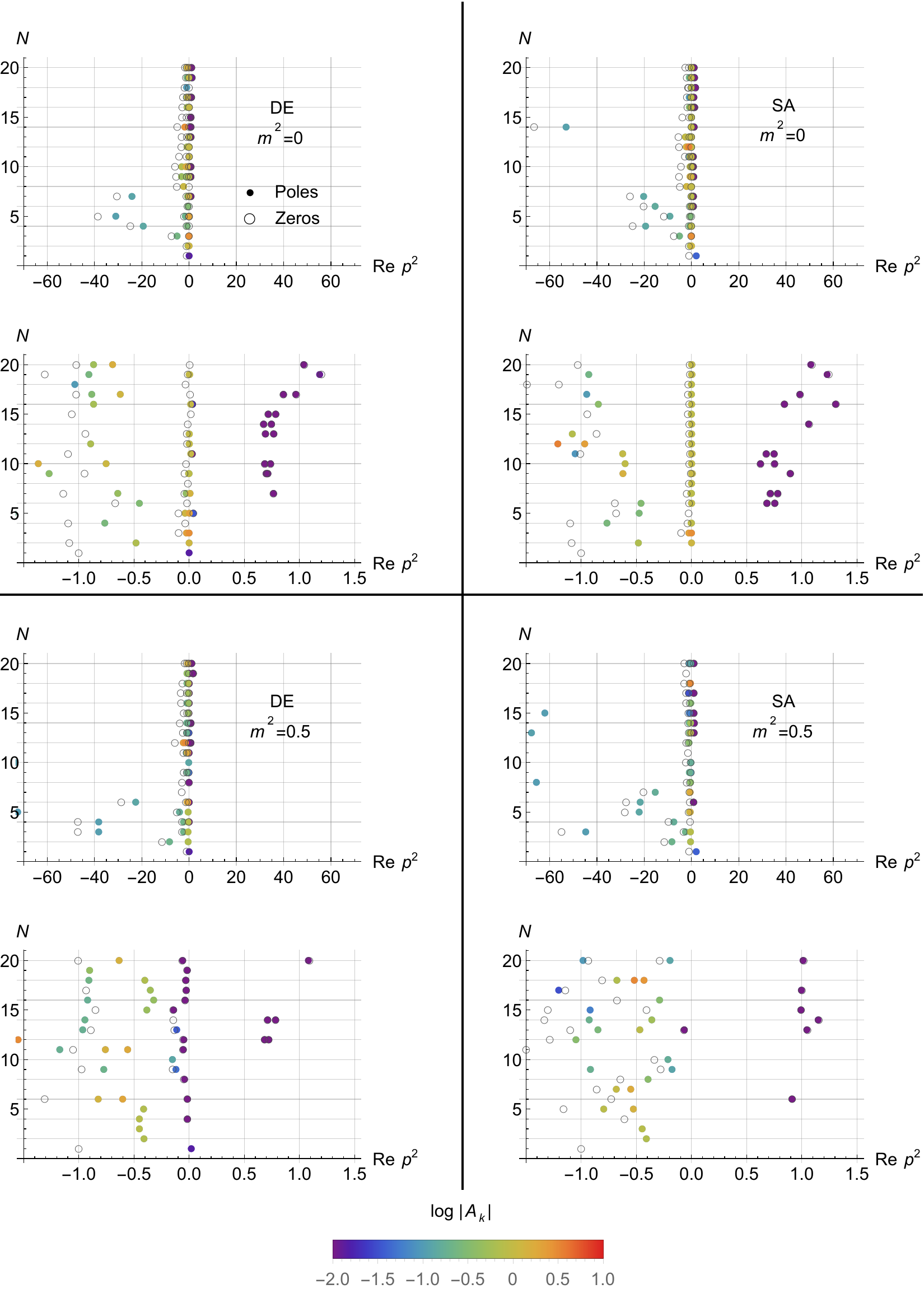}
	\caption[{Evolution of the on-axis poles and zeros with $N$ in the PA sequence, obtained from the test data generated from $f_3(p^2)$, using $m^2=0$ and $m^2=0.5$, and for both methods of minimisation, DE and SA.}]{Evolution of the on-axis poles and zeros with $N$ in the PA sequence, obtained from the test data generated from $f_3(p^2)$, using $m^2=0$ and $m^2=0.5$, and for both methods of minimisation, DE and SA. The colour scheme codes the residue's absolute value of each pole.}
	\label{fig:Tt3onaxis}
\end{figure}

With respect to the expected pole at $p^2=-m^2$, for the case where $m^2=0$, it is well identified within the PA sequence by a stable pole for both DE and SA, clearly seen in the on-axis poles and zeros representation, in Figure \ref{fig:Tt3onaxis}. Additionally, the pole at the origin is also identified by poles that appear near the origin in the off-axis representation obtained for the DE method (Figure \ref{fig:Tt3offaxis}). The case where $m^2=0.5$ does not present so clear results. The pole at $p^2=-0.5$ is more difficult to find in the on-axis representation (Figure \ref{fig:Tt3onaxis}). Although we can perceive the existence of a pole between $p^2=-0.6$ and $p^2=-0.3$, its exact position cannot be read with high precision. By comparing the two minimisation methods, we notice that this identification is more difficult for the SA than for the DE, where there are some intervals of $N$ ($N\in[2,5]$ and $[15,18]$) for which the identified pole appears to be almost stable.

As for the reproduction of the branch cut by the PA sequence, its presence in the on-axis representation (Figure \ref{fig:Tt3onaxis}) is not so clear as it was for $f_1(p^2)$. Nonetheless, alternating poles and zeros can be seen in the interval $p^2\in[-50,0[$, for lower orders of approximation. Similarly to $f_1(p^2)$, unimportant poles appear near the origin for the DE method, whereas for the SA this does not happen. This indicates that, despite the fact that the position of the branch point can hardly be found, it is not at $p^2=0$, but somewhere else in the real negative $p^2$-axis, as anticipated.

\section{Guide-lines for analytic structure identification}
Throughout Chapters \ref{Chap:PA} and \ref{Chap:AnalyticTests}, we performed numerical tests to investigate the analytic structure of a function and studied the distributions of poles and zeros obtained form the PA sequences. In this way, we arrive to the following guide-line list of how the analytic structure of a function is reproduced by a PA and how it can be identified:
\begin{itemize}
	\item Poles in the complex plane are reproduced by stable poles in the distributions of poles and zeros throughout a PA sequence. A pole at the origin is well identified by poles from the PAs that have relatively high residues. In this case, poles with residues of the same order of magnitude can appear close to the origin in the off-axis representation, for the DE method. On the other hand, a pole in a position other than the origin is reproduced by poles with higher residues. However, their position is less stable, specially for higher orders of approximation. In this case, a better stability is usually obtained when using the SA minimisation method.
	\item The branch cuts are reproduced by alternating poles and zeros. These may be easier to identify at lower orders of approximation. The exact position of the branch point is very difficult to identify, particularly if it is not at the origin. However, if that is not the case, poles with a low residue in the distributions of poles and zeros appear near the origin throughout the PA sequence when the DE method is used. This does not happen for the sequence obtained with the SA minimisation.
	\item As $N$ grows in the PA sequence, spurious poles, as well as Froissart doublets, tend to emerge and gather around the origin, forming what looks like a circumference of unit radius. A quick residue analysis, together with the study on the poles' stability, seems be enough to discard the undesired poles from the analytic structure.
\end{itemize}

\chapter{Results and discussion}
Now that we have a good idea of what seems to be a good guide to interpret the evolution of the distribution of poles and zeros within a PA sequence, and how the analytic structures of functions are reproduced by it, we can proceed and apply the PA analysis to investigate the gluon and ghost non-perturbative propagators.

\section{The ghost propagator}
\begin{figure}[tp]
	\centering
	\includegraphics[width=.8\textwidth]{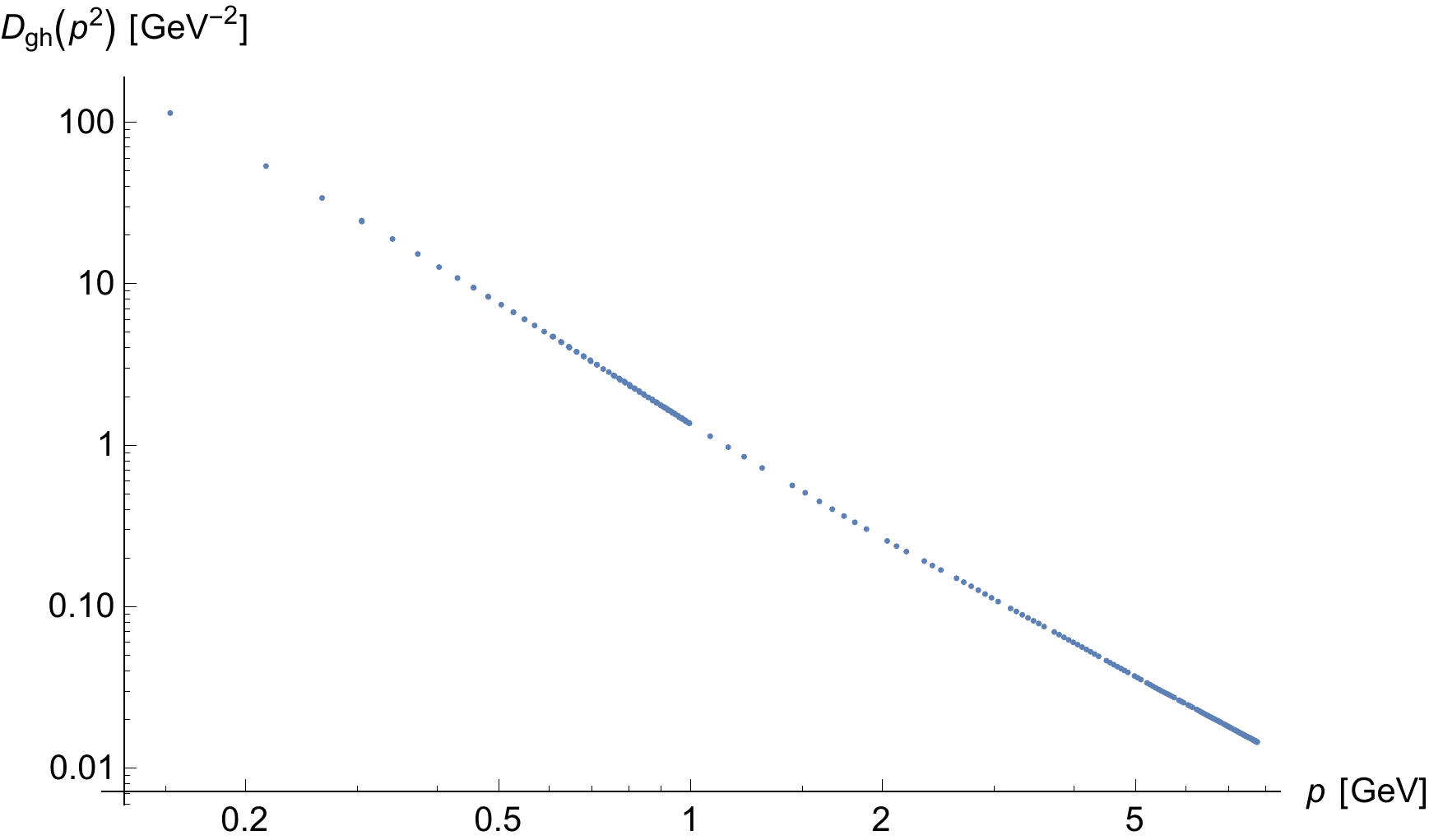}
	\caption[Landau gauge lattice ghost propagator used in the analysis.]{Landau gauge lattice ghost propagator used in the analysis.
		The statistical errors are smaller than the size of the points.}
	\label{fig:GhostProp}
\end{figure}

The data for the non-perturbative ghost propagator investigated here was published in \cite{Duarte2016}. It was obtained via lattice simulation in the Landau gauge performed on an hypercubic spacetime lattice of volume $80^4$ with 70 gauge configurations, using the Wilson gauge action for $\beta=6.0$ and renormalised in the MOM-scheme at $\mu=3~\si{GeV}$, according to
\begin{equation}
D(\mu^2)\Big|_{\mu=3~\si{GeV}}=\frac{1}{\mu^2}.
\end{equation}
The propagator data can be seen in Figure \ref{fig:GhostProp}.

Following the analysis of the previous two chapters, the lattice data for the ghost propagator was adjusted by PAs, for various orders of approximation within a PA sequence, using the two minimisation methods: DE and SA. On a first stage, various relations between $L$ and $M$ were tried. Following the conclusions of Section \ref{Sec:Hint}, the use of PAs of order $[N-1|N]$ showed to be the best choice\footnote{This is in accordance with the fact that, in the limit of high momenta, the full propagator has the same behaviour as the one given by the perturbative analysis.}.

\begin{figure}[tp]
	\centering
	\includegraphics[width=.8\textwidth]{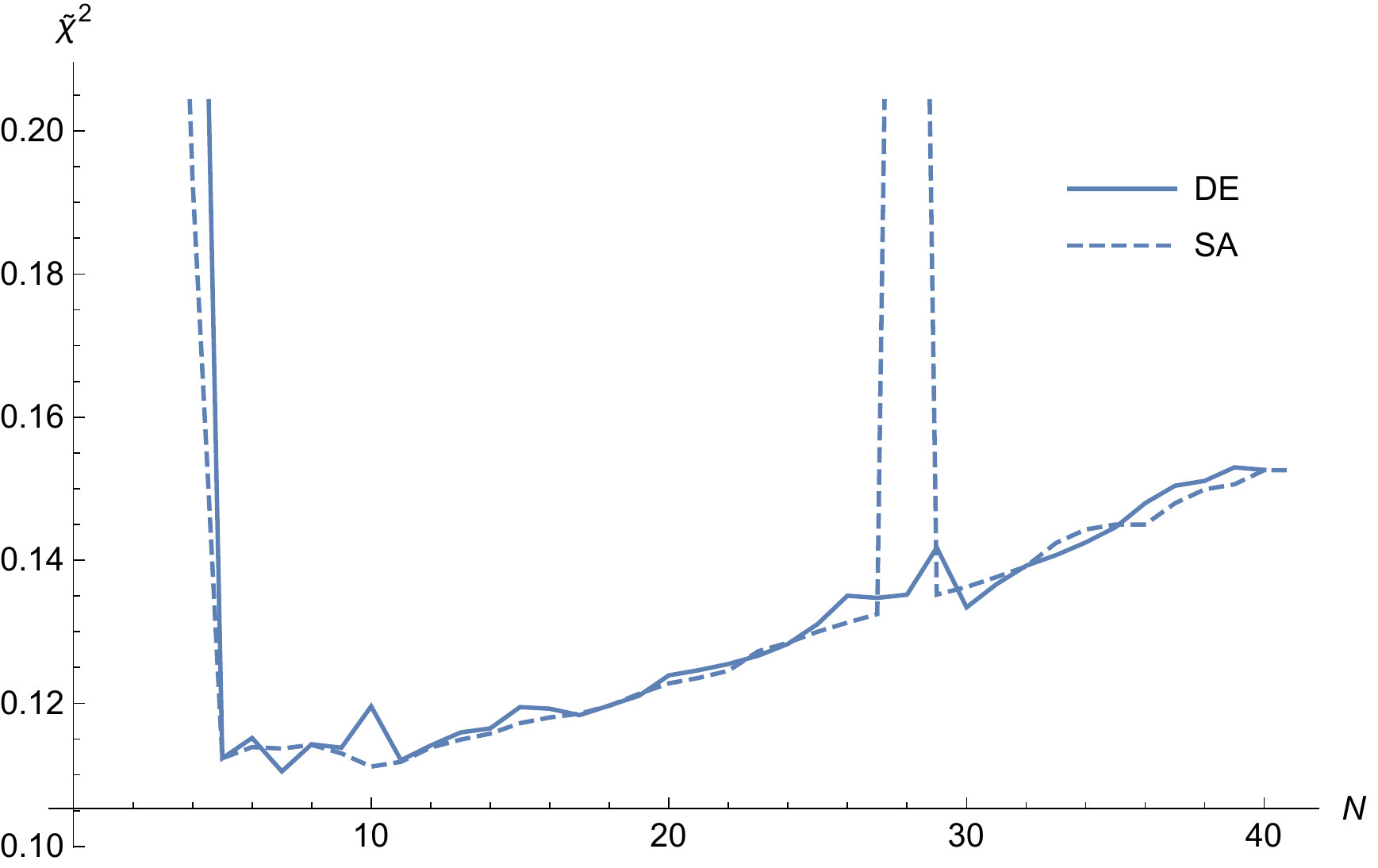}
	\caption{Achieved values of $\widetilde{\chi}^2$ for the Landau gauge lattice ghost propagator, for both methods of minimisation, DE and SA.}
	\label{fig:GhChi}
\end{figure}

The values of $\widetilde{\chi}^2$ at the minima are represented in Figure \ref{fig:GhChi}, for $N\in[1,40]$. The values obtained are below the unit, and show the quality of the minimisation, and also show that the data is well reproduced by PAs.

\begin{figure}[tp]
	\centering
	\includegraphics[width=\textwidth]{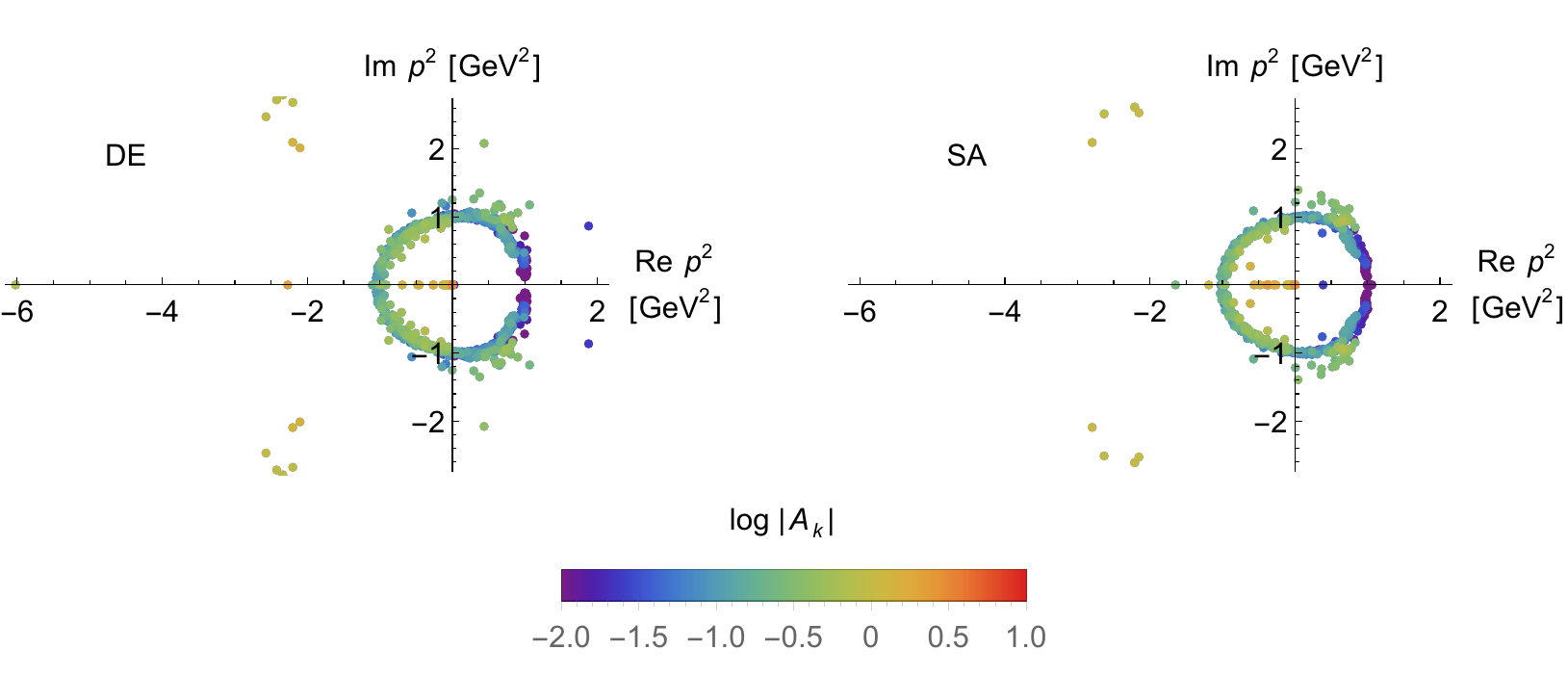}
	\caption[All-poles representation obtained for the ghost propagator data, for both methods of minimisation, DE and SA.]{All-poles representation obtained for the ghost propagator data, for both methods of minimisation, DE and SA. The colour scheme codes the residue's absolute value of each pole.}
	\label{fig:GhAll}
\end{figure}

In Figure \ref{fig:GhAll}, the all-poles representation is shown for both minimisation methods. The same artefact that already appeared in the last chapter (the circumference of unit radius formed by Froissart doublets) is clearly visible. An accumulation of poles with high residues on the real negative $p^2$-axis, near the origin, can also be observed, suggesting the presence of a branch cut.

\begin{figure}[tp]
	\centering
	\includegraphics[width=\textwidth]{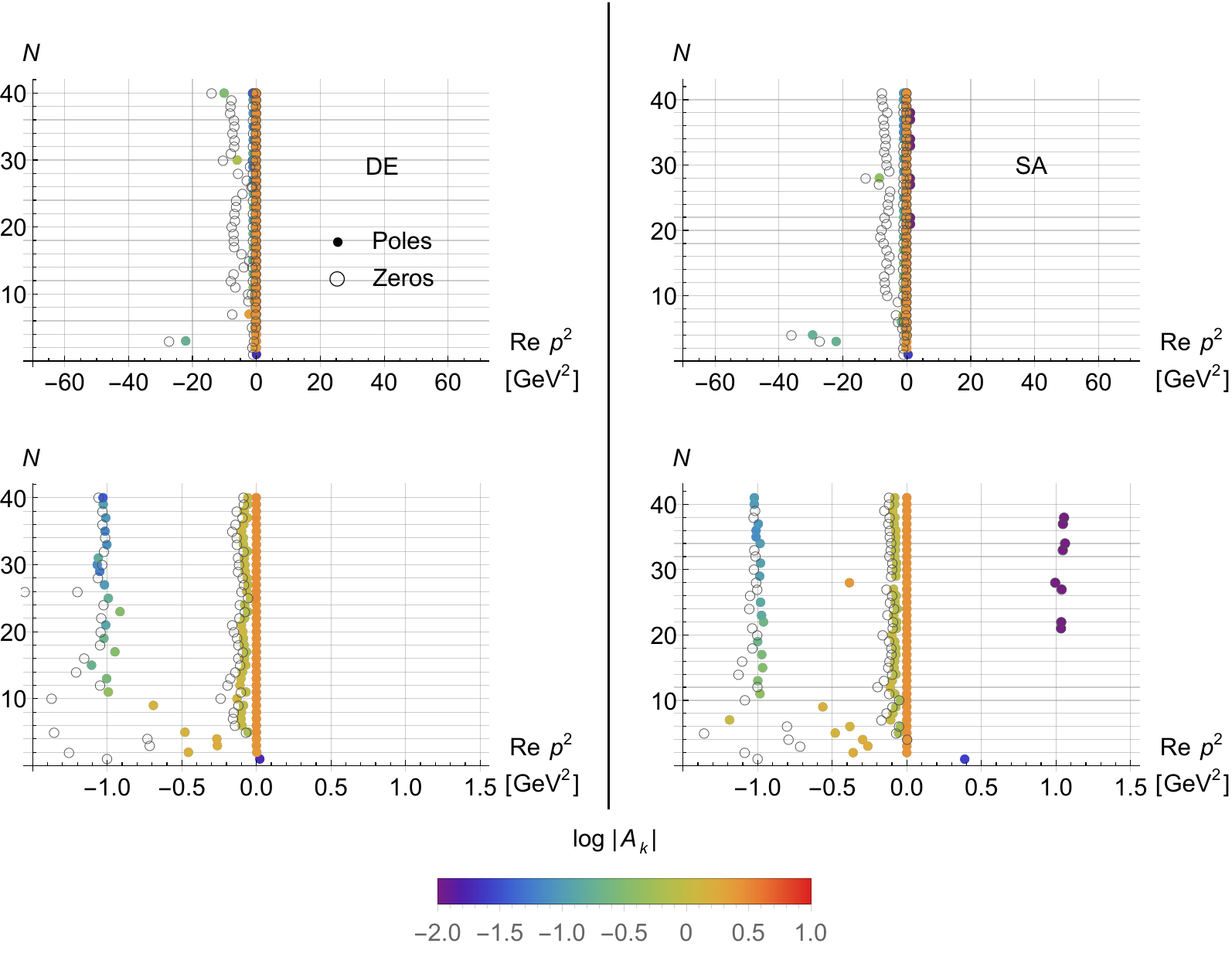}
	\caption[Evolution of the on-axis poles and zeros within the PA sequence obtained for the ghost propagator data, for both methods of minimisation, DE and SA.]{Evolution of the on-axis poles and zeros within the PA sequence obtained for the ghost propagator data, for both methods of minimisation, DE and SA. The colour scheme codes the residue's absolute value of each pole.}
	\label{fig:Ghonaxis}
\end{figure}

An analysis of the on-axis poles and zeros, which is represented in Figure \ref{fig:Ghonaxis}, shows a very stable pole with high residue at the origin, for both the DE and the SA methods. This is a clear indication of the presence of a pole at $p^2=0$ in the analytic structure of the ghost propagator. The remaining poles and zeros seem to indicate the presence of a branch cut on the real negative $p^2$-axis, with a branch point at $p^2\sim-0.1~\si{GeV^2}$ \footnote{We must not forget the numerical tests made in Chapter \ref{Chap:Discrete}, where we saw that the branch point is hard to identify if it is not at the origin. Nonetheless, in this case, the stability of the on-axis poles may allow us to correctly identify the branch point's position.}. 

\begin{figure}[tp]
	\centering
	\includegraphics[width=\textwidth]{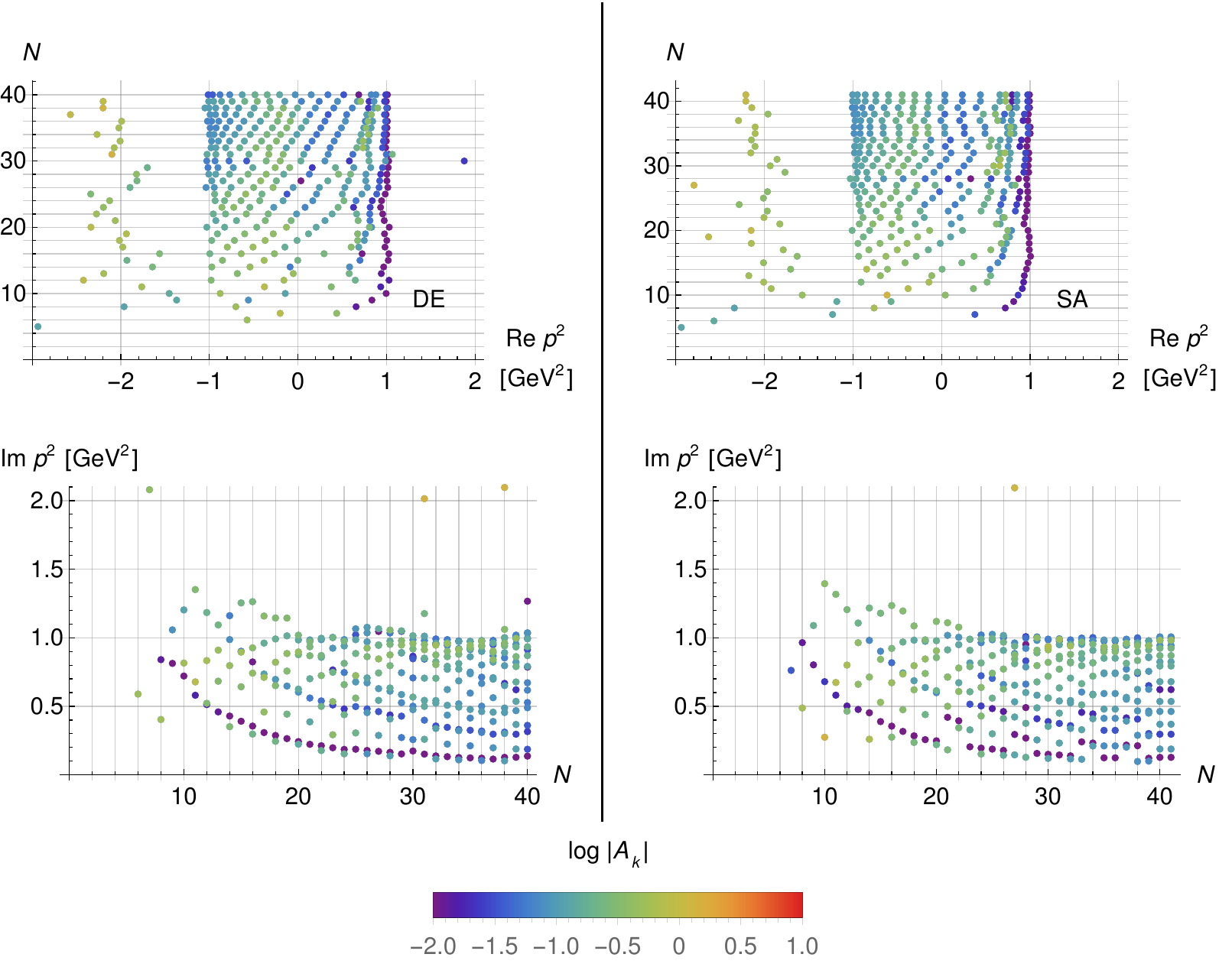}
	\caption[Evolution of the off-axis poles within the PA sequence obtained for the ghost propagator data, for both methods of minimisation, DE and SA.]{Evolution of the off-axis poles within the PA sequence obtained for the ghost propagator data, for both methods of minimisation, DE and SA. The colour scheme codes the residue's absolute value of each pole.}
	\label{fig:Ghoffaxis}
\end{figure}

On the other hand, by looking at the evolution of the off-axis poles, in Figure \ref{fig:Ghoffaxis}, we see that no relevant stable poles appear throughout the PA sequence, in the remaining complex $p^2$-plane.

\vspace{1em}

These results are in agreement with \cite{Binosi2020,Hayashi2020}, where the analytic structure of the ghost propagator suggested to be similar to the perturbative result, with no complex singularities. The results also support the no-pole condition for the ghost propagator, as proposed in \cite{Gribov1978}.

\section{The gluon propagator}
\begin{figure}[tp]
	\centering
	\includegraphics[width=.8\textwidth]{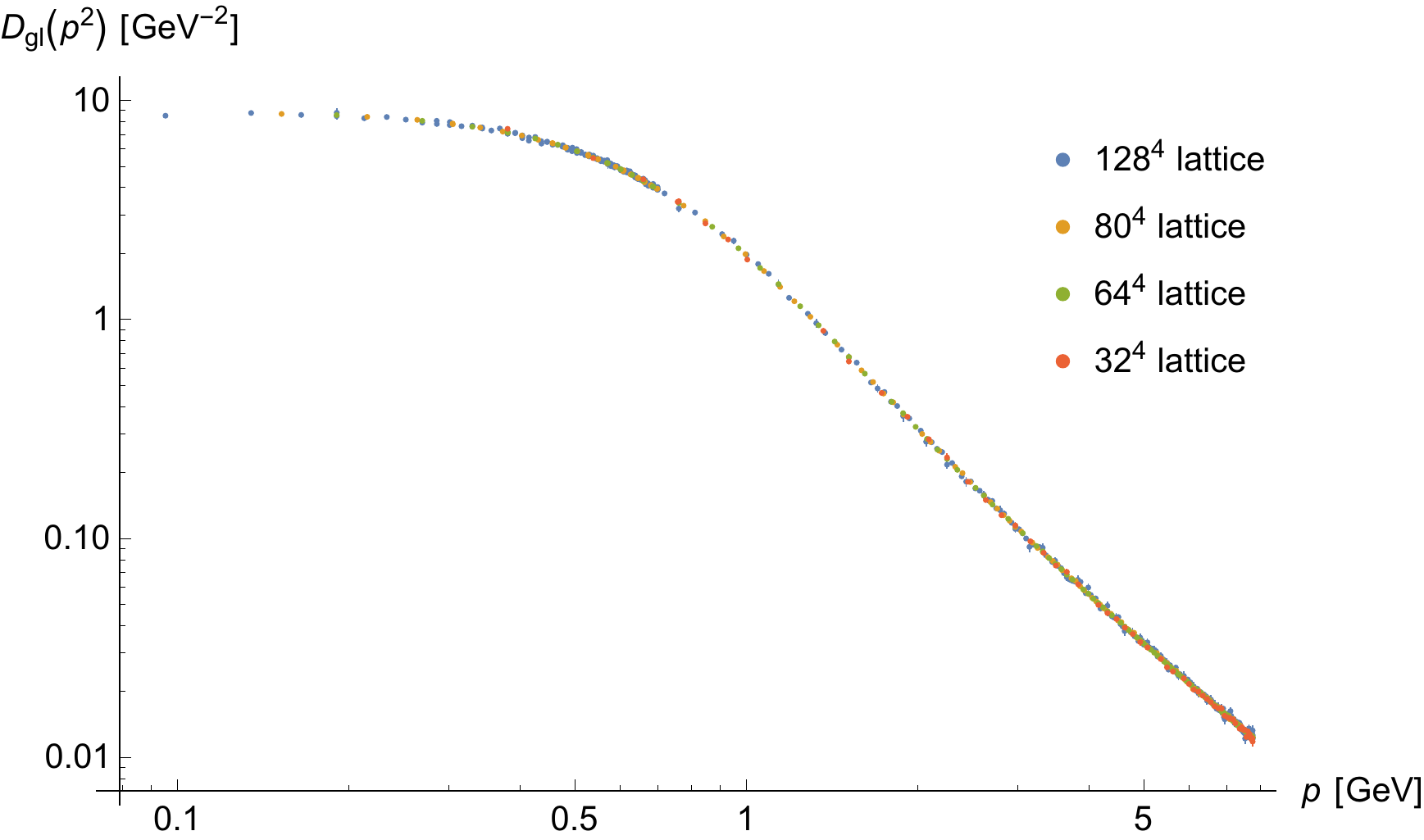}
	\caption{Gluon propagator used in the analysis, obtained with four different lattice volumes.}
	\label{fig:GluonProp}
\end{figure}

For the gluon propagator, we use the results of lattice simulations in the Landau gauge using the Wilson gauge action for $\beta=6.0$, renormalised in the MOM-scheme at $\mu=3~\si{GeV}$, similarly to the ghost propagator in the last section. Several data sets are considered, which were obtained using: a $32^4$ lattice, physical volume of $(3.25~\si{fm})^4$, with 50 gauge configurations, from \cite{Bicudo2015}; a $64^4$ lattice, physical volume of $(6.50~\si{fm})^4$, with 2000 gauge configurations, from \cite{Dudal2018}; a $80^4$ lattice, physical volume of $(8.13~\si{fm})^4$, with 550 gauge configurations, from \cite{Dudal2018}; and a $128^4$ lattice, physical volume of $(13.01~\si{fm})^4$, with 35 gauge configurations, from \cite{Duarte2016}. The mentioned data sets for the gluon propagator are represented in Figure \ref{fig:GluonProp}\footnote{The zero momentum propagator is not considered.}. All of them are essentially compatible with each other at one standard deviation level, and so they define a unique curve.

As seen in Figure \ref{fig:GluonProp}, the $128^4$ lattice simulation has more information in the IR region of momentum, \ie, $p\lesssim 1~\si{GeV}$. In fact, a larger lattice simulation means more information in the IR. In this sense, by considering these four data sets, a better sensitivity to the different regions of momentum is achieved. 

\vspace{1em}

As it was done for the ghost propagator, various relations between $L$ and $M$ were tried. According to the conclusion of Section \ref{Sec:Hint}, the best choice revealed to be, once more, the use of PAs of order $[N-1|N]$.

\begin{figure}[tp]
	\centering
	\includegraphics[width=.9\textwidth]{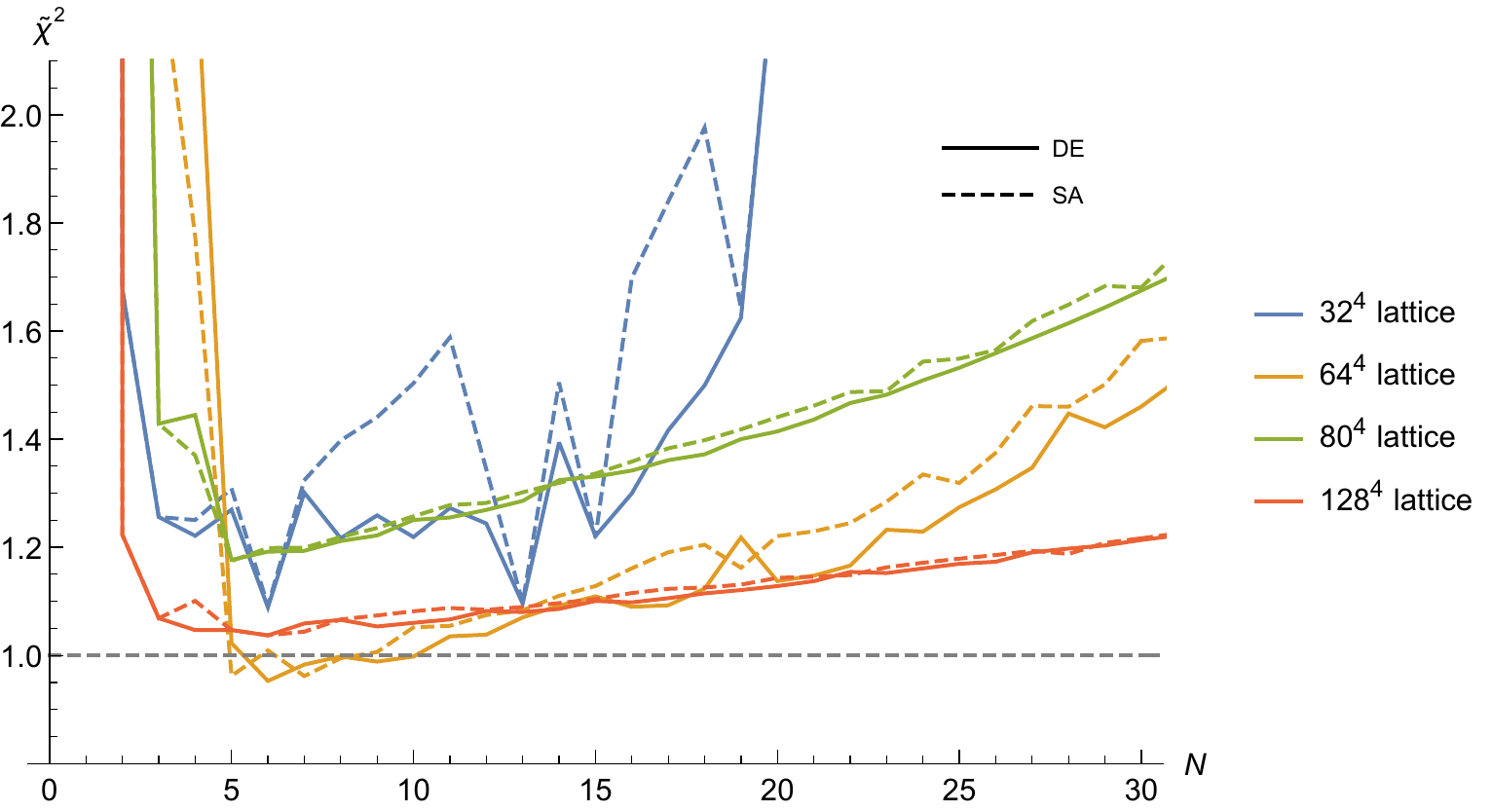}
	\caption{Achieved values of $\widetilde{\chi}^2$ for the Landau gauge lattice gluon propagator, and for both methods of minimisation, DE and SA.}
	\label{fig:GlChi}
\end{figure}

In Figure \ref{fig:GlChi}, the $\widetilde{\chi}^2$ at the minima are shown. The several obtained $\widetilde{\chi}^2$ are essentially the same for both minimisation methods. We also observe that the quality of the approximation increases ($\widetilde{\chi}^2$ decreases) with the increase of the lattice volume. 

\begin{figure}[tp]
	\centering
	\includegraphics[width=.97\textwidth]{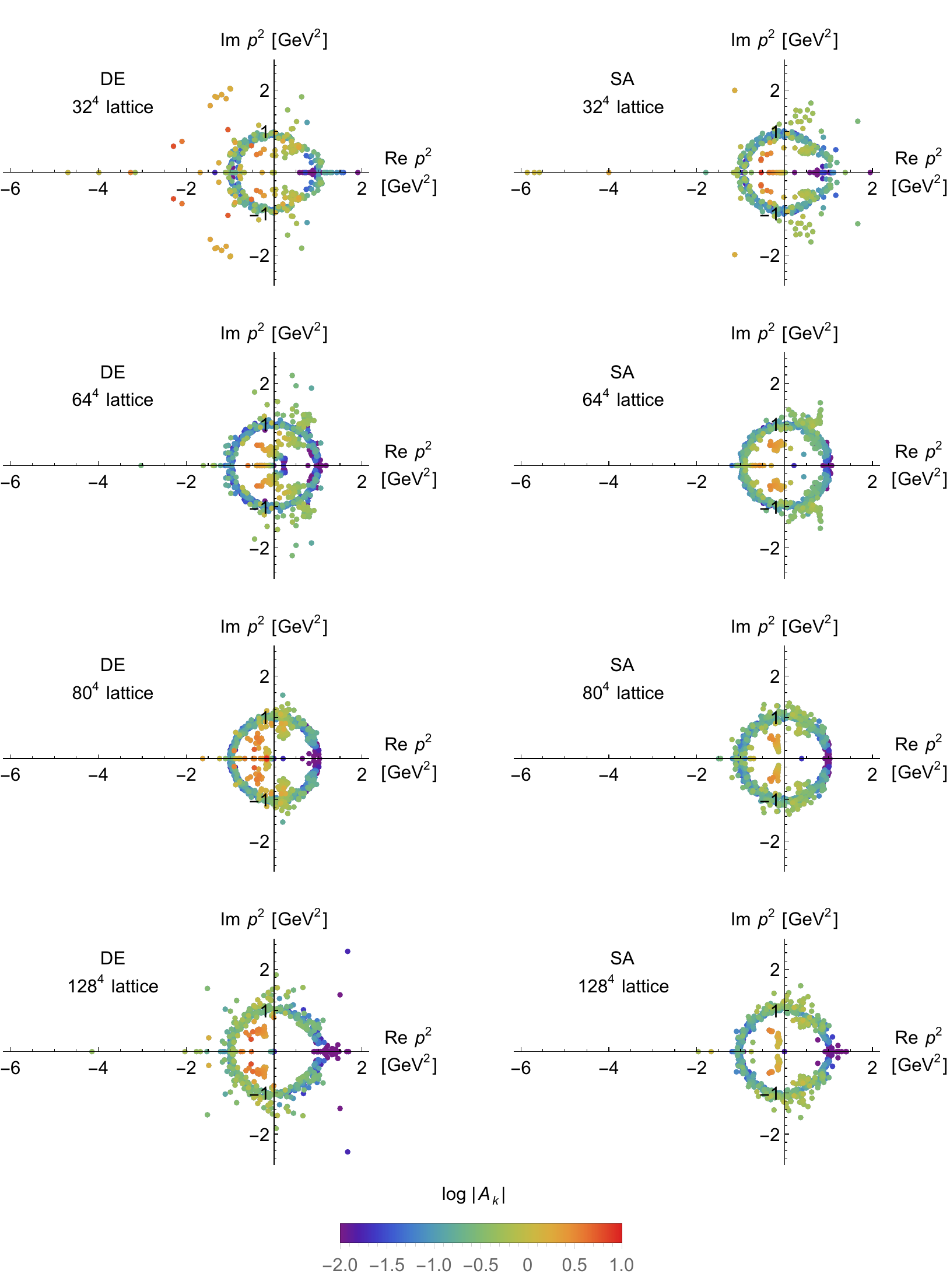}
	\caption[All-poles representation obtained for the gluon propagator data with $32^4$, $64^4$, $80^4$ and $128^4$ lattices, for both methods of minimisation, DE and SA.]{All-poles representation obtained for the gluon propagator data with $32^4$, $64^4$, $80^4$ and $128^4$ lattices, for both methods of minimisation, DE and SA. The colour scheme codes the residue's absolute value of each pole.}
	\label{fig:GlAll}
\end{figure}

In Figure \ref{fig:GlAll}, the all-poles representation is shown for the four lattice volumes, and for both minimisation methods. Apart from the usual artefact, a new structure emerges in the complex $p^2$-plane: a conjugate pair of complex poles of high residue at $\text{Re}(p^2)<0$. This structure becomes clearer and well defined for higher lattice volumes, which indicates that this pair of poles is associated with the IR structure of the theory. On the other hand, for smaller lattice volumes, a branch cut may be identified on the real negative $p^2$-axis. Also in Figure \ref{fig:GlAll}, the slight suggestion of the existence of another conjugate pair of complex poles may be seen at $\text{Re}(p^2)>0$. However, these poles have a lower residue and are not present for all simulations. Additionally, for some cases, the pair is not identified with both minimisation methods. In this sense, further studies are needed to see if these poles are meaningful or artefacts of the method. Herein, we will not consider the poles at $\text{Re}(p^2)>0$.

In the following subsections we examine with more detail the positions of the identified poles and branch cut.

\subsection{Complex poles}

\begin{figure}[tp]
	\centering
	\includegraphics[width=.9\textwidth]{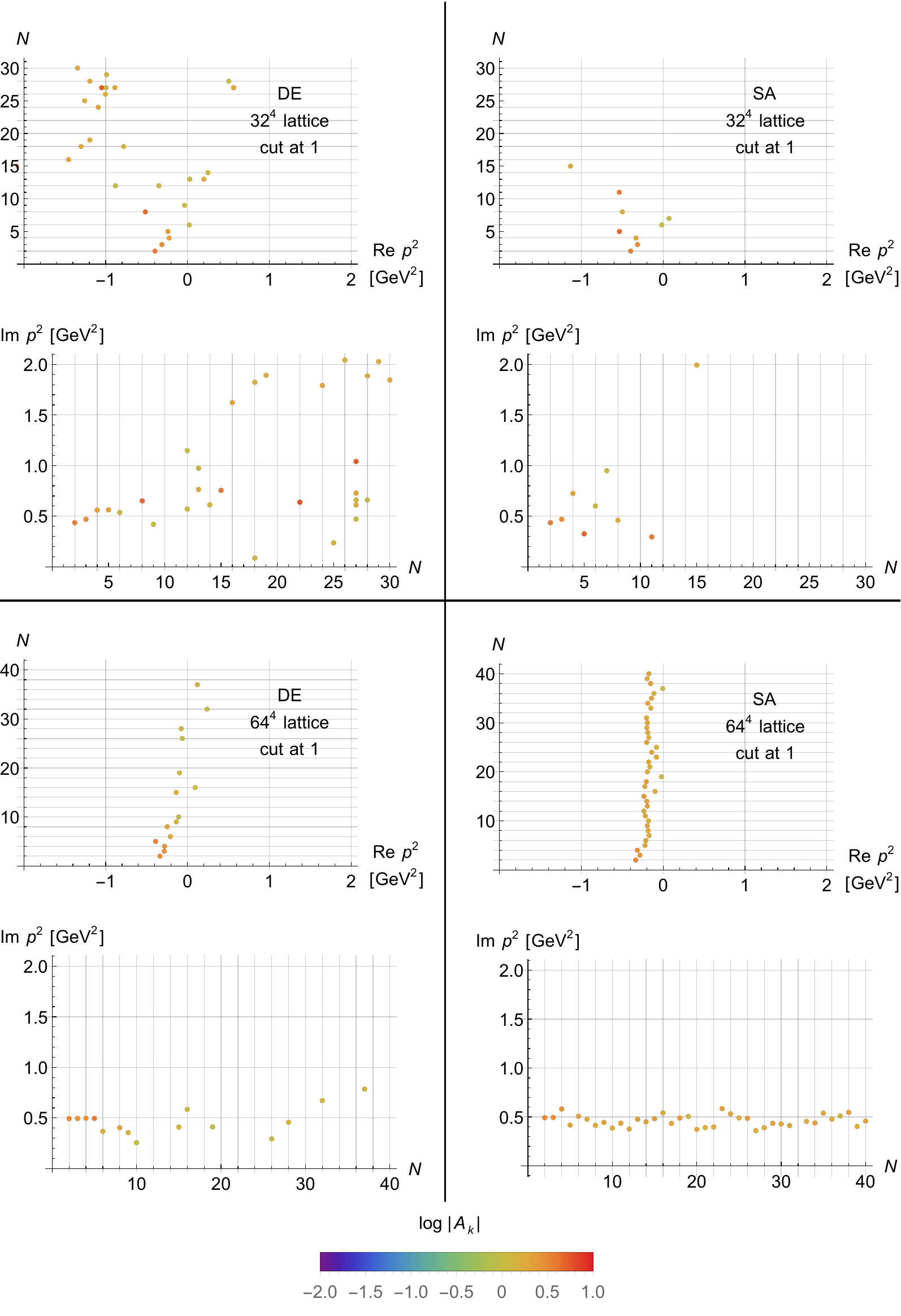}
	\caption[Evolution of the off-axis poles within the PA sequence, obtained for the gluon propagator data with $32^4$ and $64^4$ lattices, for both methods of minimisation, DE and SA.]{Evolution of the off-axis poles within the PA sequence, obtained for the gluon propagator data with $32^4$ and $64^4$ lattices, for both methods of minimisation, DE and SA. A cut in the residues at $\log|A_k|=0$ has been performed. The colour scheme codes the residue's absolute value of each pole.}
	\label{fig:Gloffaxis1}
\end{figure}

\begin{figure}[tp]
	\centering
	\includegraphics[width=.9\textwidth]{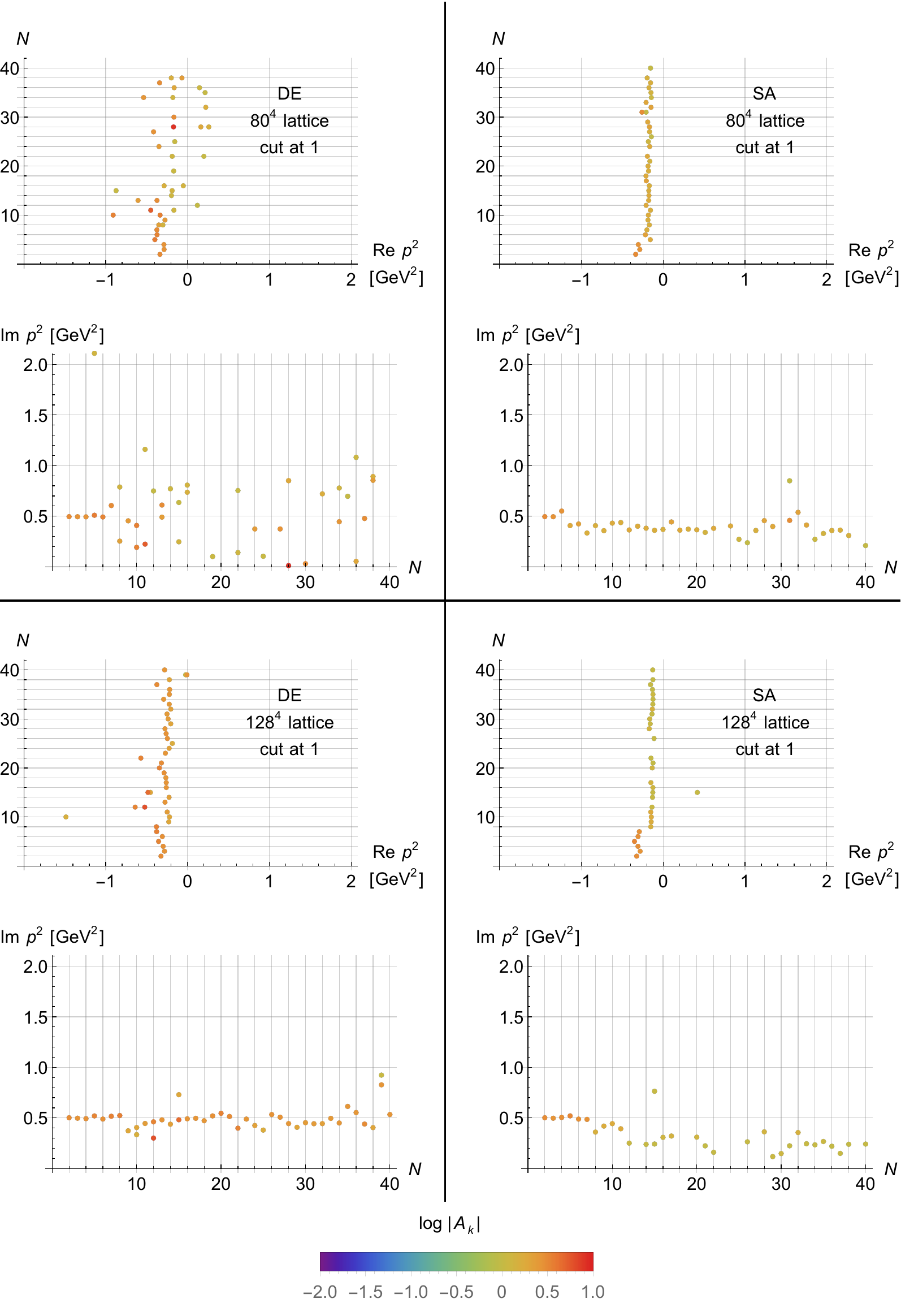}
	\caption[Evolution of the off-axis poles within the PA sequence, obtained for the gluon propagator data with $80^4$ and $128^4$ lattices, for both methods of minimisation, DE and SA.]{Evolution of the off-axis poles within the PA sequence, obtained for the gluon propagator data with $80^4$ and $128^4$ lattices, for both methods of minimisation, DE and SA. A cut in the residues at $\log|A_k|=0$ has been performed. The colour scheme codes the residue's absolute value of each pole.}
	\label{fig:Gloffaxis2}
\end{figure}

In Figures \ref{fig:Gloffaxis1} and \ref{fig:Gloffaxis2}, the off-axis poles are represented for the four lattice volumes, and for both minimisation methods. A cut in the residues at $|A_k|=1$ was already performed, and only the relevant poles appear. As seen already in the all-poles representations, the complex poles are more stable throughout the PA sequence for higher lattice volumes, particularly for lower values of $N$. For this reason, the results for the simulation with the largest lattice volume, \ie, the $128^4$ lattice, are those appropriate to read out the position of the poles.

In the last chapter, we saw that the position of a pole is identified with precision only for lower orders of approximation. In this case, despite the fact that the pole is reproduced throughout the whole PA sequence by poles with high residue, only the ones obtained for lower orders of $N$ should be used to estimate the position of this singularity in the analytic structure of the propagator. In fact, by looking at the off-axis poles for the DE method, in Figure \ref{fig:Gloffaxis2}, we observe that the poles are more stable for $N\in[2,8]$ than for the remaining orders. As for the SA method, although in the representation of Re$(p^2)$ the pole is very stable in one position for $N\in[2,7]$, and in another position for $N\in[8,40]$, the representation of Im$(p^2)$ shows otherwise: for $N\in[8,40]$, the imaginary part is much less stable. This behaviour is in accordance with the results of the last chapter. There, we saw that, starting at $N\sim9$ and only for the SA method, the identified pole began to falsely move toward the origin, followed by a decrease in the respective residue.

The poles obtained with $N\in[2,8]$ for the DE method, and the ones obtained with $N\in[2,7]$ for the SA method, are used to estimate the position of these singularities in the analytic structure\footnote{Although the remaining poles are not used, their appearance is important to confirm the presence of these poles in the analytic structure of the propagator.}. An arithmetic average of the respective poles' positions gives the following results for the position of the poles present in the analytic structure of the gluon propagator, for both minimisation methods:
\begin{align*}
\text{DE:}\quad &p^2=-0.332(30) \pm i 0.506(11)~\si{GeV^2};\\
\text{SA:}\quad &p^2=-0.311(20) \pm i 0.500(10)~\si{GeV^2}.
\end{align*}

\begin{figure}[tp]
	\centering
	\includegraphics[width=.8\textwidth]{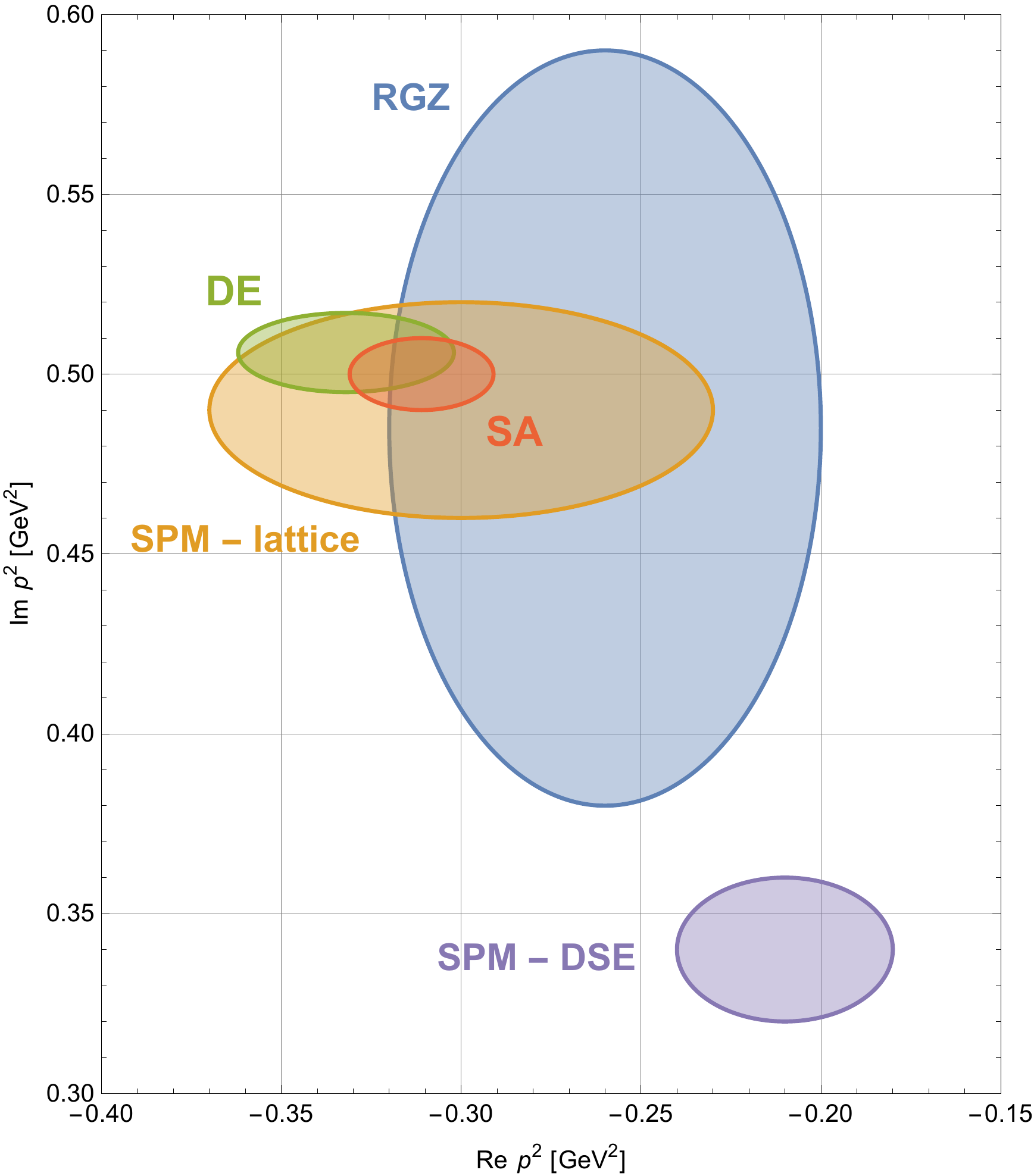}
	\caption[Position of the poles present in the analytic structure of the gluon propagator obtained in the present work and in other studies.]{Position of the poles present in the analytic structure of the gluon propagator obtained in the present work - DE and SA -, and in other studies (see text). The confidence region for the positions is defined as an ellipse, for which the semiaxes correspond to the errors associated with each result. Only one of the complex poles in the conjugate pair is shown.}
	\label{fig:ResultsComp}
\end{figure}

In Figure \ref{fig:ResultsComp}, the above results are represented, together with the following results, obtained in previous studies. In \cite{Dudal2018}, the tree level prediction of the RGZ action, to describe the lattice data up to $p\sim1~\si{GeV}$, was used to obtain the position of the singularity in the gluon propagator. There, a global fit identified a pole at Re$(p^2)\in[-0.32,-0.20]~\si{GeV^2}$ and Im$(p^2)\in\pm[0.38,0.59]~\si{GeV^2}$. In \cite{Binosi2020}, a fixed order PA, computed with the Schlessinger Point Method (SPM) mentioned in Section \ref{Sec:SimpleFit}, found the singularity at: $p^2=-0.30(7) \pm i 0.49(3)~\si{GeV^2}$, for the same $64^4$ lattice data here; and at $p^2=-0.21(3) \pm i 0.34(2)~\si{GeV^2}$ for the decoupling solution of the DSE.

By comparing the above results (see Figure \ref{fig:ResultsComp}), we can conclude that there is a discrepancy between the results obtained via DSE and the ones obtained using lattice data (DE, SA, SPM - lattice, and RGZ), even though the latter were obtained with different approaches to reproduce the lattice data. Indeed, the pole found with the DSE is at smaller absolute value of $\text{Re}(p^2)$ and $\text{Im}(p^2)$, and, thus, closer to the origin.

Regarding the results based on the lattice data (DE, SA, SPM - lattice, and RGZ), we see that the identified pole is at a slightly smaller $\text{Re}(p^2)$, but at a similar $\text{Im}(p^2)$, when compered to the result from the DSE. Nonetheless, the good overall compatibility in the position of the conjugate pair of complex poles is reassuring. Furthermore, they match with the analysis inspired by the Gribov-Zwanziger type of actions \cite{Dudal2018}, for which the solution for the gluon propagator is itself a ratio of polynomials and, thus, a type of PA.

\subsection{Branch cut and branch point}

\begin{figure}[tp]
	\centering
	\includegraphics[width=.9\textwidth]{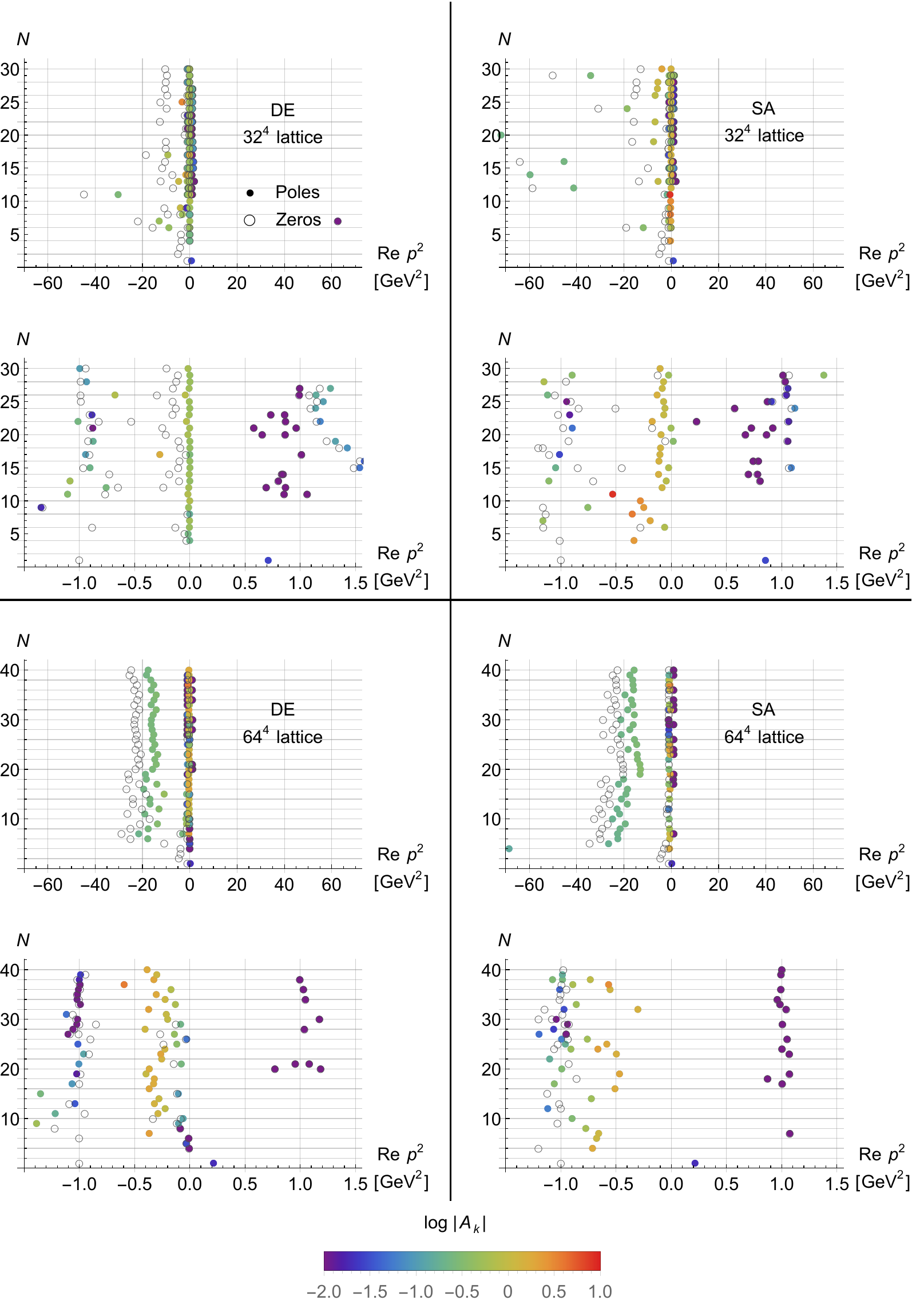}
	\caption[Evolution of the on-axis poles and zeros within the PA sequence, obtained for the gluon propagator data with $32^4$ and $64^4$ lattices, for both methods of minimisation, DE and SA.]{Evolution of the on-axis poles and zeros within the PA sequence, obtained for the gluon propagator data with $32^4$ and $64^4$ lattices, for both methods of minimisation, DE and SA. The colour scheme codes the residue's absolute value of each pole.}
	\label{fig:Glonaxis1}
\end{figure}

\begin{figure}[tp]
	\centering
	\includegraphics[width=.9\textwidth]{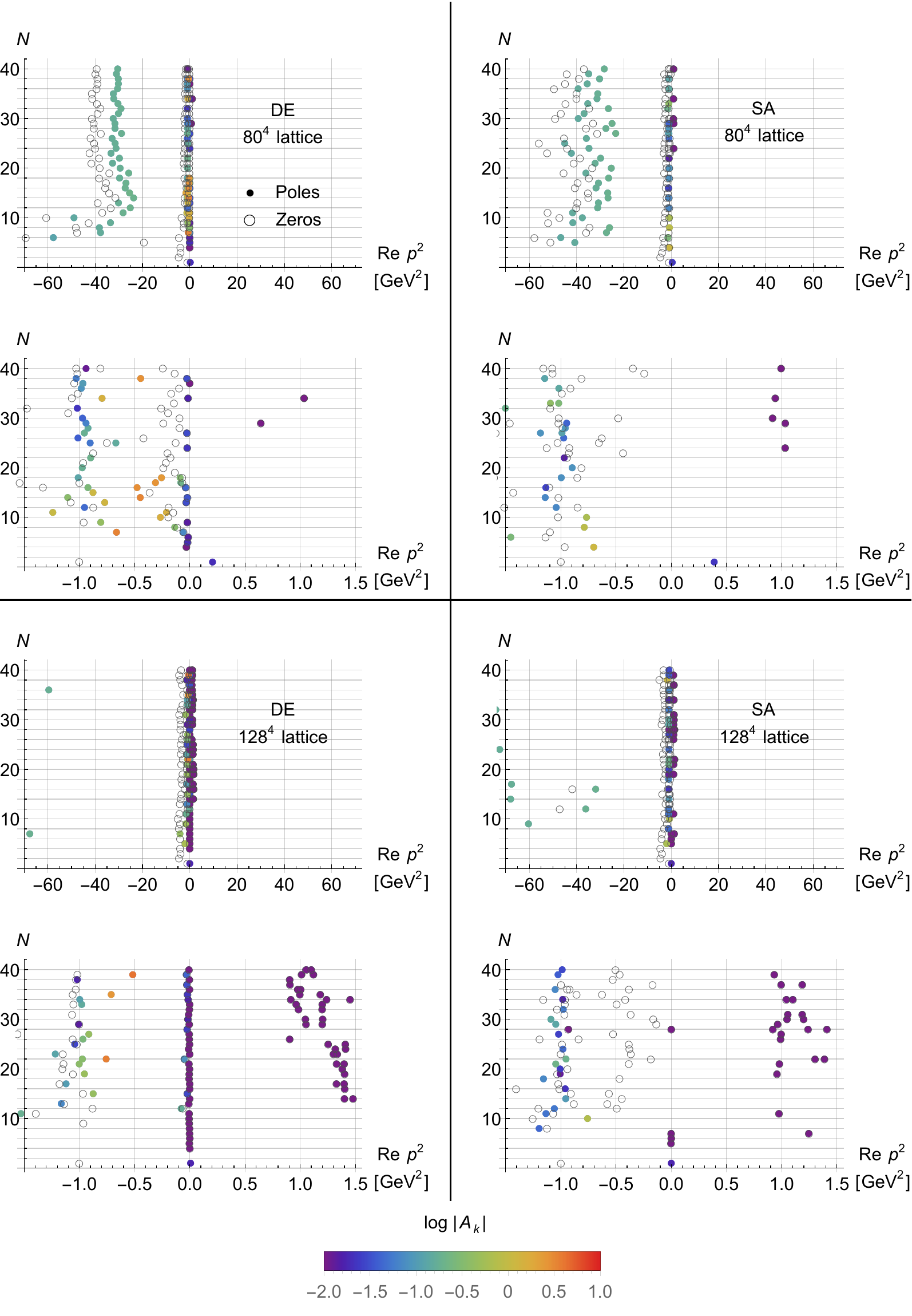}
	\caption[Evolution of the on-axis poles and zeros within the PA sequence, obtained for the gluon propagator data with $80^4$ and $128^4$ lattices, for both methods of minimisation, DE and SA.]{Evolution of the on-axis poles and zeros within the PA sequence, obtained for the gluon propagator data with $80^4$ and $128^4$ lattices, for both methods of minimisation, DE and SA. The colour scheme codes the residue's absolute value of each pole.}
	\label{fig:Glonaxis2}
\end{figure}

The evolution of the on-axis poles and zeros within the PA sequence is shown in Figures \ref{fig:Glonaxis1} and \ref{fig:Glonaxis2}, for both minimisation methods, DE and SA, applied to the four simulation data sets. In contrast to the complex poles identified in the last subsection, the presence of a branch cut on the real negative $p^2$-axis is more evident for the smaller lattices, where the alternating poles and zeros are more noticeable.

Regarding the branch point, we have concluded, in Chapter \ref{Chap:Discrete}, that its position is difficult to identify using the present methodology. Notwithstanding, we can infer, from Figures \ref{fig:Glonaxis1} and \ref{fig:Glonaxis2}, that the branch point is not at the origin, but somewhere between $p^2=0$ and $p^2=-0.5~\si{GeV^2}$. A possible way to better identify the branch cut and the branch point might be the use of a much larger ensemble of gauge configurations.

Another approach to this problem could be to use approximants inspired by the perturbative gluon propagator, for example,
\begin{equation}
D_{gl}(p^2)\approx\frac{Q_L(p^2)}{R_M(p^2)}\left[\omega \ln \frac{S_F(p^2)}{T_G(p^2)}+1\right]^{-\gamma}
\end{equation}
where $Q_L(p^2)/R_M(p^2)$ and $S_F(p^2)/T_G(p^2)$ are the usual PAs of orders $[L|M]$ and $[F|G]$, respectively. In this way, the information about the branch cut and the branch point could be accessed from the analysis of the poles and zeros of $S_F(p^2)/T_G(p^2)$. Several tests were made in the context of this work. However, the obtained results were very unstable. Thus, such results are not reported here.

Notwithstanding, the interval of momenta identified above for the position of the branch point is in agreement with the naive identification of the latter with the quoted ``gluon mass'' term found in \cite{Siringo2016} and \cite{Gracey2019}, which is $0.12~\si{GeV^2}$ and $0.36~\si{GeV^2}$ for each, respectively. Additionally, in \cite{Dudal2018}, it was obtained the mass scale of $0.216~\si{GeV^2}$. Thus, we undoubtedly see the connection between the position of the branch point and the mass scale that regularises the logarithm correction to the perturbative result. This mass scale prevents the IR logarithmic divergence of the propagator, making it finite at zero momentum.

\chapter{Conclusions and future work}
Throughout this work, we explored the use of Padé Approximants to compute the analytic structure of the Landau gauge gluon and ghost propagators. The approximants were build as a global optimisation problem that minimises a chi squared function, resorting to two different numerical minimisation methods. The PAs showed to faithfully reproduce the original functions, as well as their analytic structure. This allowed to have a first glimpse of the analytic structure of the propagators, \ie, to identify its singularities and branch cuts, without relying on a particular theoretical or empirical model to describe the lattice data.

Our methodology revealed the existence of a conjugate pair of poles in the complex $p^2$-plane, for the gluon propagator, clearly stemming from the IR structure of the theory. The presence of these complex singularities supports their connection with the non-perturbative phenomenon of confinement. Regarding the ghost propagator, a unique pole was found at $p^2=0$, in agreement with the respective perturbative result. A branch cut on the real negative $p^2$-axis was identified in the analytic structure of both propagators, with the branch points at $\text{Re}(p^2)<0$. Unfortunately, a precise value for their positions could not be achieved yet.

\vspace{1em}

In the future, it would be important to refine the methodology used here, in order to improve the sensitivity of the branch point position. Larger lattice volumes could be simulated, in order to enhance the IR structure of QCD. It would be interesting, as well, to reconstruct and explore the spectral functions for the gluon and the ghost, using the PAs obtained here for the respective propagators. Additionally, the general properties of the propagators in different gauges could be investigated by considering lattice simulations performed in other gauges.

\addcontentsline{toc}{chapter}{References}
\printbibliography[title=References]

\end{document}